\documentclass[11pt,leqno]{amsart}

\usepackage[top=2.5 cm,bottom=2 cm,left=2.5 cm,right=2 cm]{geometry}
\usepackage[utf8]{inputenc}
  \usepackage{color}

\usepackage{esint}
\usepackage{graphicx}
\usepackage[pagebackref]{hyperref}
\usepackage[normalem]{ulem}
\usepackage{amsmath}

\newtheorem*{maintheorem*}{Main Theorem}
\allowdisplaybreaks
\numberwithin{equation}{section}

\renewcommand{\i}{\ifmmode\mathit{\mathchar"7010 }\else\char"10 \fi}
\renewcommand{\j}{\ifmmode\mathit{\mathchar"7011 }\else\char"11 \fi}
\newcommand{\R}{\mathbb{R}}

\newcommand{\p}{\partial}

\def\begi{\begin{itemize}}
\def\endi{\end{itemize}}
\def\bega{\begin{array}}
\def\enda{\end{array}}

{%

\begin{enumerate}}%
{\end{enumerate}}

{%

\begin{enumerate}}%
{\end{enumerate}}
\usepackage{cleveref}
\usepackage{etoolbox}
\makeatletter
\newif\ifdavid@number
\preto\equation{\david@numberfalse}
\preto\endequation{\ifdavid@number\else\notag\fi}
\patchcmd\label@in@display{\@empty}{\@empty\david@numbertrue}{}{}
\makeatother

\begin{document}

\title{Qualitative aspects in nonlocal dynamics}

\author{G. M. Coclite}
\author{S. Dipierro}
\author{G. Fanizza}
\author{F. Maddalena}
\author{M. Romano}
\author{E. Valdinoci}

\address[Giuseppe Maria Coclite, Marzia Romano, and Francesco Maddalena]{\newline
   Dipartimento di Meccanica, Matematica e Management, Politecnico di Bari,
  Via E.~Orabona 4,I--70125 Bari, Italy.}
\email[]{giuseppemaria.coclite@poliba.it}
\email[]{marzia.romano@poliba.it}
\email[]{francesco.maddalena@poliba.it}

\address[Serena Dipierro and Enrico Valdinoci]{\newline Department of Mathematics and Statistics,
University of Western Australia, 35 Stirling Highway, WA6009 Crawley, Australia}
\email[]{serena.dipierro@uwa.edu.au}
\email[]{enrico.valdinoci@uwa.edu.au}

\address[Giuseppe Fanizza]{\newline
  Instituto de Astrofis\'\i ca e Ci\^encias do Espa\c co, Faculdade de Ci\^encias, Universidade de Lisboa,
Edificio C8, Campo Grande, P-1740-016, Lisbon, Portugal.}
\email[]{gfanizza@fc.ul.pt}

\date{\today}


\thanks{GMC, SD, FM, MR and EV are members of the Gruppo Nazionale per l'Analisi Matematica, la Probabilit\`a e le loro Applicazioni (GNAMPA) of the Istituto Nazionale di Alta Matematica (INdAM). GF is member of the Gruppo Nazionale per la Fisica Matematica (GNFM) of the Istituto Nazionale di Alta Matematica (INdAM). 
SD and EV are members of the AustMS.
GMC, FM and MR have been partially supported by the  Research Project of National Relevance ``Multiscale Innovative Materials and Structures'' granted by the Italian Ministry of Education, University and Research (MIUR Prin 2017, project code 2017J4EAYB and the Italian Ministry of Education, University and Research under the Programme Department of Excellence Legge 232/2016 (Grant No. CUP - D94I18000260001). 
GF acknowledges support by FCT under the program {\it Stimulus} with the grant no. CEECIND/04399/2017/CP1387/CT0026. EV has been supported by
the Australian Laureate Fellowship FL190100081 ``Minimal surfaces, free boundaries and partial differential equations''.}

\begin{abstract} 
In this paper we investigate, through numerical studies,  the dynamical evolutions encoded in a linear 
one-dimensional nonlocal equation arising in peridynamcs. The different propagation regimes ranging from the hyperbolic to the dispersive,  induced by the nonlocal feature of the equation, are carefully analyzed.
The study of an initial value  Riemann-like  problem suggests the formation of a singularity.
\end{abstract}

\maketitle

\section{Introduction}
In the  previous works \cite{Coclite_2018,CDFMV} some relevant mathematical features characterizing 
the dynamical evolution ruled by a  peridynamic model introduced by S.A. Silling (\cite{Sill4,Sill, Sill1, Sill2, Sill3}) have been studied in detail by focusing
on  the mathematical consistency 
(see also \cite{EP}) of this new formulation of
continuum  mechanics. The nonlocal effects related to this model
represent a very interesting problem in mathematical physics 
which has captured the interest  of many researchers during the past decades (see e.g. \cite{E1, E2, Ku, Kr, StrG1}). 

In the present paper we exploit a qualitative analysis of a one-dimensional linear peridynamical  model to 
highlight the behavior of the solutions of an initial value problem in dependence of the characteristics affecting the nonlocal properties of the equation.
More precisely, we show through numerical investigations the way nonlocality rules a wide range of modes of wave propagation (see also \cite{CFLMP, LP}).  Indeed, the dynamics predicted by the equation under study
exhibits a behavior ranging from the hyperbolic-like
to the dispersive-like propagation (\cite{Wh}), according to its
highly nontrivial  dispersive   relation.
These new phenomena, unpredictable in classical continuum mechanics (\cite{G}), are ruled by two constitutive parameters both related to the nonlocal character of the equation.

The paper is organized as follows. In Section \ref{linper} we introduce the main initial value problem in terms of a linear peridynamic model and the related dispersion relation focusing on the asymptotic limits. Morever, the important notion of group velocity ruling the transport of energy is analyzed.
In Section \ref{gaus} we study the evolution of an initial  localized profile in the range of  the main parameters 
  characterizing the problem. The shape of the initial datum is contrasted with the parameters affecting 
  nonlocality, thus revealing an entire catalogue of different interesting behaviors. In Section \ref{hyperb} we show how a suitable choice of parameters delivers the case of purely hyperbolic propagation exhibiting traveling wave solution.
  Eventually, in Section \ref{sing} we explore the possible occurrence of singularities (\cite{CDMV}),  allowed in principle for energy finite solutions. To this end, the study of a Riemann-like problem allows room for the formation of a singularity.

\section{One-dimensional peridynamic model}
\label{linper}
We perform a qualitative study regarding the phenomenology of the linear peridynamic model
\begin{equation}
\begin{cases}
\rho\,u_{tt}=-2\,\kappa
\displaystyle\int_{-\delta}^{\delta}\frac{u(t,x)-u(t,x-y)}{|y|^{1+2\,\alpha}}dy
=: K(u),&\quad t>0,\,x\in\R,\\[10pt]
u(0,x)=v_0(x),&\quad x\in\R,\\[5pt]
u_t(0,x)=v_1(x),&\quad x\in\R\,,
\end{cases}
\label{eq:linear_per}
\end{equation}
for $\alpha \in (0,1)$.

As shown in \cite{CDFMV}, this problem admits the unique formal solution
\begin{equation}
u(t,x)
=\int_{\mathbb{R}} e^{-i\xi x}\left[\widehat{v_0}(\xi)\cos\left(\omega(\xi)\,t\right)
+\frac{\widehat{v_1}(\xi)}{\omega(\xi)}\sin\left(\omega(\xi)\,t\right)\right]d\xi\,,
\label{eq:sol}
\end{equation}
where $\widehat{v_0}(\xi)$ and $\widehat{v_1}(\xi)$ are respectively the Fourier transforms of the initial conditions $v_0(x)$ and $v_1(x)$ and
\begin{equation}
\omega(\xi):=\left(\frac{2\kappa}{\rho\,\delta^{2\alpha}}\,\int_{-1}^{1}\frac{1-\cos (\xi\delta z)}{|z|^{1+2\alpha}}dz\right)^{1/2}\,,
\label{eq:disp_rel}
\end{equation}
is the dispersive relation actually governing the dynamics. Eq. \eqref{eq:disp_rel} admits solution for a wide class of initial conditions (see \cite{CDFMV} for the rigorous discussion in this regard). However, in this manuscript, we limit our phenomenological investigation to the particular case of initial conditions formally corresponding to a (possibly approximate)
traveling wave. This case is described with any loss of generality by the constraint
\begin{equation}
v_1(x)=-v\,v'_0(x)\,,
\label{eq:tr_wav}
\end{equation}
where $v$ is a number denoting the initial velocity of the perturbations
(or, in frequency space, one can recast Eq.~\eqref{eq:tr_wav}
in the form~$\widehat{v_1}(\xi)=iv\xi\widehat{v_0}(\xi)$). Eq. \eqref{eq:tr_wav} can be easily understood as the limit for $t\rightarrow 0$ of the velocity of a traveling waveform $f(x- v t)$: indeed, its velocity is just $\p_t f(x-vt)=-v\,\p_q f(q)$, where $q=x-vt$. Hence, in the above-mentioned limit, $q\rightarrow x$ and the constraint \eqref{eq:tr_wav} follows.

The physical interest in the traveling wave initial conditions stands in the comparison with the linear 1-D wave equation. Indeed, since the latter admits non-dispersive propagations, any traveling wave initial conditions will evolve as at most two independent traveling waves, according to the value of $v$. On the opposite, Eq. \eqref{eq:disp_rel} exhibits a highly non-trivial relation between $\omega$ and $\xi$. More precisely, it has been proven in \cite{CDFMV} that Eq. \eqref{eq:disp_rel} admits two characteristic behaviors discriminated by the length scale $\delta$, namely
\begin{align}
&\text{if}\quad\xi \ll 1/\delta\quad\text{then}\quad\omega\sim\xi,\\
&\text{if}\quad\xi \gg 1/\delta\quad\text{then}\quad\omega\sim\xi^\alpha\,.
\end{align}
More precisely
\begin{align*}
\lim_{\xi\to 0^+}\frac{\omega(\xi)}{\xi} =\delta^{1-\alpha}\sqrt{\frac{\kappa}{\rho\left( 1-\alpha\right)}},\\
\lim_{\xi\to \infty}\frac{\omega(\xi)}{\xi^\alpha} =\sqrt{\frac{4\kappa}{\rho}\int_0^\infty\frac{1-\cos\tau}{\tau^{1+2\alpha}}d\tau}=\sqrt{-\frac{4\kappa}{\rho}\cos(\pi\alpha)\Gamma(-2\alpha)}\,,
\end{align*}
where $\Gamma$ is the Euler gamma function. Hence, for large scale modes, the dynamics is practically non-dispersive and these modes travel with group velocity $v_g\equiv \omega'(\xi)\approx\delta^{1-\alpha}\sqrt{\frac{\kappa}{\rho\left( 1-\alpha\right)}}$. On the other hand, small scale modes experience a sub-linear dispersive dynamics with group velocity $v_g\approx \xi^{\alpha-1}\sqrt{\frac{4\kappa}{\rho}\int_0^\infty\frac{1-\cos\tau}{\tau^{1+2\alpha}}d\tau}$. According to the initial conditions, qualitative differences with respect to the classic case are expected when small scales are switched on along the dynamics. In the following section, we will explore several scenarios to point out this peculiarity.

Let us remark that $\alpha\in(0,1)$ plays a crucial role in the subsequent analysis. Indeed, besides ruling the smoothness of the admissible solutions for the problem \eqref{eq:linear_per}, it is responsible in tuning the effects of the nonlocal feature in the present theory. More precisely, the higher the $\alpha$, the more hyperbolic the propagation, whereas the lower the $\alpha$, the stronger the nonlocal effects, leading to dispersive propagation.

We conclude this section by proving for the Cauchy problem in \eqref{eq:linear_per} that
the group velocity given by
\begin{equation}
v_g:=\omega'(\xi)
\end{equation}
represents (see \cite{Bi}) the velocity of propagation of the energy density
of monochromatic waves.
To this end, we define the energy density $e$
\begin{equation}
e:=\frac{\rho}{2}|u_t|^2+W
\end{equation}
and the energy flux $G$
\begin{equation}
G:=\omega \p_\xi W\,,
\end{equation}
where $W$ is the potential energy density
\begin{equation}
W:=\frac{\kappa}{2}\int_{-\delta}^{\delta}\frac{|u(t,x)-u(t,x-y)|^2}{|y|^{1+2\alpha}}dy\,.
\end{equation}
In this setting, we 
consider a monochromatic wave
\begin{equation}
u(t,x)=Ae^{i\omega(\xi)t}e^{i\xi x}\,.
\end{equation}
Here, for simplicity, up to a phase shift, we assume that~$A\in\R$.
At a formal level, this wave would correspond to picking a~$\xi\in\R$
and choosing as initial conditions~$v_0$ such that~$\widehat{v_0}$
is the Dirac delta at $\xi$ and~$v_1$ such that~\eqref{eq:tr_wav} is satisfied.
We claim that
\begin{equation}\label{3PGSB}
v_g=\frac{G}{e}\,.
\end{equation}
To check this,
we observe that the potential energy density of the
monochromatic wave is given by
\begin{align}
W=&\,\frac{\kappa}{2}\int_{-\delta}^{\delta}\frac{|u(t,x)-u(t,x-y)|^2}{|y|^{1+2\alpha}}dy
\nonumber\\
=&\,\frac{\kappa\,A^2}{2}\int_{-\delta}^{\delta}\frac{|e^{i\xi x}-e^{i\xi(x-y)}|^2}{|y|^{1+2\alpha}}dy
\nonumber\\
=&\,\frac{\kappa\,A^2}{2}\int_{-\delta}^{\delta}\frac{|1-e^{-i\xi y}|^2}{|y|^{1+2\alpha}}dy
\nonumber\\
=&\,\frac{\kappa\,A^2}{2}\int_{-\delta}^{\delta}\frac{\left[1-\cos(\xi y)\right]^2+\sin^2(\xi y)}{|y|^{1+2\alpha}}dy
\nonumber\\
=&\,\kappa\,A^2\int_{-\delta}^{\delta}\frac{1-\cos(\xi y)}{|y|^{1+2\alpha}}dy
\nonumber\\
=&\,\kappa\,A^2\,\delta^{-2\alpha}\int_{-1}^1\frac{1-\cos(\xi\delta z)}{|z|^{1+2\alpha}}dy\nonumber
\\
=&\,\frac{\rho\,A^2}{2}\omega^2(\xi)\,.
\end{align}
On the other hand, the total energy density is given by
\begin{equation}
e=\frac{\rho}{2}|u_t|^2+W
=\rho A^2\omega^2(\xi)\,.
\end{equation}
It then follows that
\begin{equation}
\frac{G}{e} =\omega(\xi)\frac{\p_\xi W}{e}=\frac{\omega^2(\xi)\rho\,A^2\omega'(\xi)}{\rho\,A^2\omega^2(\xi)}=\omega'(\xi)\equiv v_g\,,
\end{equation}
hence the relation in~\eqref{3PGSB}
is proven. This result is interesting since it extends
the classical result for the wave equation \cite{Bi,Wh}
to the linear peridynamic model.

\section{Localized initial conditions}
\label{gaus}
In this section we study the dispersive phenomena of the linear peridynamic model in the range of parameters of $\alpha$ and $\delta$.
As a first step, we fix our constitutive parameters as $\rho=1$ and $\kappa=1/2$ and leave them unchanged for the rest of the discussion. Aiming to study the traveling waves, we then consider Gaussian initial conditions
\begin{equation}
v_0(x)=\sqrt{2\pi}\,e^{-\frac{x^2}{2\sigma^2}}\qquad\text{and}\qquad v_1(x)=-v\,v_0(x)\,.
\label{eq:ini}
\end{equation}
The interesting feature of those initial conditions relies in their localization close to the origin. Indeed, we will provide numerical evidences that these localized initial conditions remain localized as time evolves. This suggests that the evolution experiences almost finite propagation velocity. From now on, the key point to have in mind will be always the relative amplitude between the two parameters $\delta$ and group velocity $v_g$ characterizing the nonlocality of~\eqref{eq:linear_per} and the two
parameters $\sigma$ and $v$
involved in the initial conditions.

The rest of the section is divided into three parts, devoted to the analysis of the interplay between the nonlocal range $\delta$ and the amplitude of the initial condition $\sigma$, since their ratio plays a crucial role in occurrence of dispersive effects in the wave propagation. More precisely, when $\sigma$ is much smaller than $\delta$, the solution develops high oscillations typical of the dispersive regime almost immediately, showing the deep difference between the peridynamics and the wave equation. When $\sigma$ is of the same order of $\delta$ we capture the strong transport effect in short times and the occurrence of dispersion on the long range. Finally, when $\sigma$ is much bigger than $\delta$, the dynamics experiences the (almost)
total absence of dispersion and the occurrence only of hyperbolic propagation, typical of the wave equation.

\subsection{Case~$\boldsymbol{\sigma\ll\delta}$} 
\begin{figure}[ht!]
\centering
\includegraphics[scale=0.4]{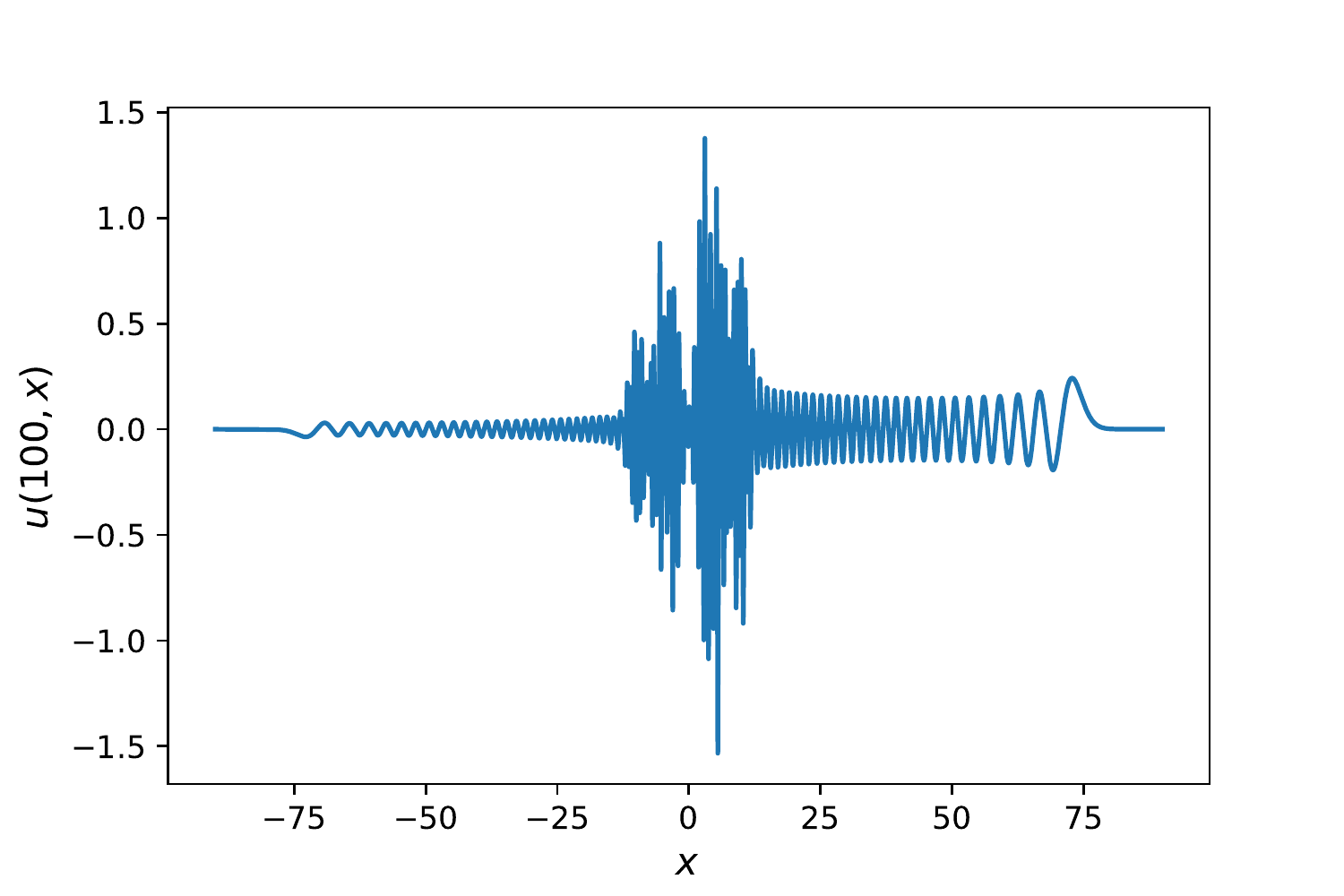}
\includegraphics[scale=0.4]{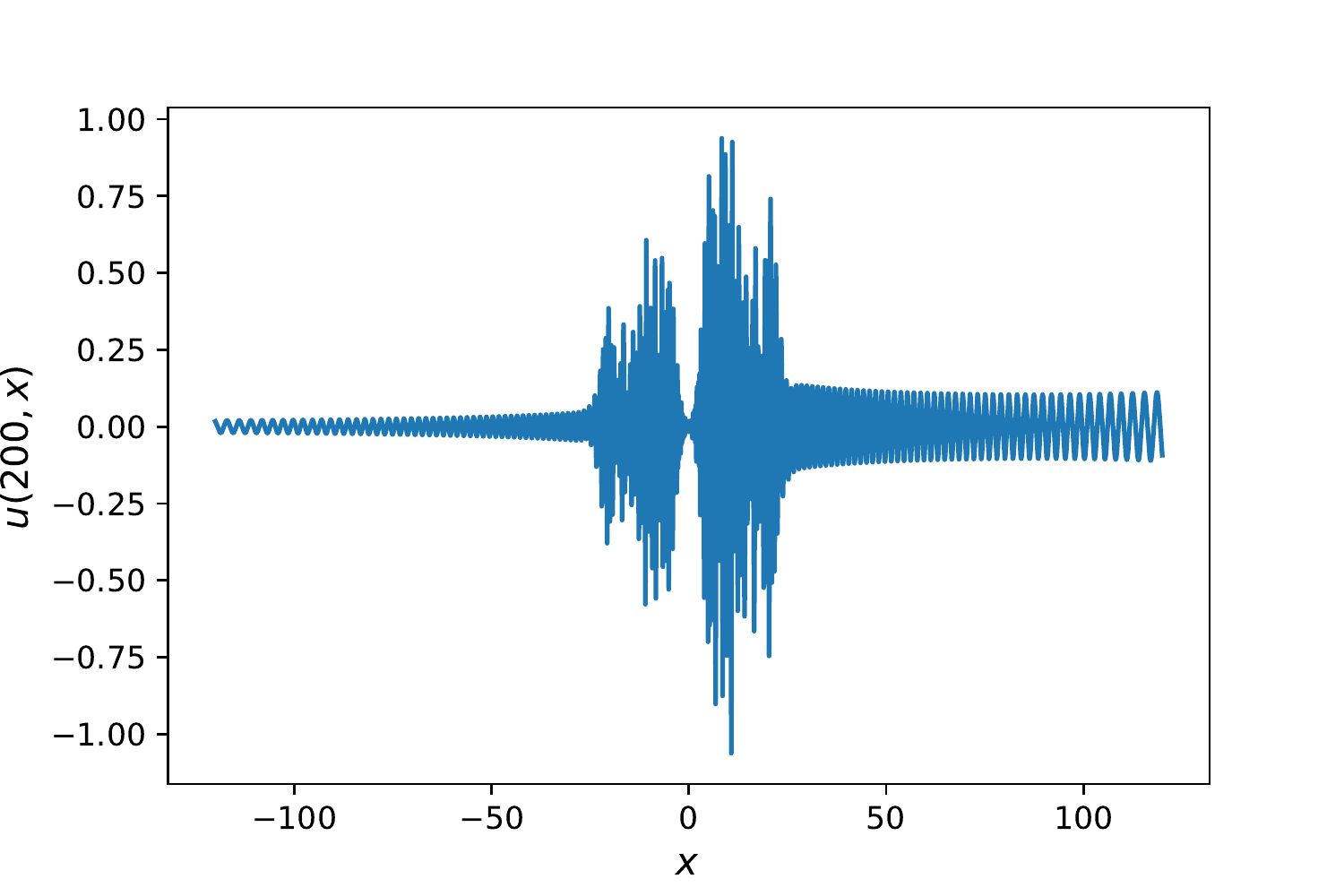}
\caption{Numerical simulations of \eqref{eq:linear_per}-\eqref{eq:ini} at different times in correspondence of $\sigma=10^{-1}< 1=\delta$, $v=1$ and $\alpha=10^{-1}$.}
\label{fig:1}
\end{figure}

\begin{figure}[ht!]
\centering
\includegraphics[scale=0.4]{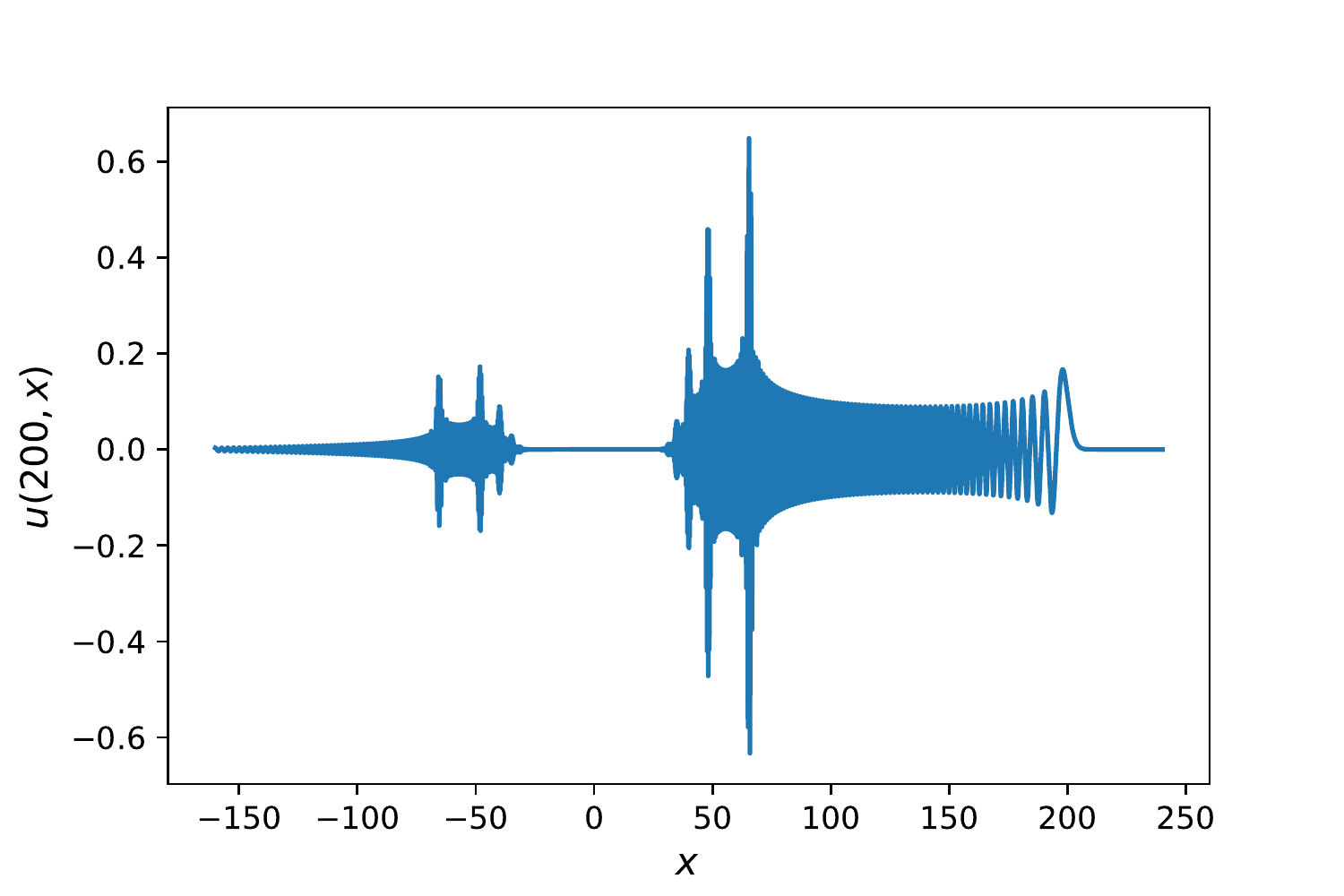}
\caption{Numerical simulations of \eqref{eq:linear_per}-\eqref{eq:ini} at different times in correspondence of $\sigma=10^{-1}<1=\delta$, $v=1$ and $\alpha=1/2$.}
\label{fig:2}
\end{figure}

\begin{figure}[ht!]
\centering
\includegraphics[scale=0.4]{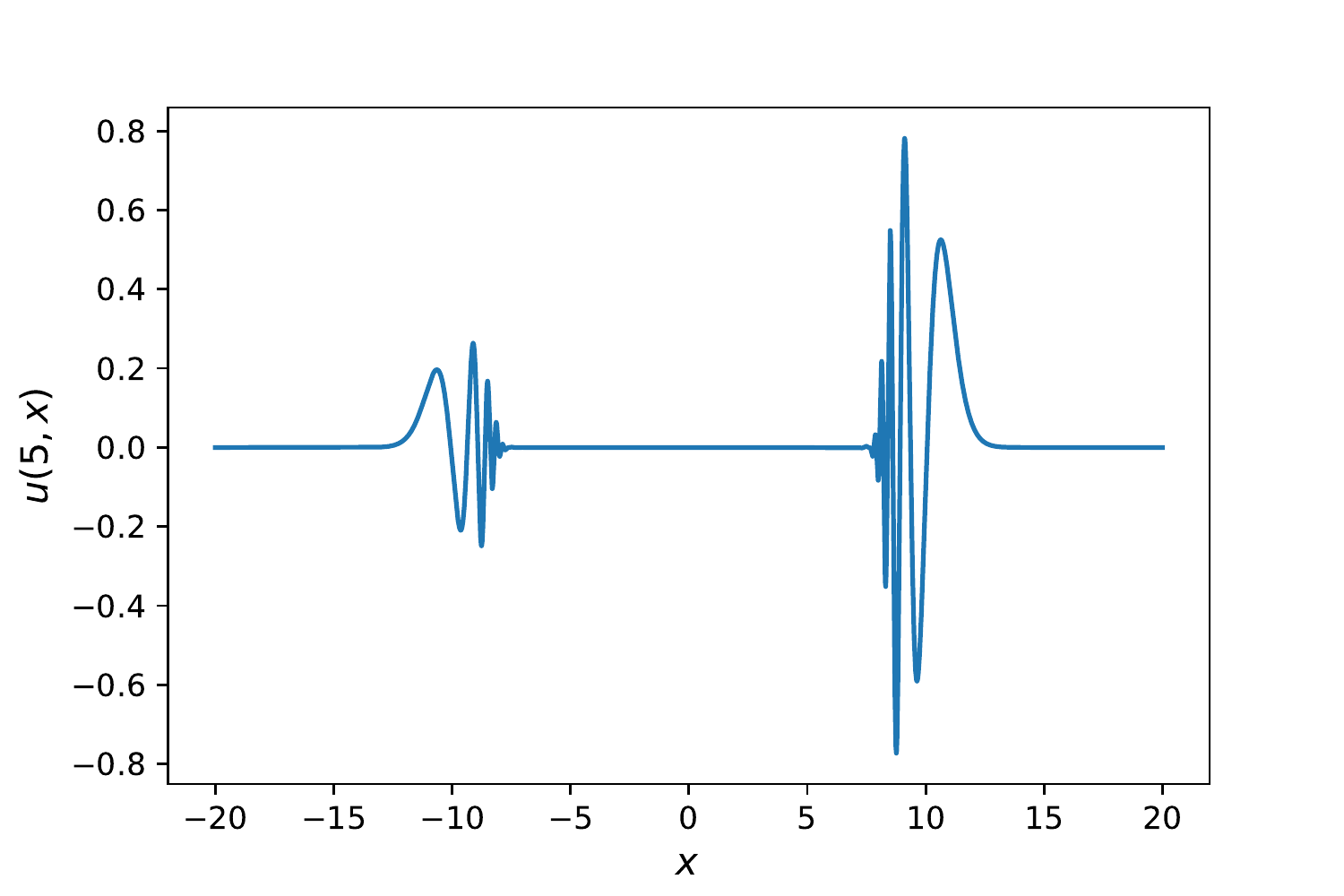}
\includegraphics[scale=0.4]{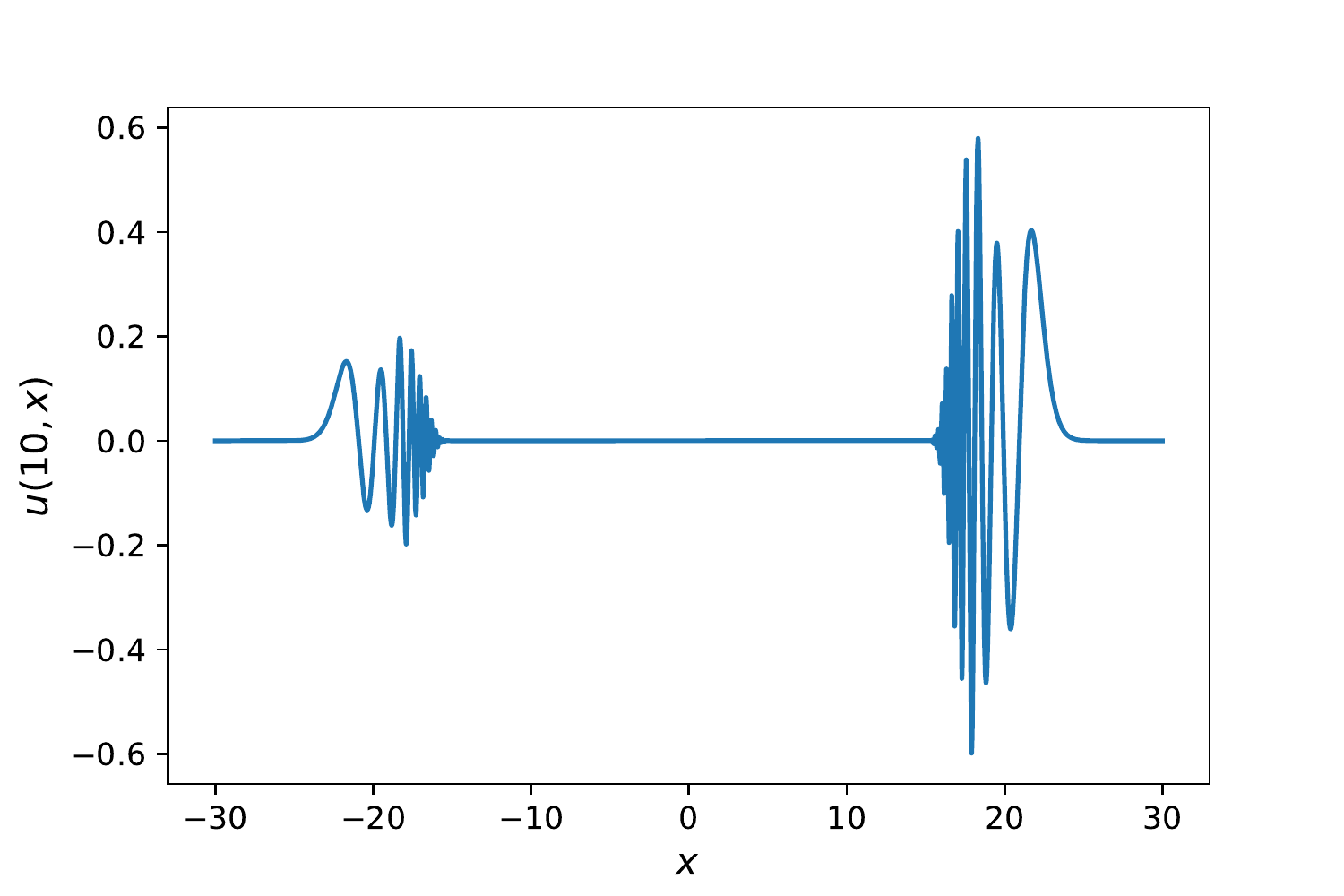}
\caption{Numerical simulations of \eqref{eq:linear_per}-\eqref{eq:ini} at different times in correspondence of $\sigma=10^{-1}<1=\delta$, $v=1$ and $\alpha=9/10$.}
\label{fig:3}
\end{figure}

This subsection is devoted to the choice
\begin{equation}
\sigma=10^{-1}\quad\text{and}\quad\delta=1.
\end{equation}
In order to support our claim about the immediate occurrence of dispersive behavior, independently of the choice of $\alpha$, we show different cases regarding three possible choices of $\alpha$.
\newpage
In Fig. \ref{fig:1}, we consider
\begin{equation}
\alpha=10^{-1}.
\end{equation}
\begin{figure}[ht!]
\centering
\includegraphics[scale=0.4]{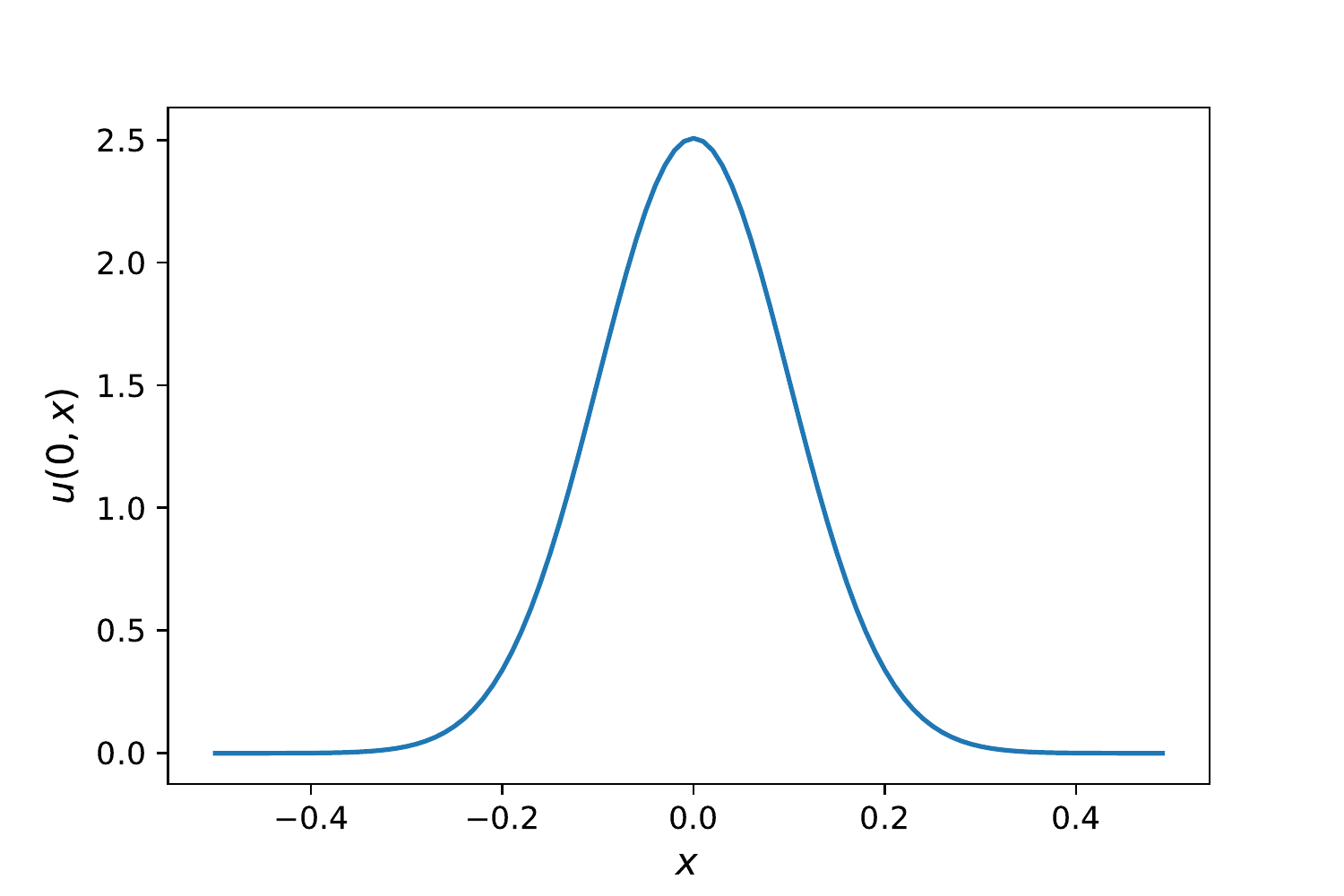}
\includegraphics[scale=0.4]{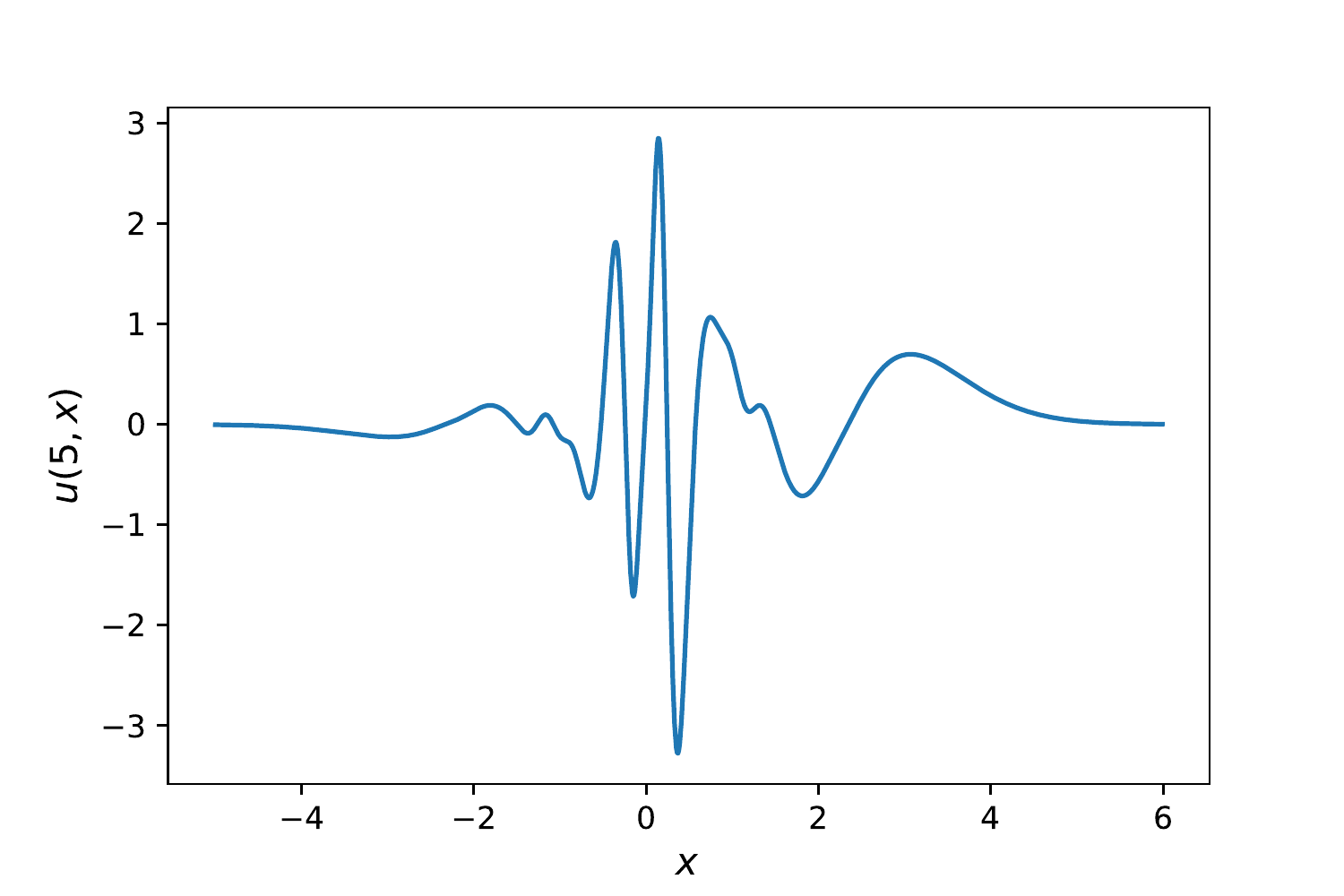}
\end{figure}
\begin{figure}[ht!]
\centering
\includegraphics[scale=0.4]{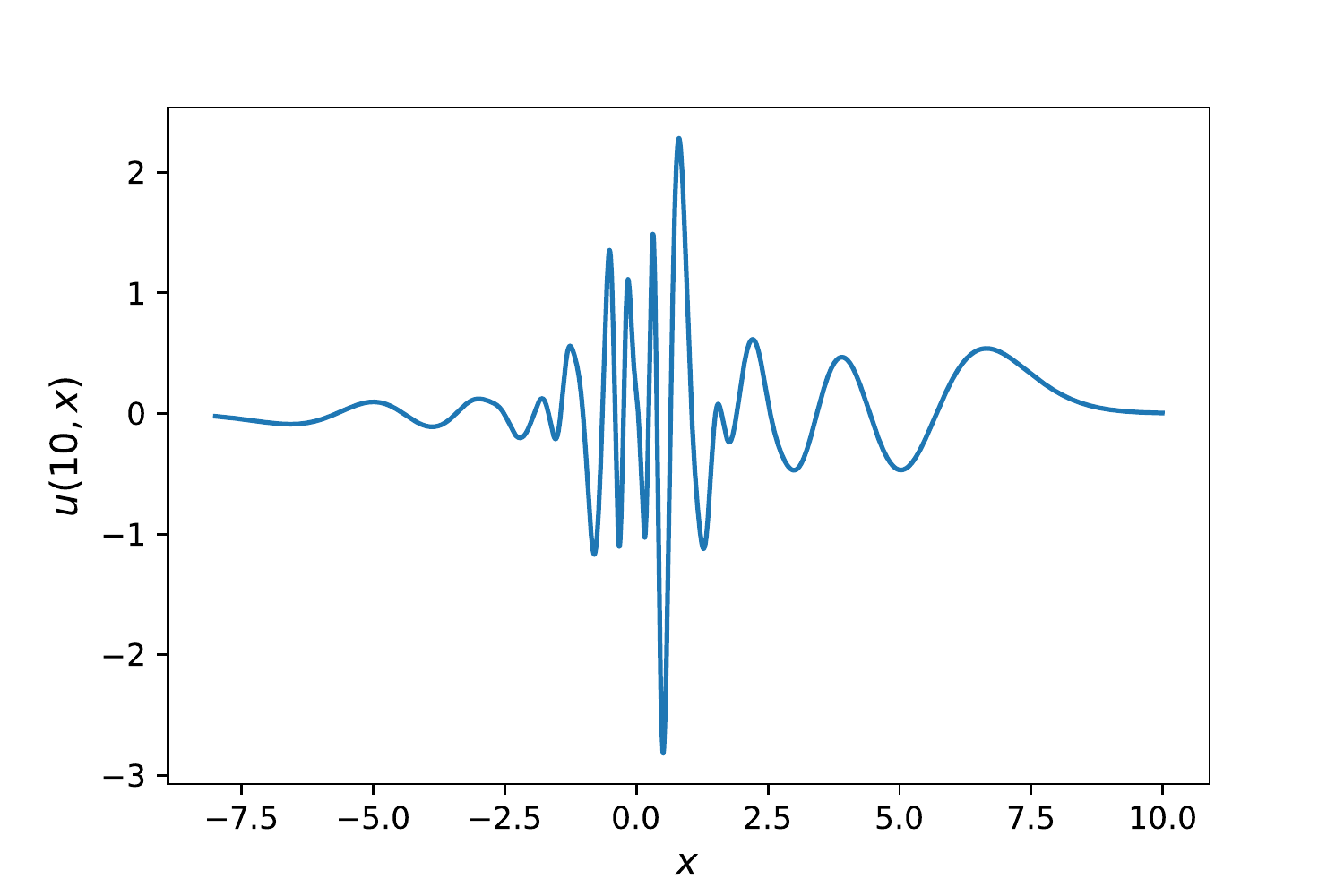}
\includegraphics[scale=0.4]{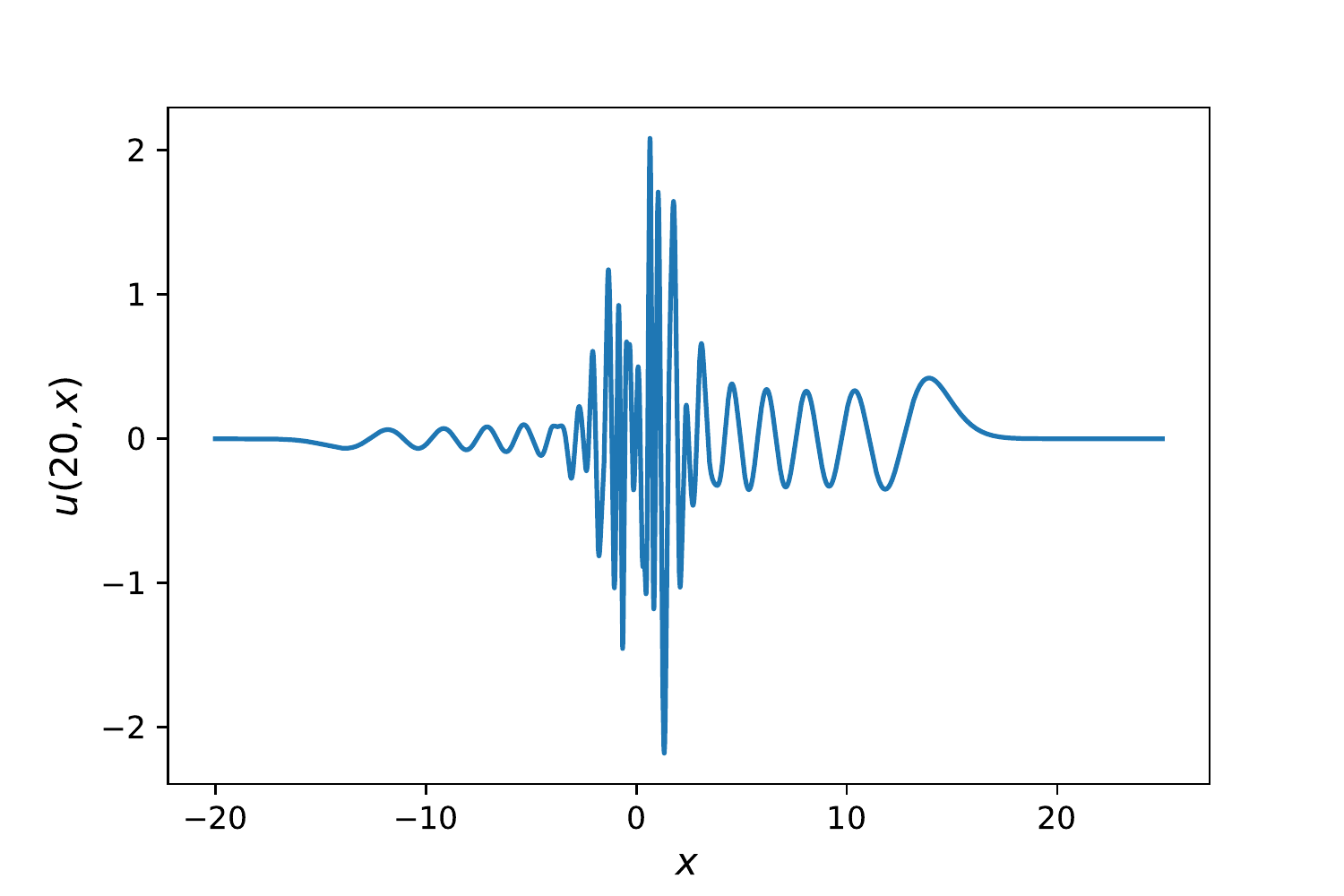}
\end{figure}
\begin{figure}[ht!]
\centering
\includegraphics[scale=0.4]{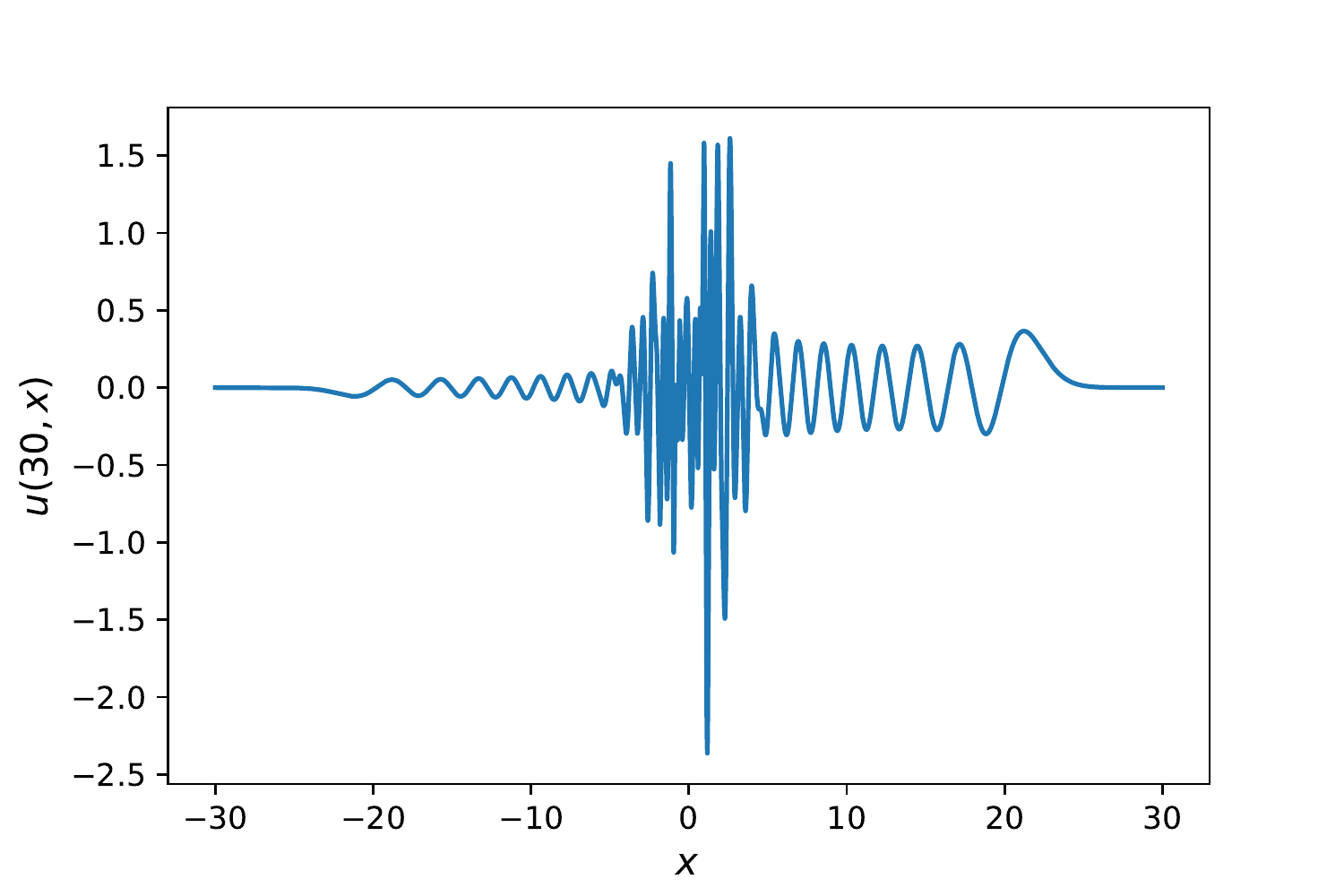}
\includegraphics[scale=0.4]{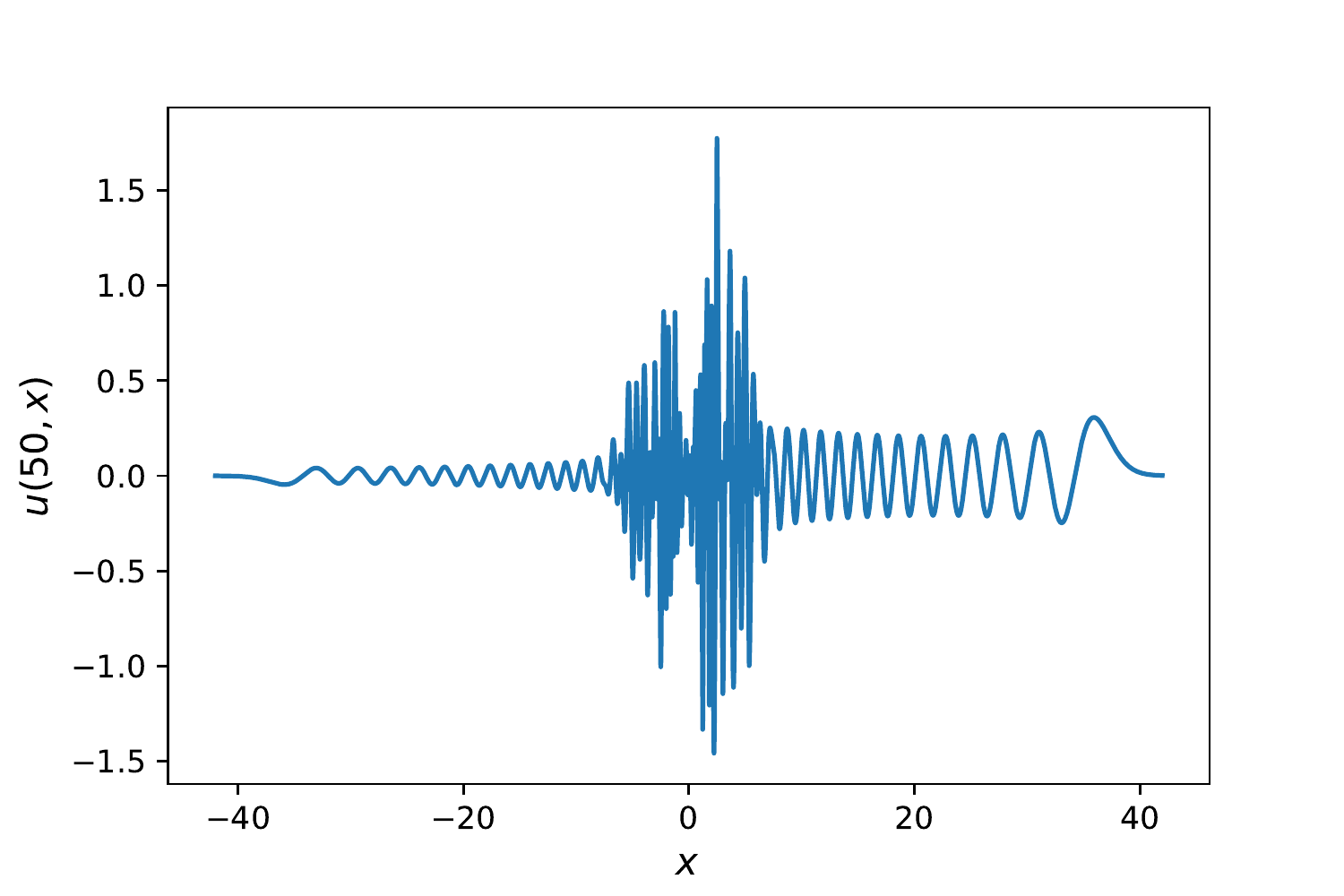}
\end{figure}
\newpage
In Fig. \ref{fig:2}
we consider
\begin{equation}
\alpha=1/2.
\end{equation}
\begin{figure}[ht!]
\centering
\includegraphics[scale=0.4]{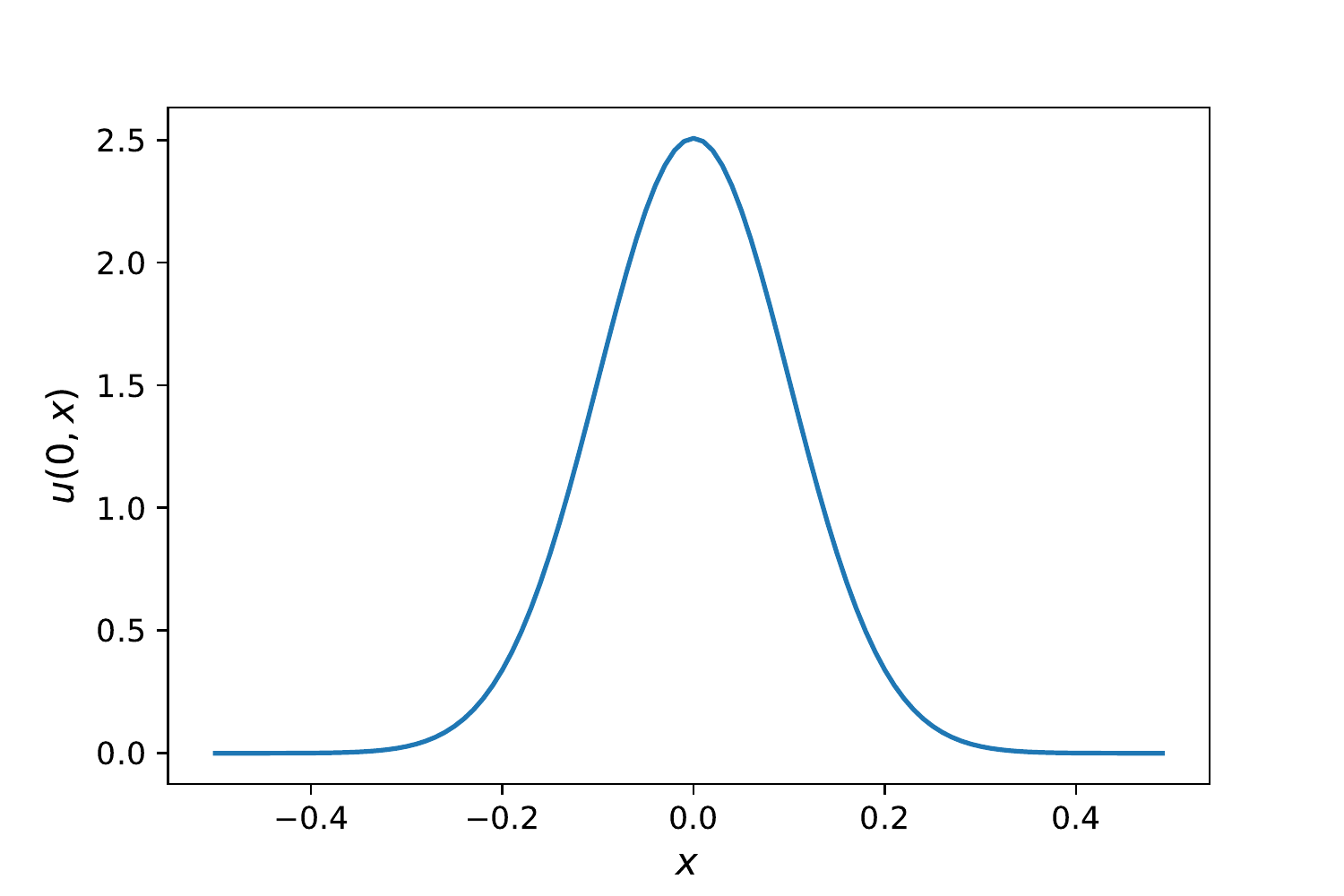}
\includegraphics[scale=0.4]{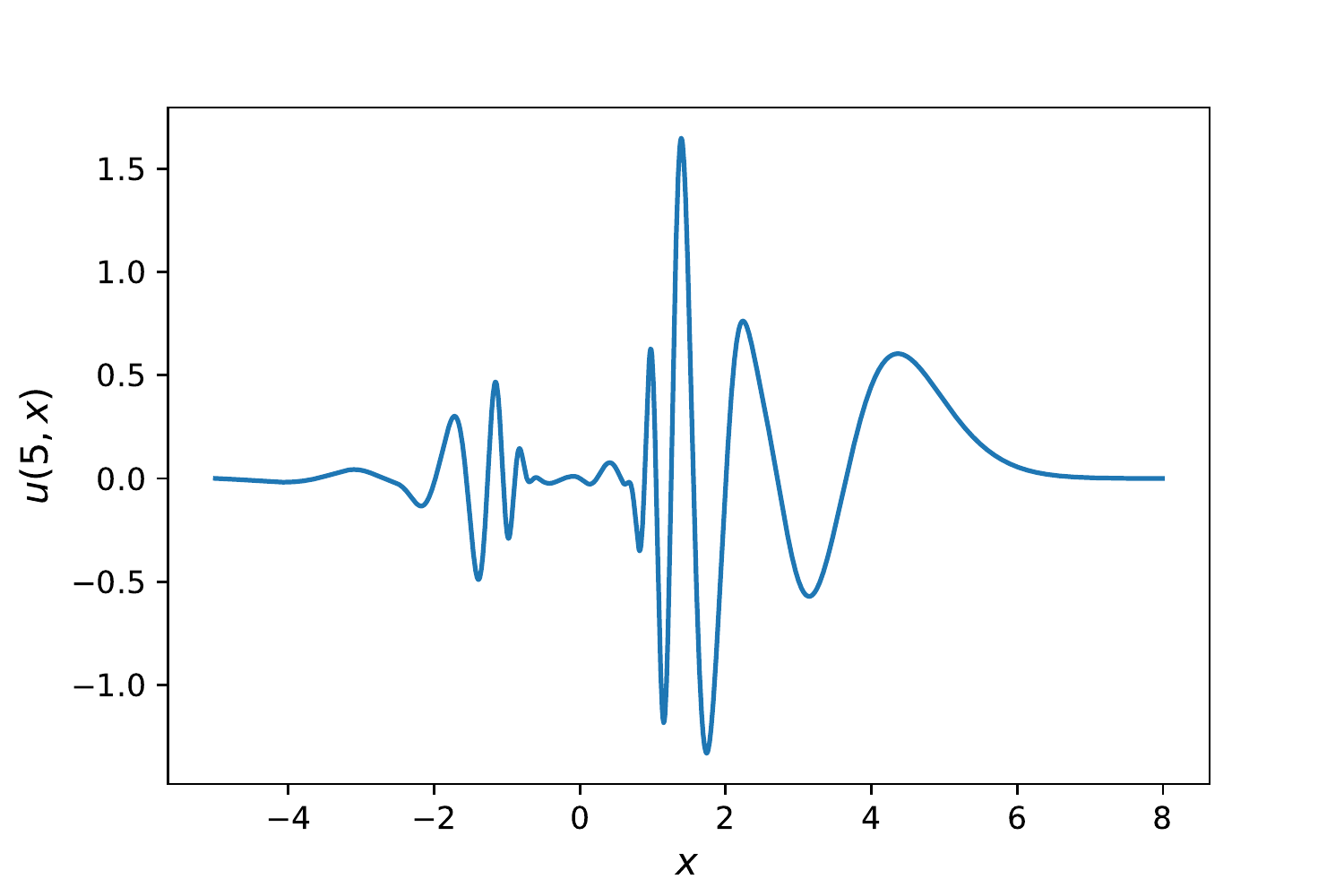}
\end{figure}
\begin{figure}[ht!]
\centering
\includegraphics[scale=0.4]{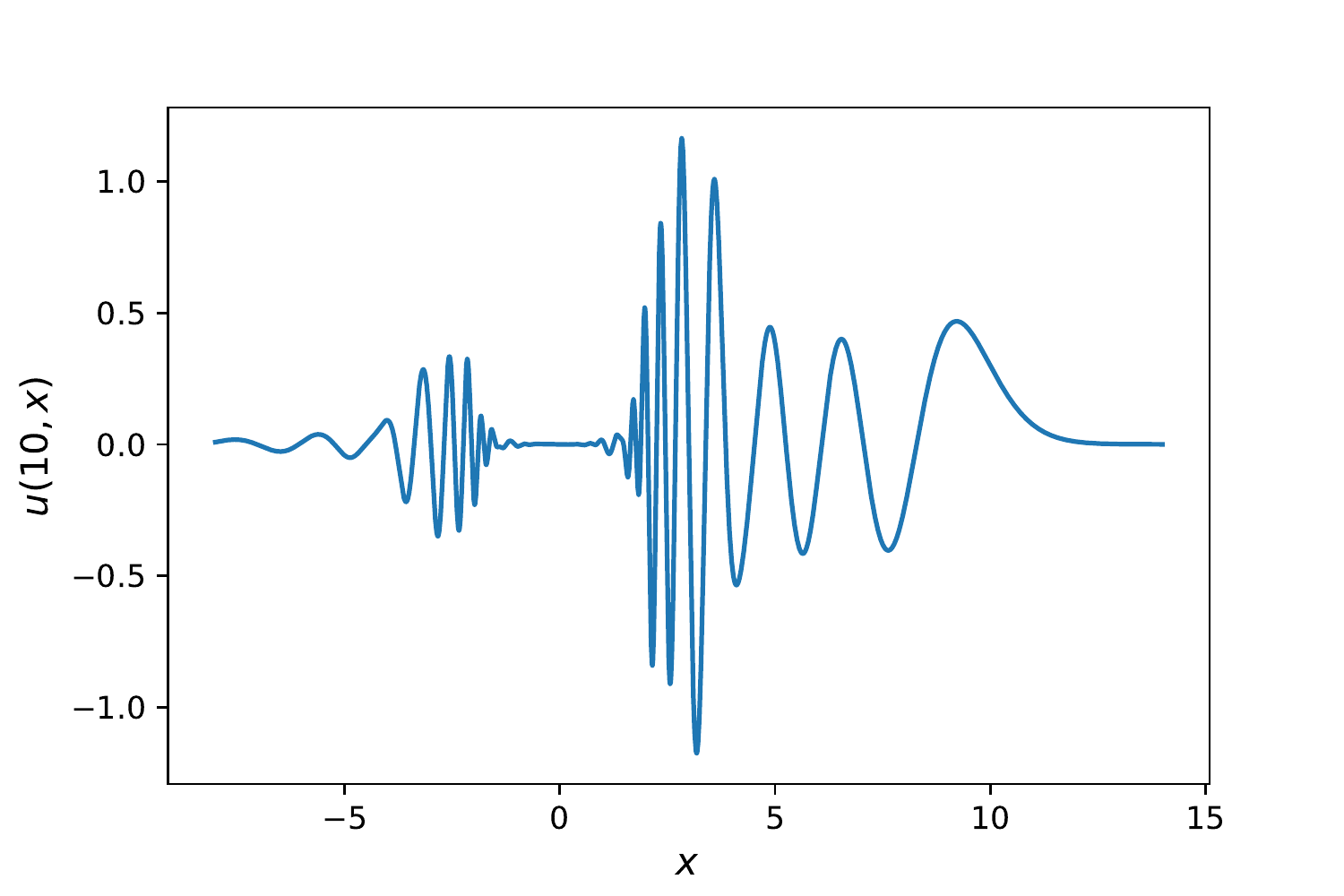}
\includegraphics[scale=0.4]{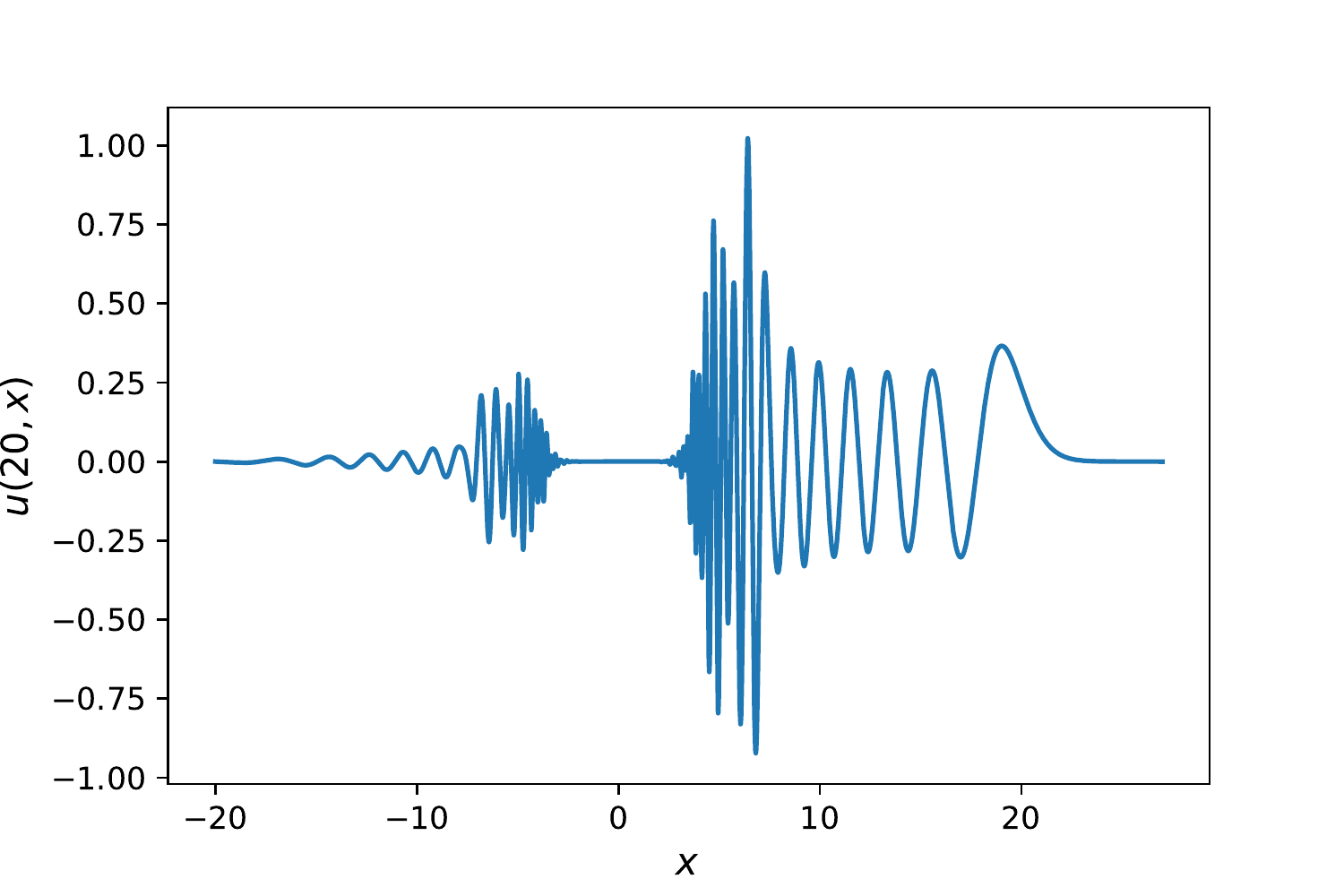}
\end{figure}
\begin{figure}[ht!]
\centering
\includegraphics[scale=0.4]{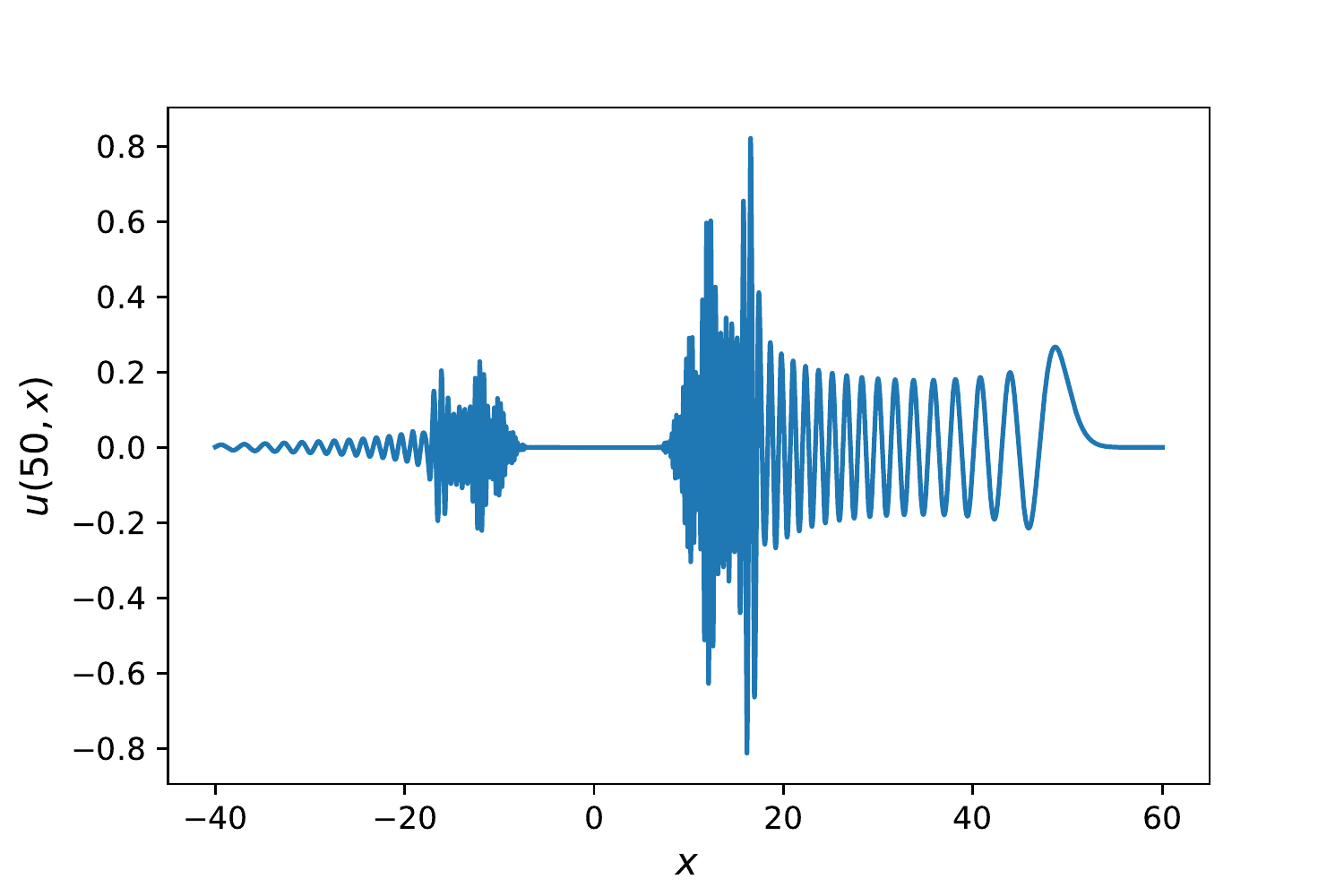}
\includegraphics[scale=0.4]{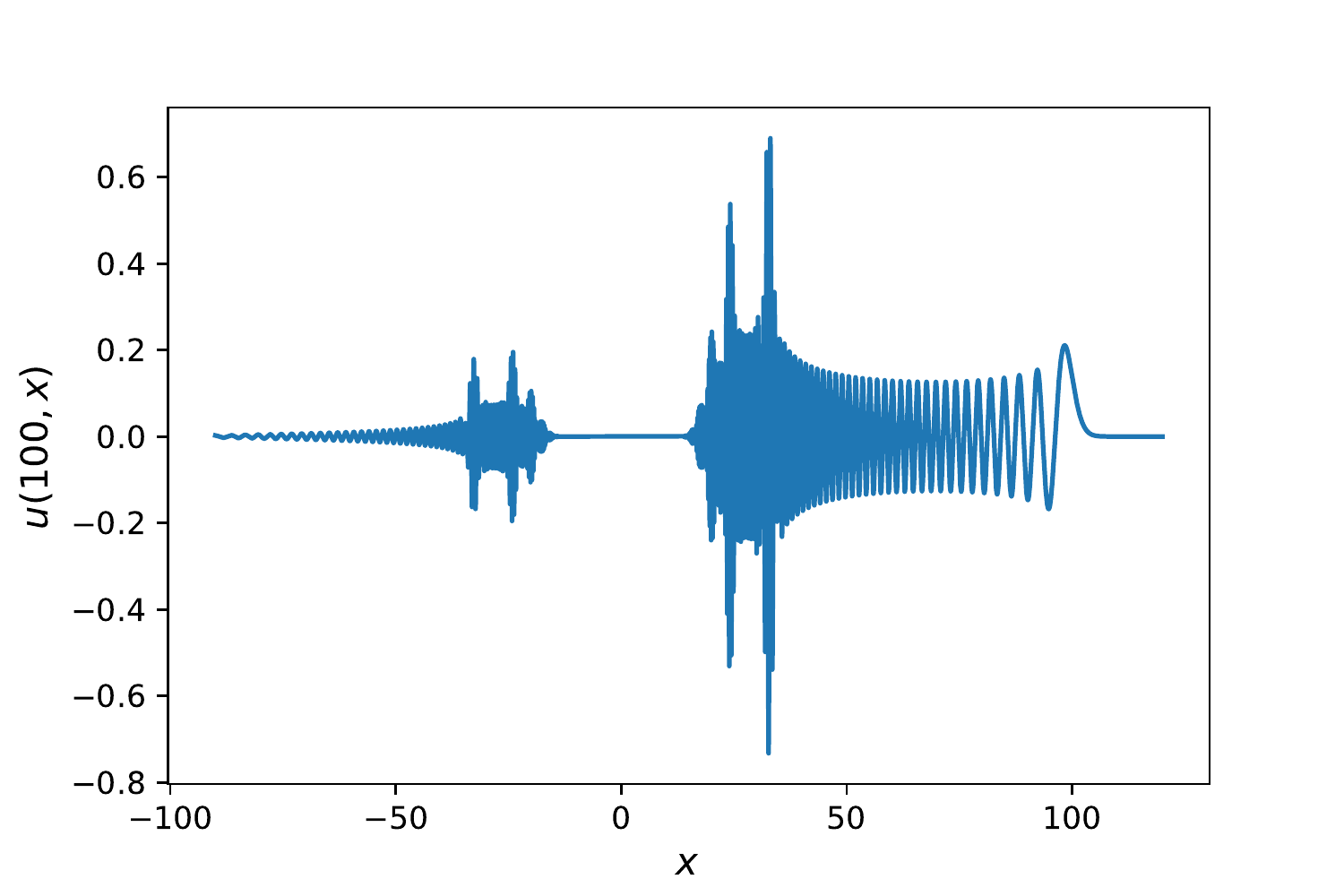}
\end{figure}
\newpage
In Fig. \ref{fig:3}
we consider
\begin{equation}
\alpha=9/10.
\end{equation}
\begin{figure}[ht!]
\centering
\includegraphics[scale=0.4]{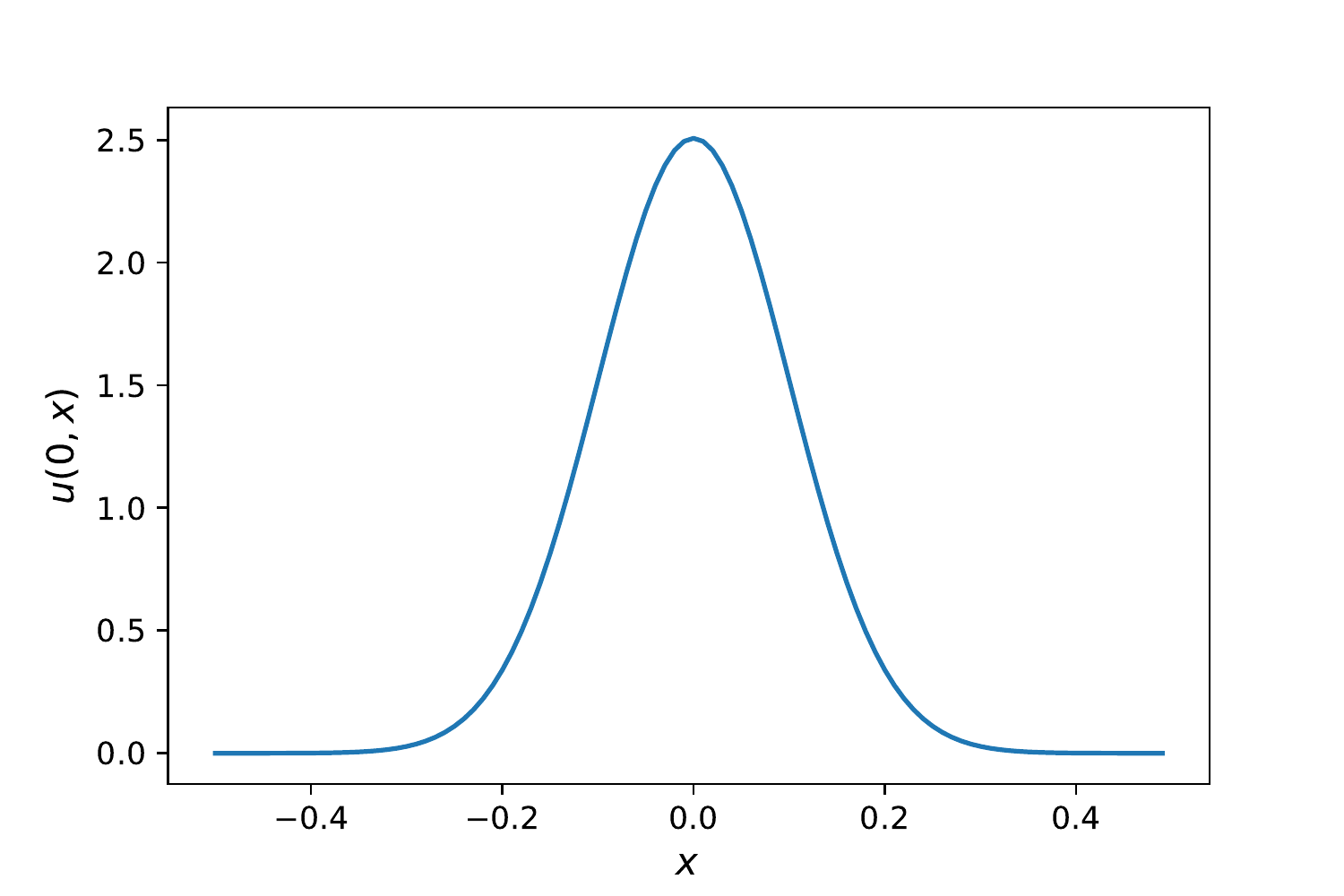}
\includegraphics[scale=0.4]{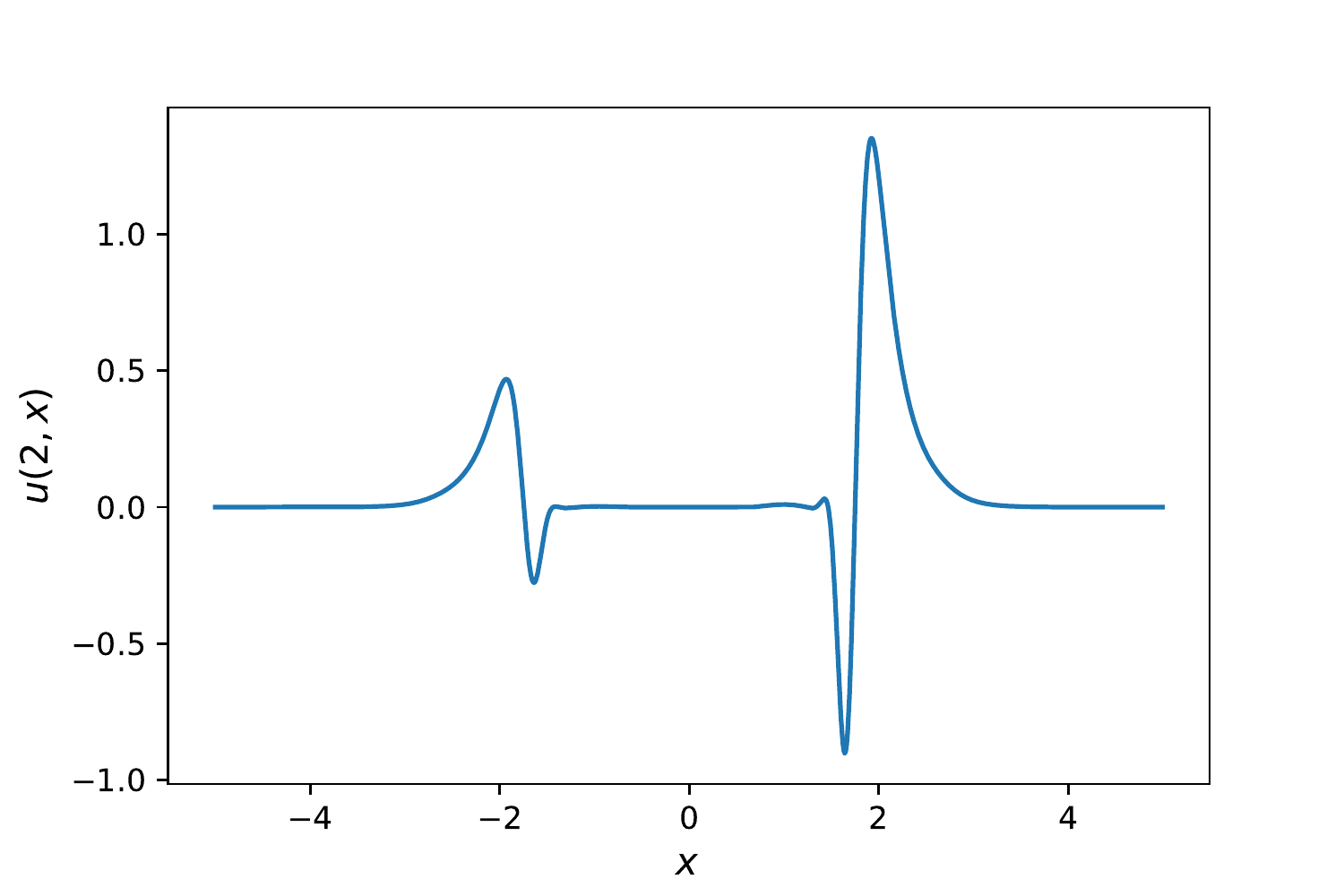}
\end{figure}

\subsection{Case~$\boldsymbol{\sigma\sim\delta}$} 
\begin{figure}[ht!]
\centering
\includegraphics[scale=0.4]{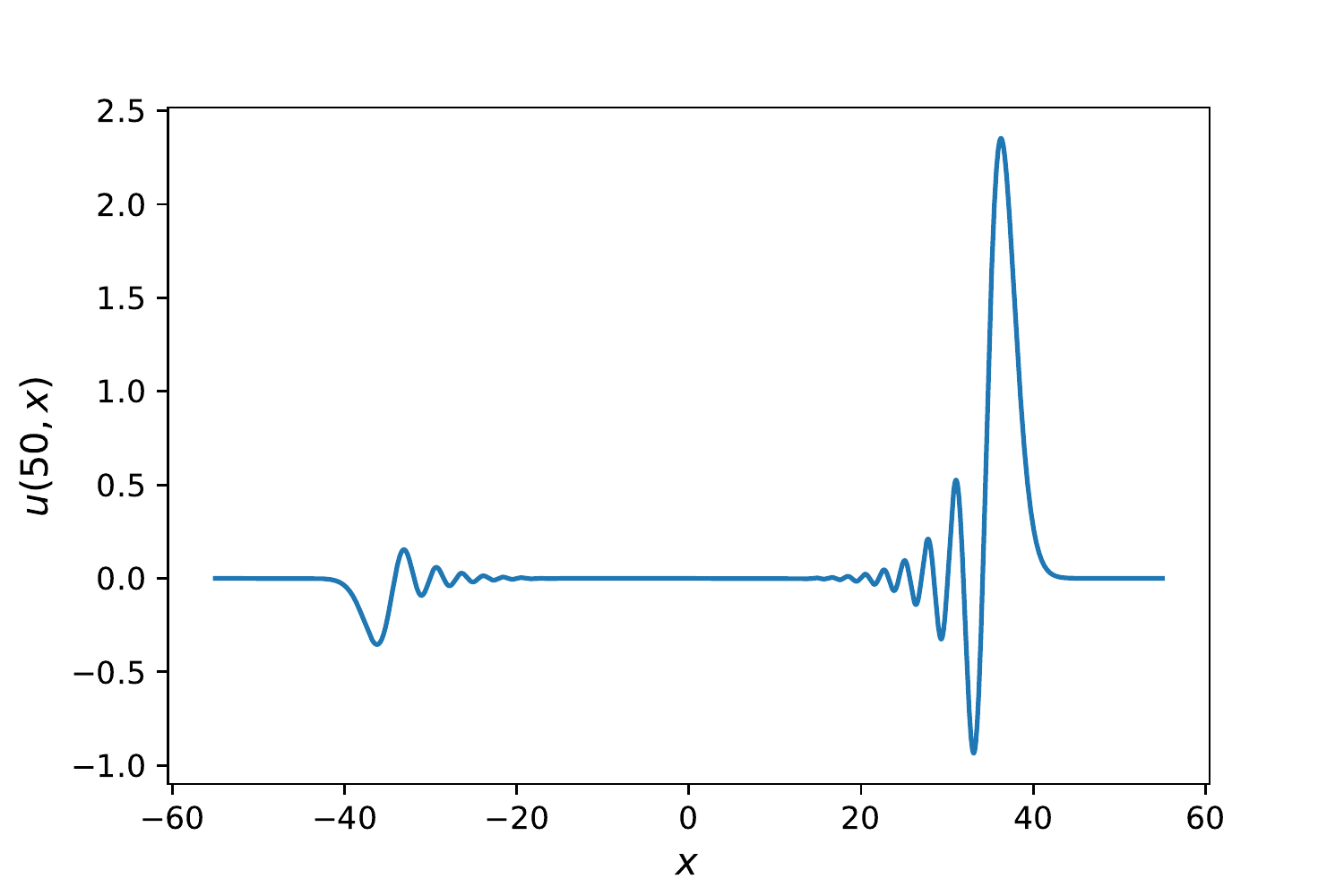}
\caption{Numerical simulations of \eqref{eq:linear_per}-\eqref{eq:ini} at different times in correspondence of $\sigma= \delta = 1$, $v=1$ and $\alpha=10^{-1}$.}
\label{fig:4}
\end{figure}

\begin{figure}[ht!]
\centering
\includegraphics[scale=0.4]{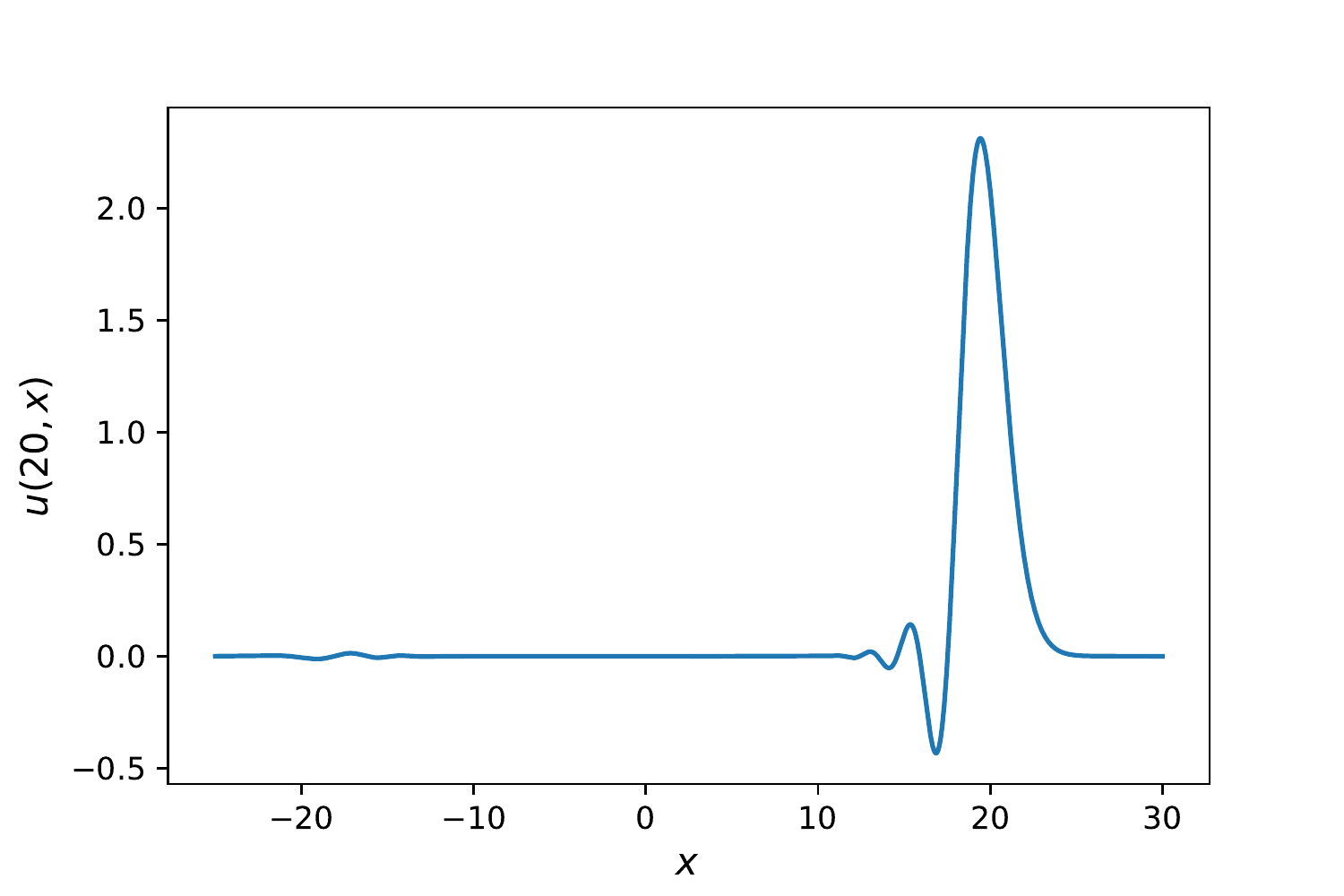}
\includegraphics[scale=0.4]{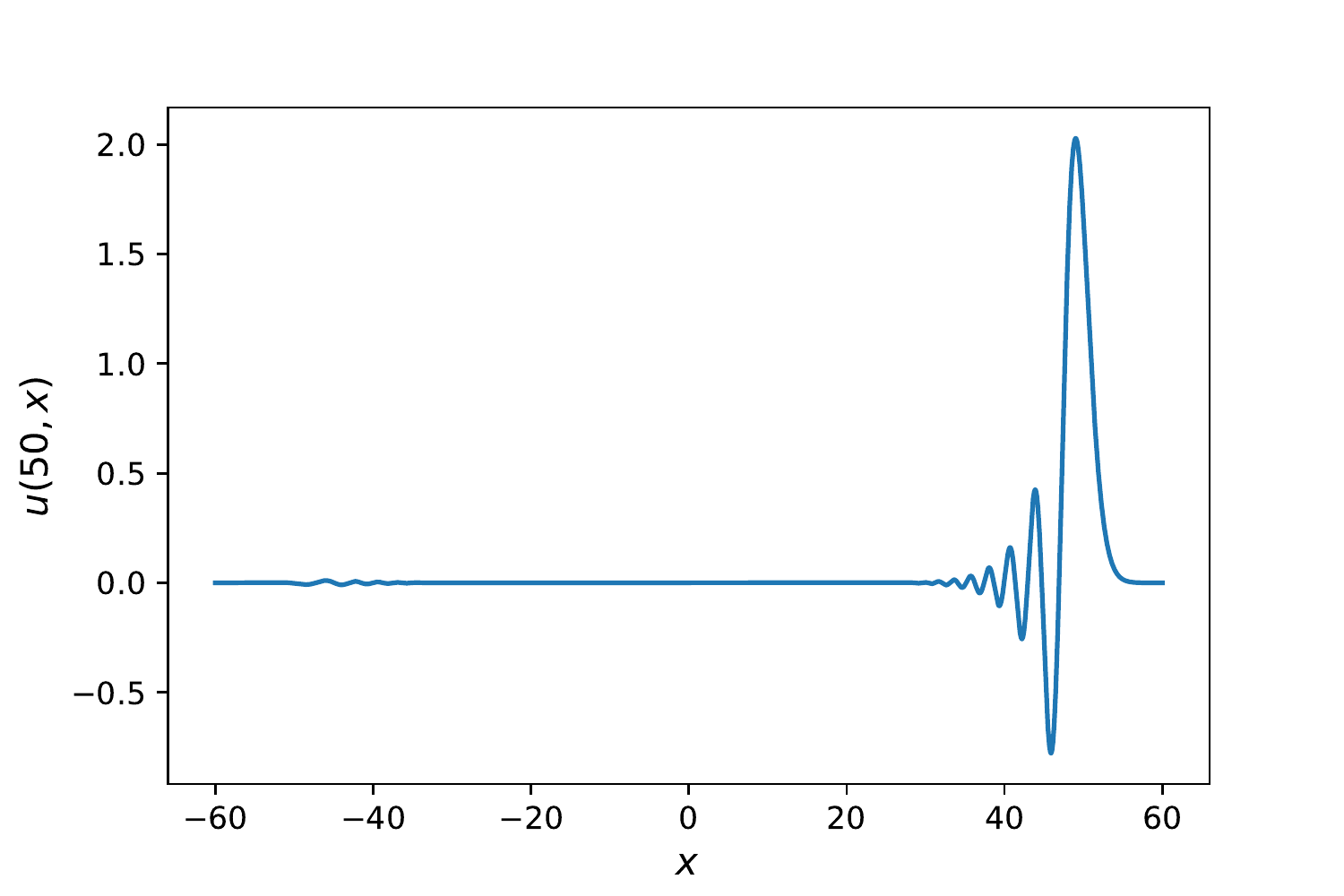}
\caption{Numerical simulations of \eqref{eq:linear_per}-\eqref{eq:ini} at different times in correspondence of $\sigma=\delta=1$, $v=1$ and $\alpha=1/2$.}
\label{fig:5}
\end{figure}

\begin{figure}[ht!]
\centering
\includegraphics[scale=0.4]{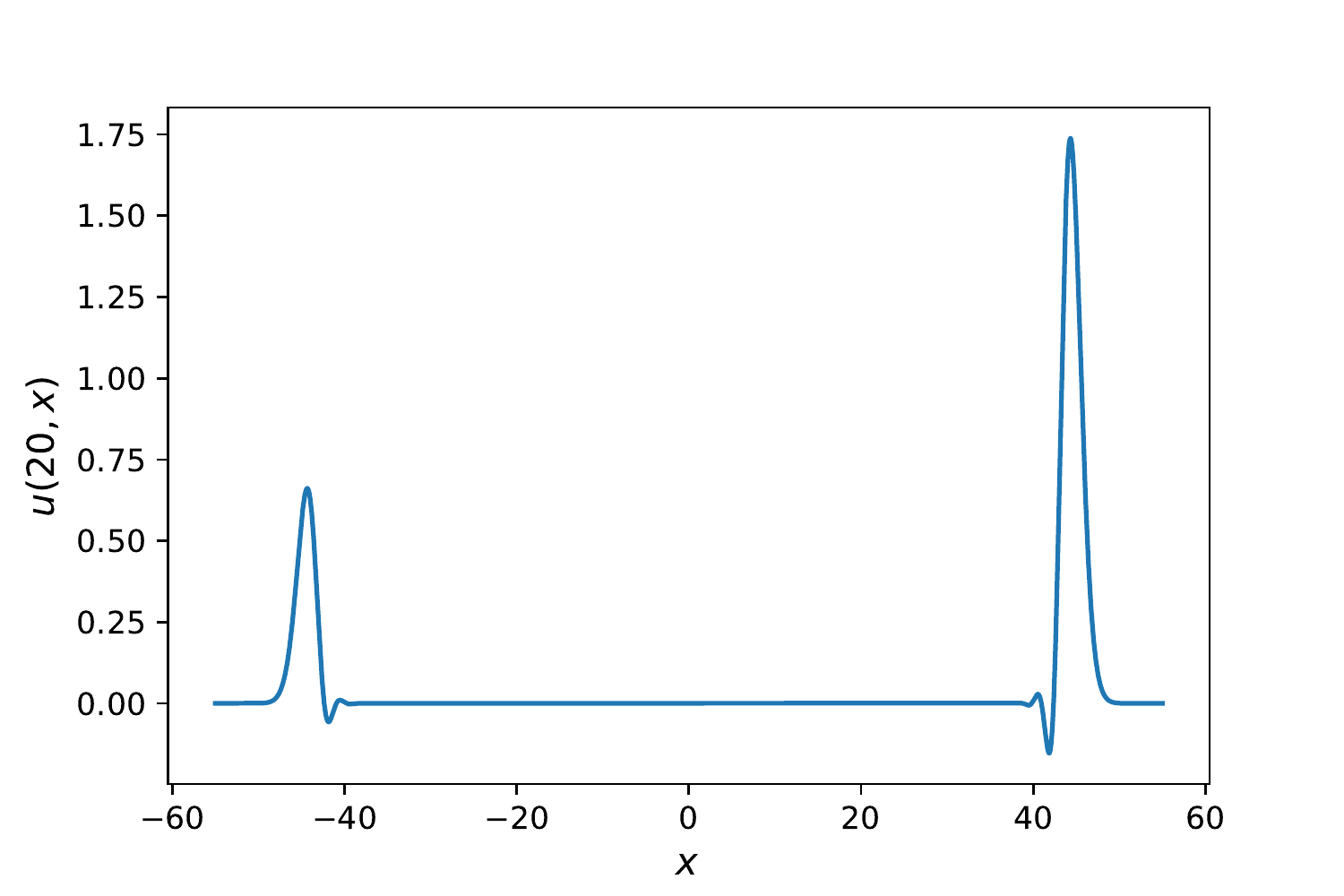}
\includegraphics[scale=0.4]{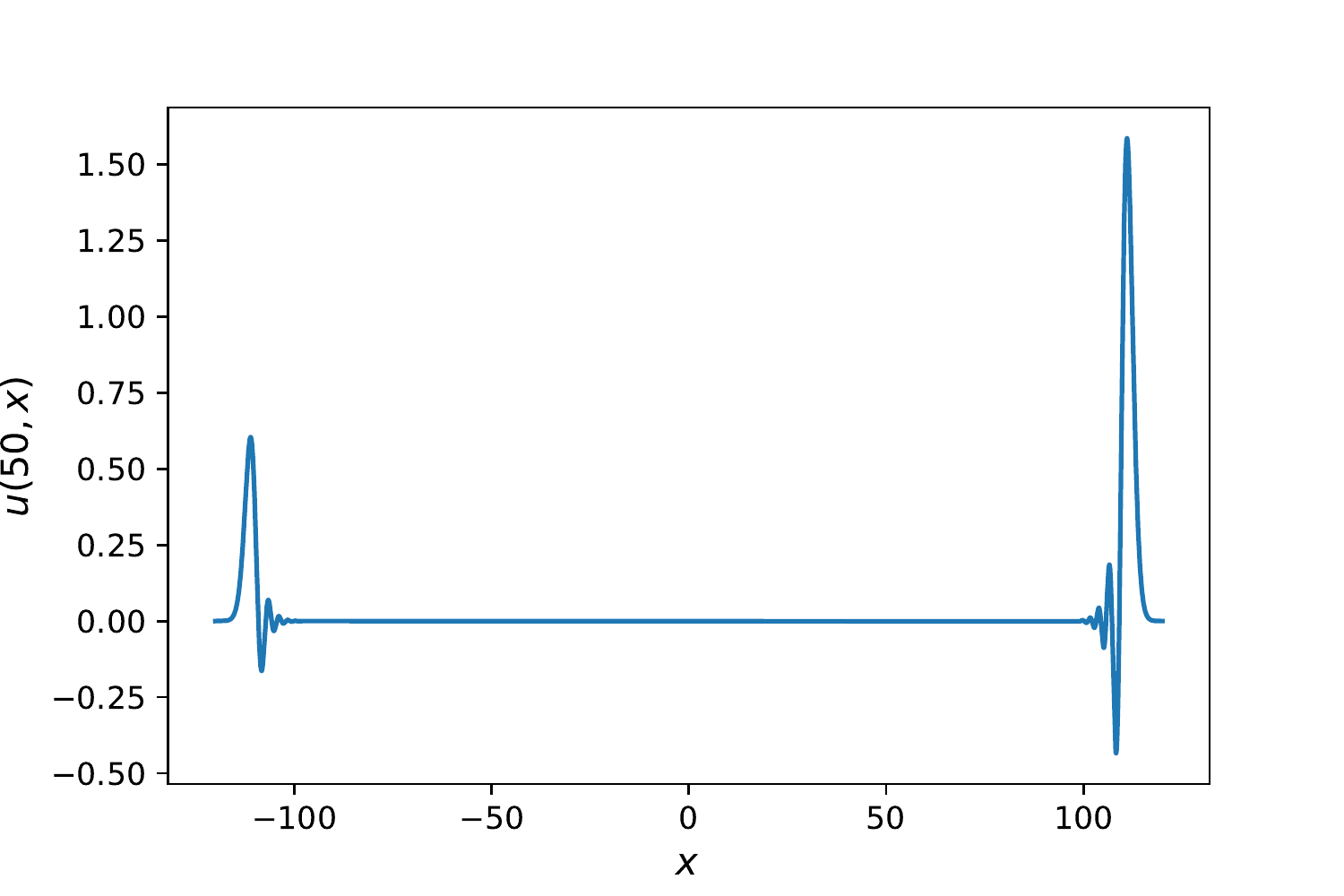}
\caption{Numerical simulations of \eqref{eq:linear_per}-\eqref{eq:ini} at different times in correspondence of $\sigma=\delta=1$, $v=1$ and $\alpha=9/10$.}
\label{fig:6}
\end{figure}

This subsection is devoted to the choice
\begin{equation}
\sigma=1\quad\text{and}\quad\delta=1.
\end{equation}
In order to support our claim about the competitive occurrence of dispersive behavior and hyperbolic propagation, according of the value of $\alpha$, we show different cases regarding three possible choices of $\alpha$.
\newpage
In Fig. \ref{fig:4}, we consider
\begin{equation}
\alpha=10^{-1}.
\end{equation}
\begin{figure}[ht!]
\centering
\includegraphics[scale=0.4]{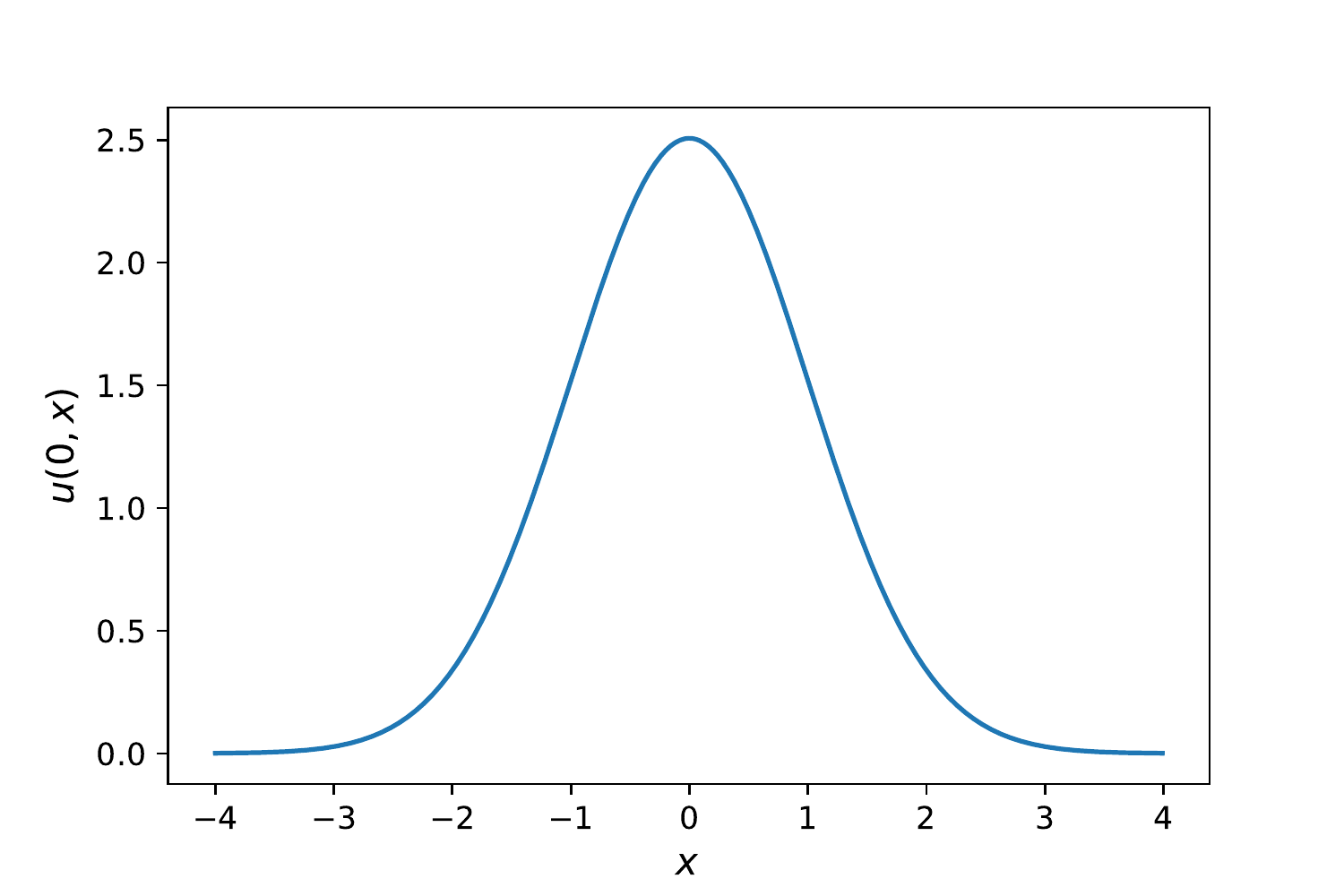}
\includegraphics[scale=0.4]{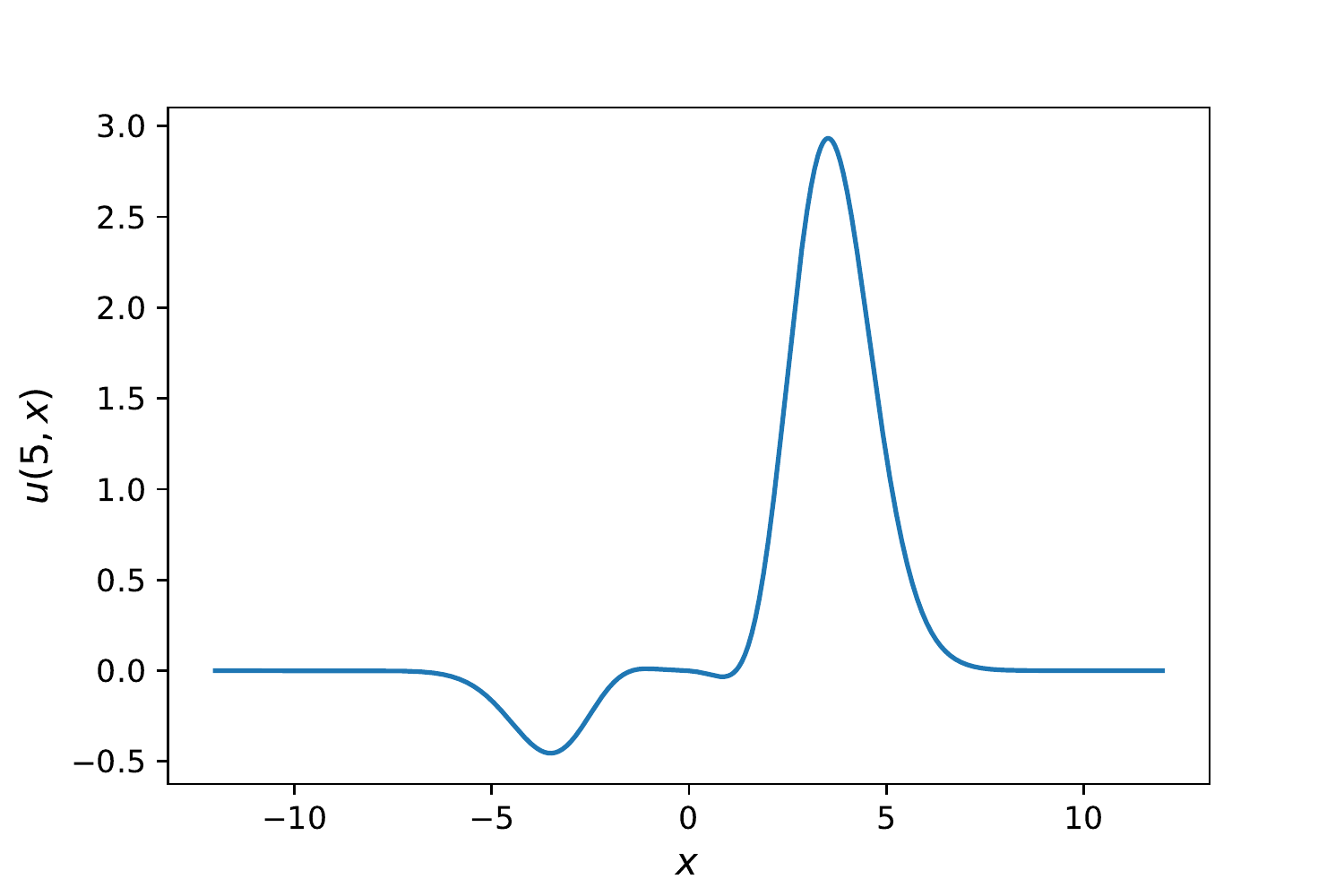}
\end{figure}
\begin{figure}[ht!]
\centering
\includegraphics[scale=0.4]{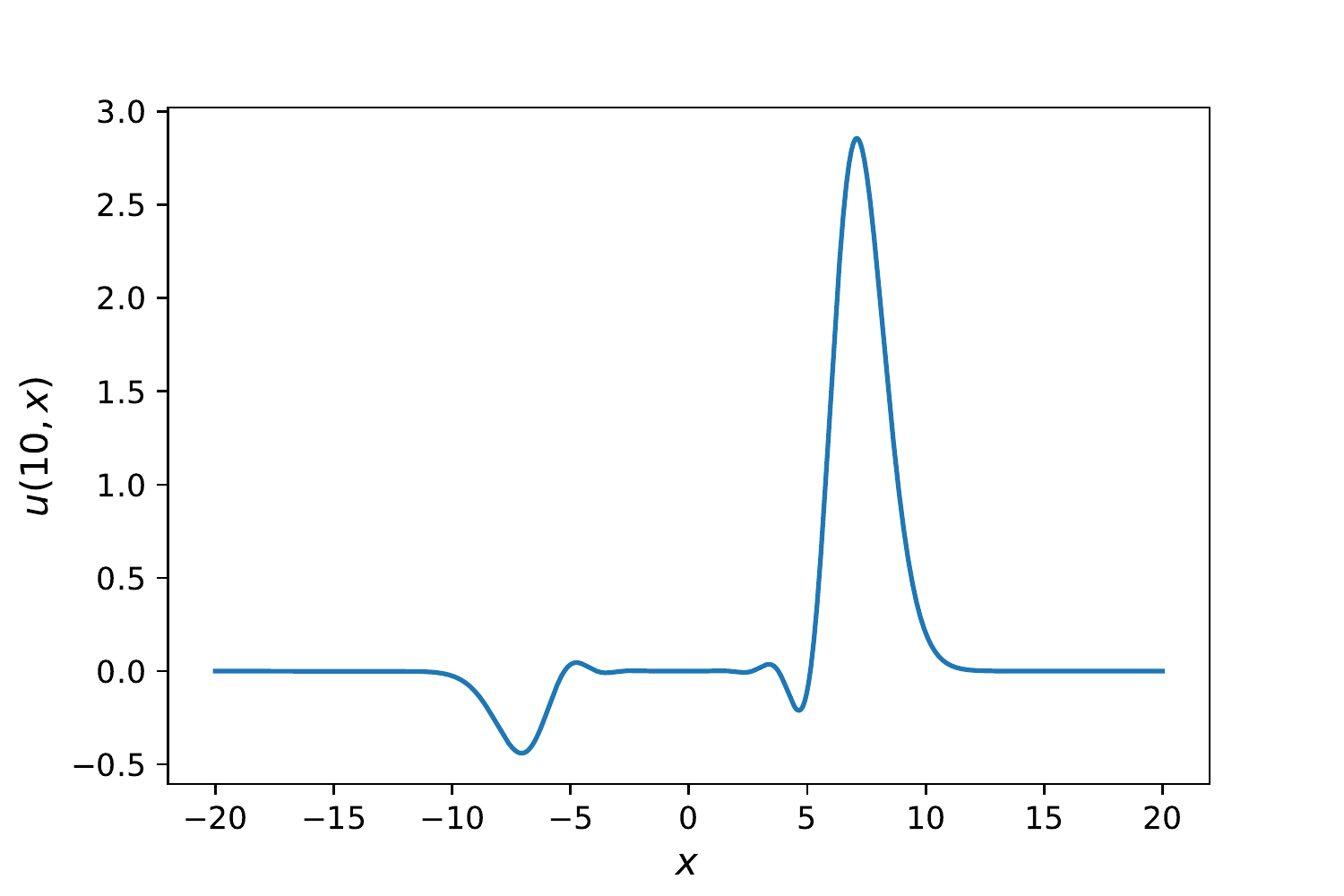}
\includegraphics[scale=0.4]{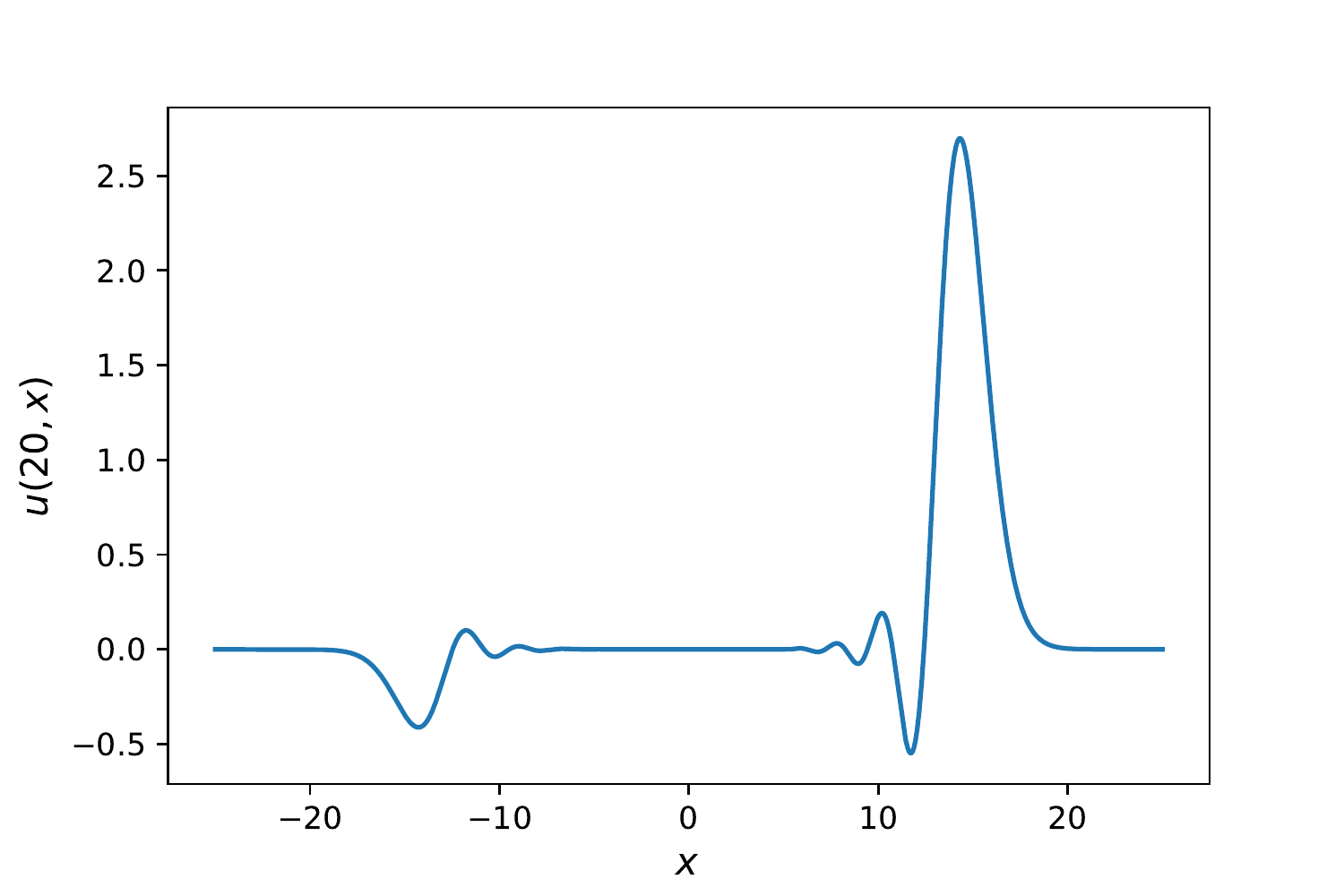}
\end{figure}
\newpage
In Fig. \ref{fig:5}
we consider
\begin{equation}
\alpha=1/2.
\end{equation}
\begin{figure}[ht!]
\centering
\includegraphics[scale=0.4]{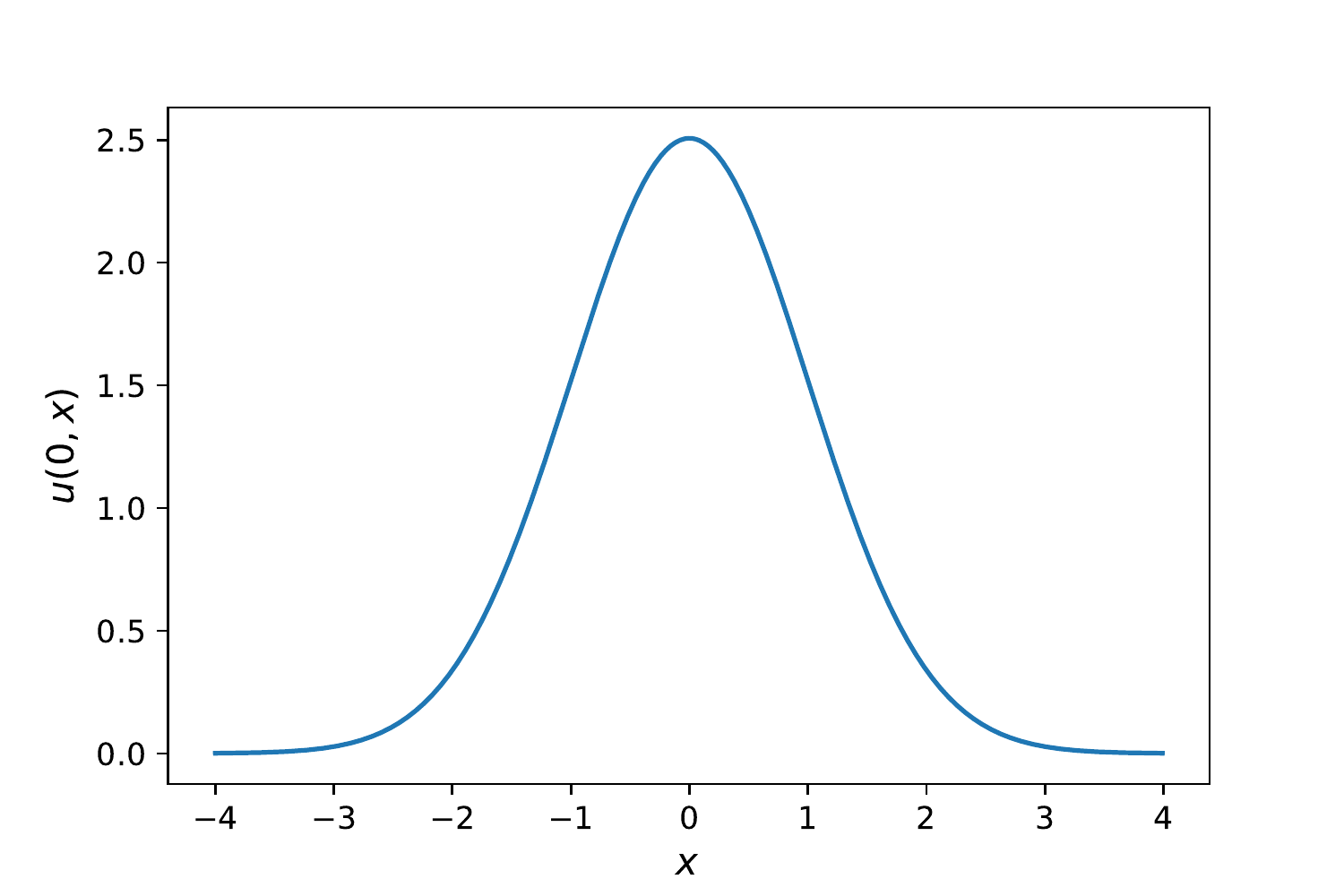}
\includegraphics[scale=0.4]{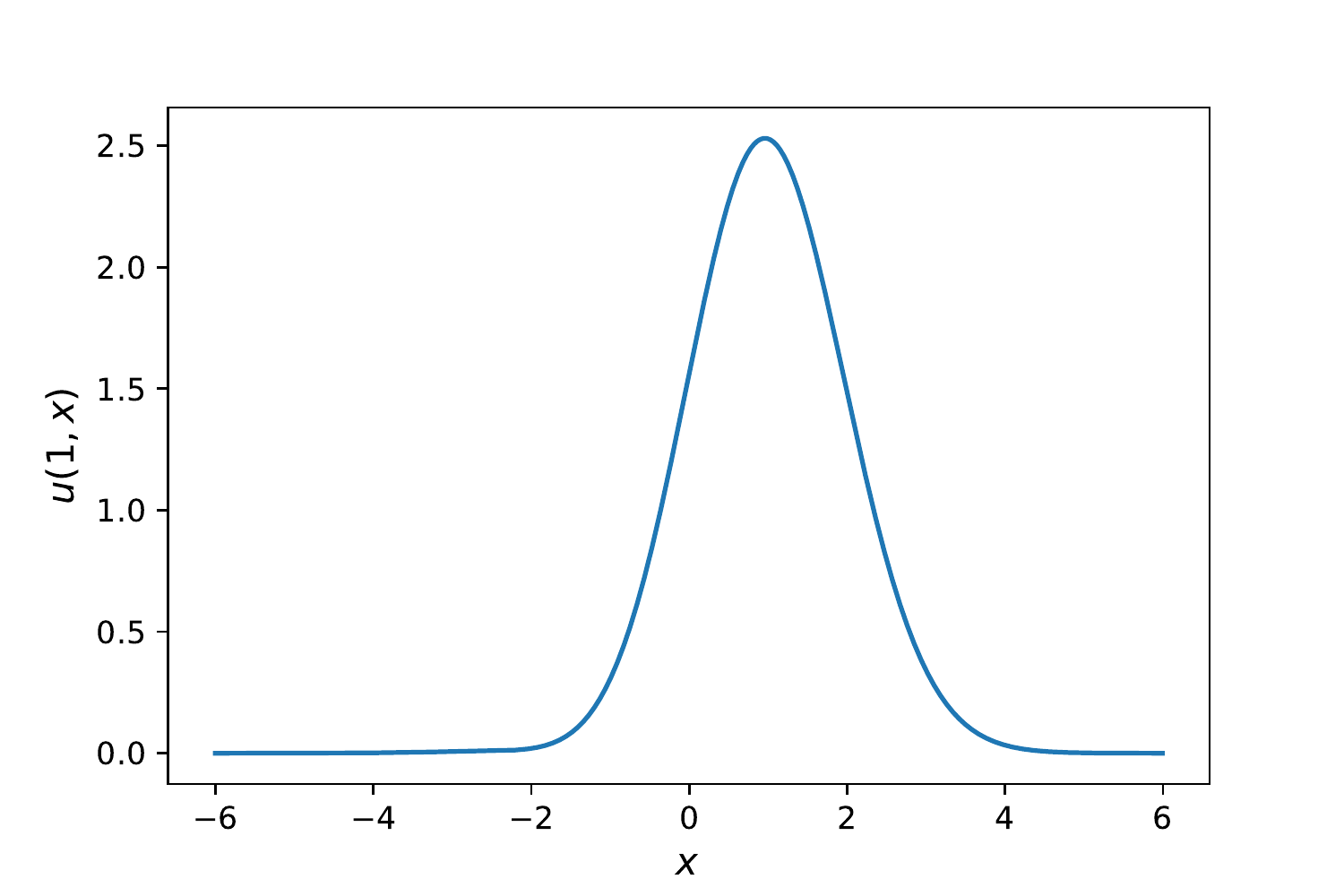}
\end{figure}
\begin{figure}[ht!]
\centering
\includegraphics[scale=0.4]{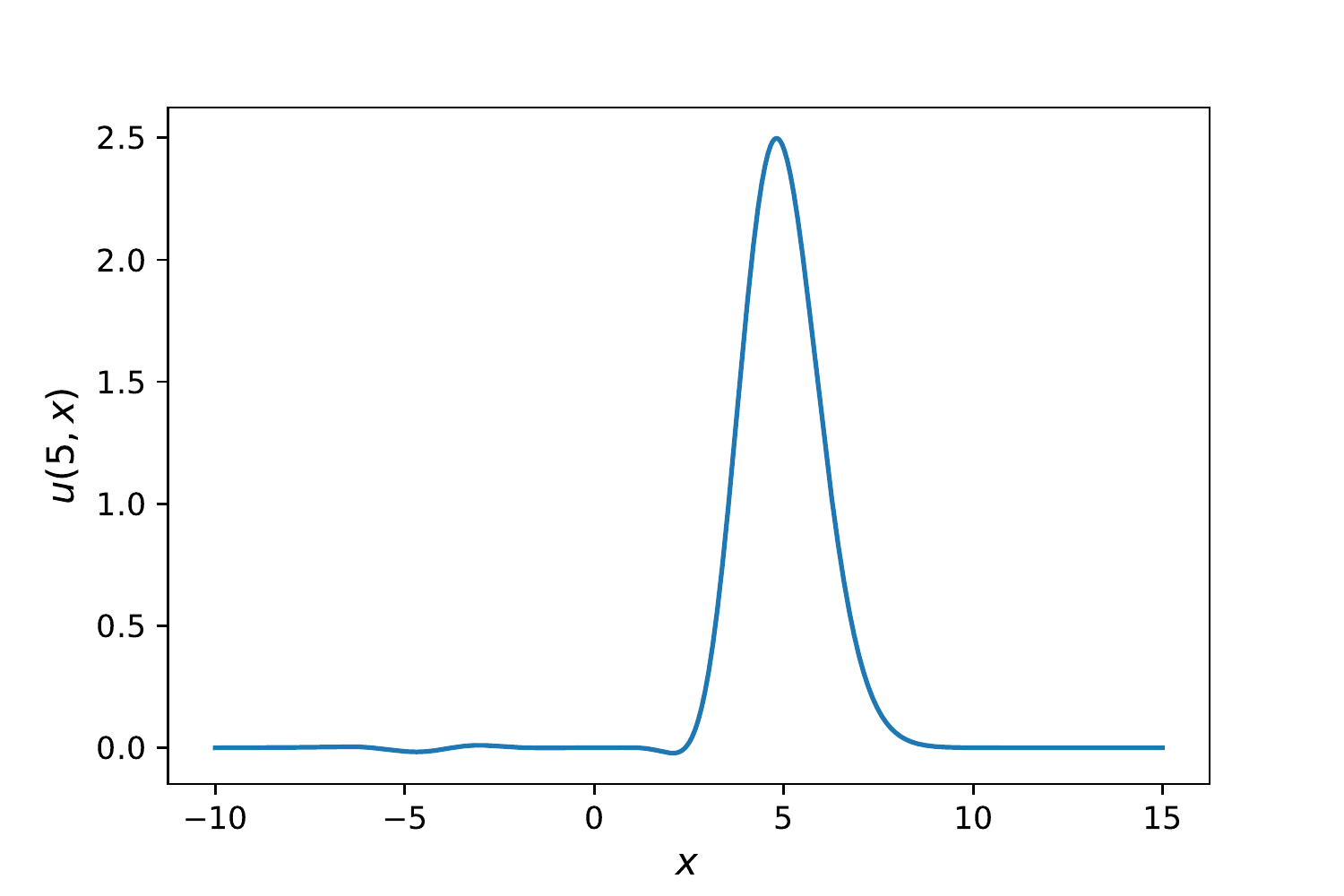}
\includegraphics[scale=0.4]{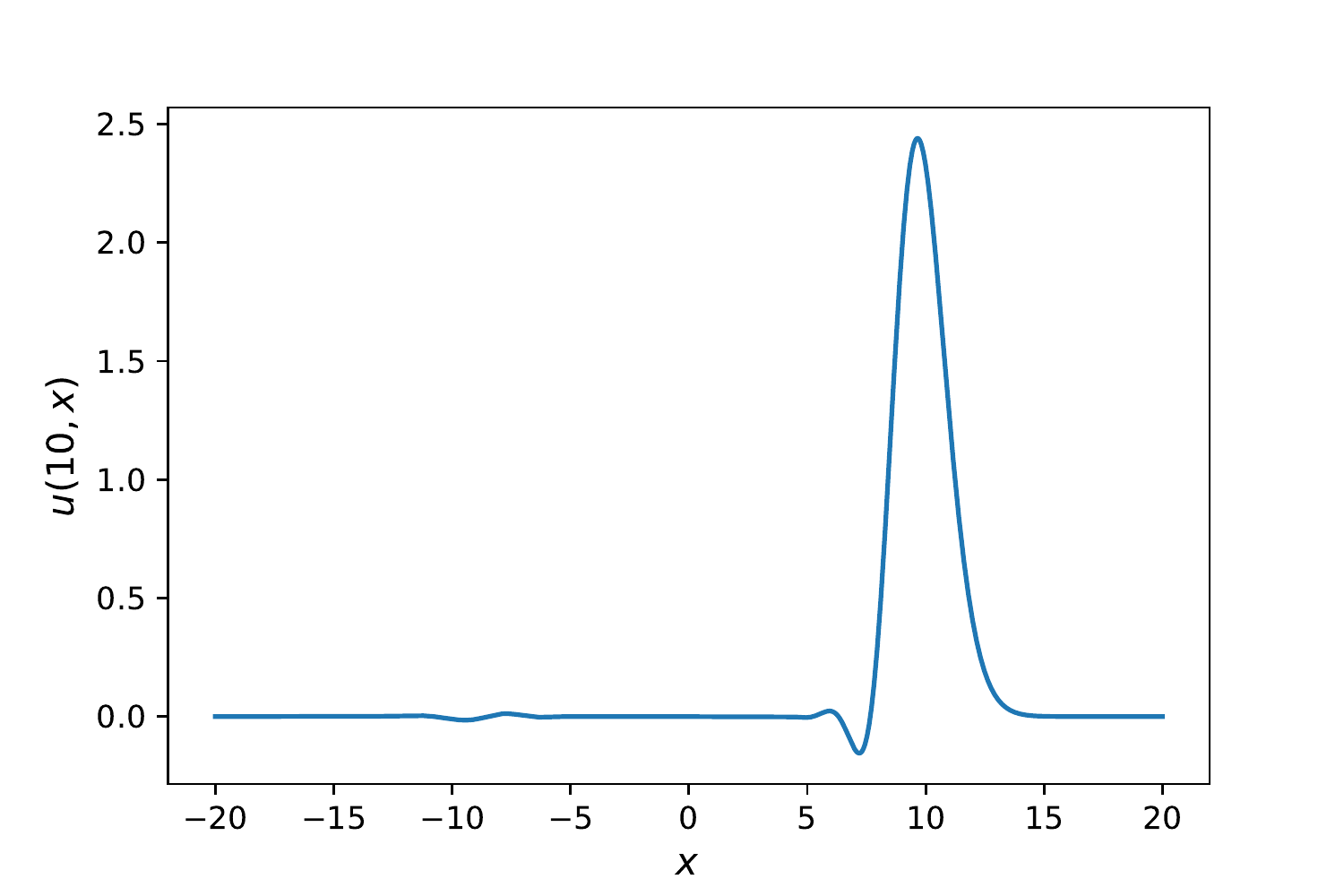}
\end{figure}
\newpage
In Fig. \ref{fig:6}
we consider
\begin{equation}
\alpha=9/10.
\end{equation}
\begin{figure}[ht!]
\centering
\includegraphics[scale=0.4]{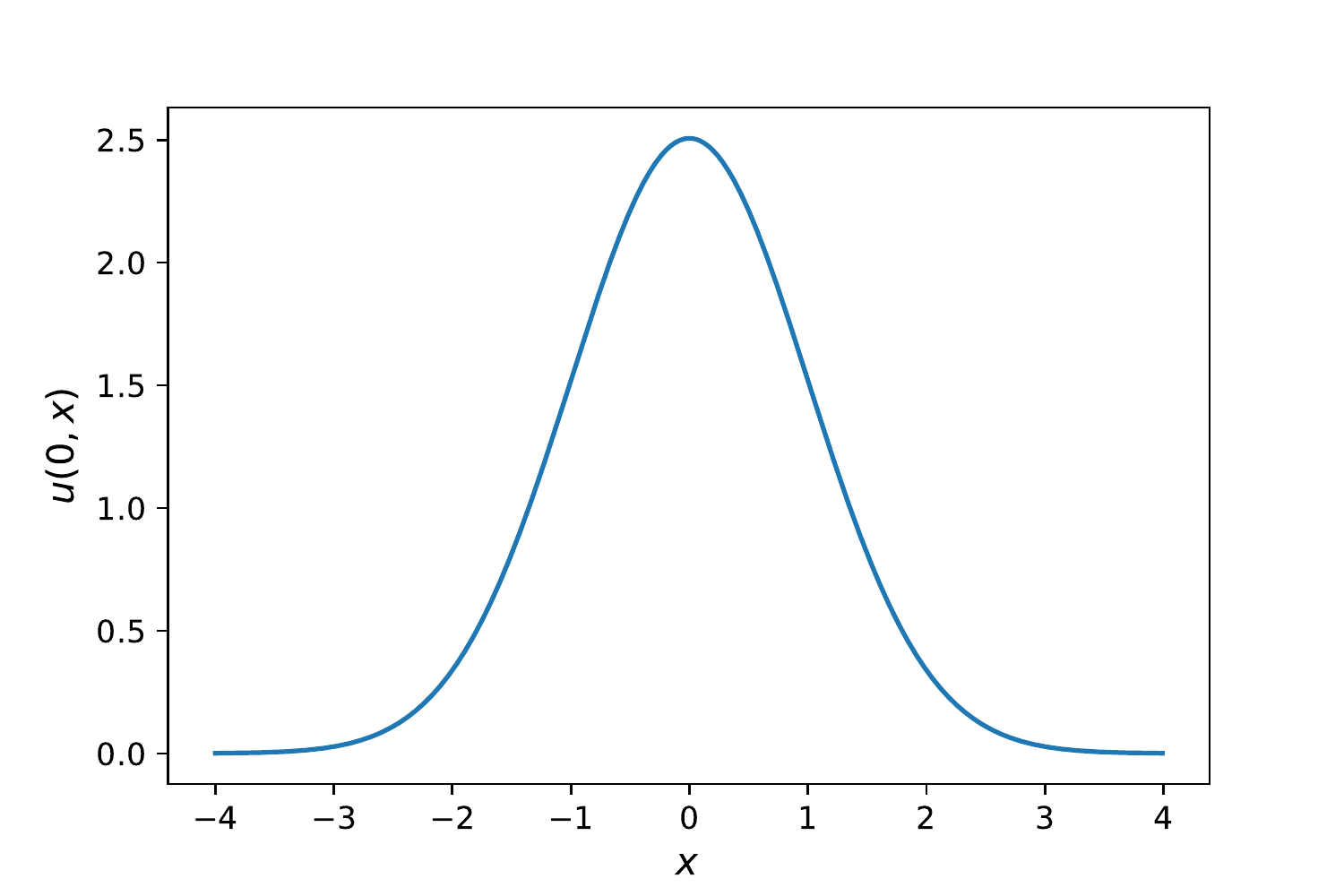}
\includegraphics[scale=0.4]{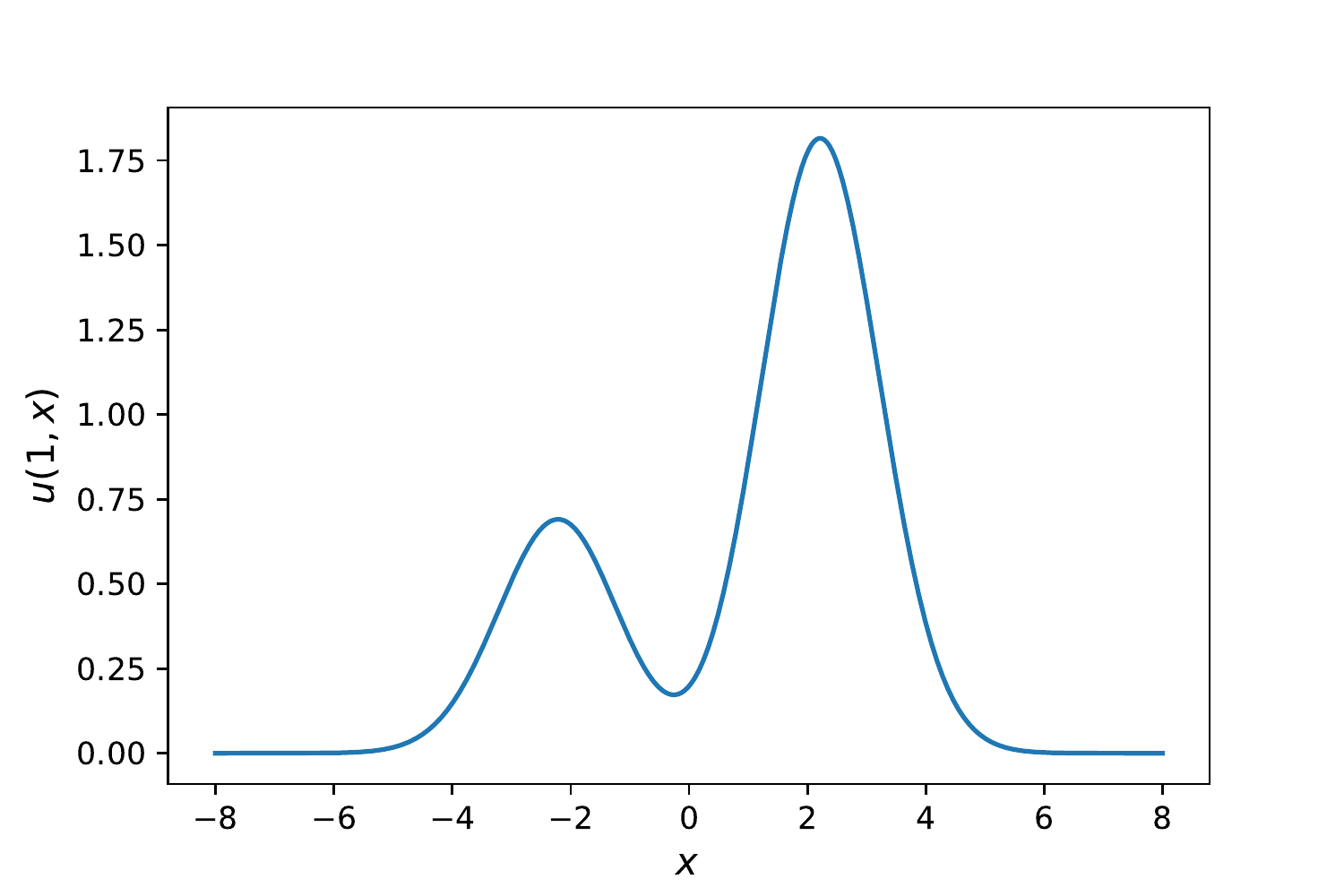}
\end{figure}
\begin{figure}[ht!]
\centering
\includegraphics[scale=0.4]{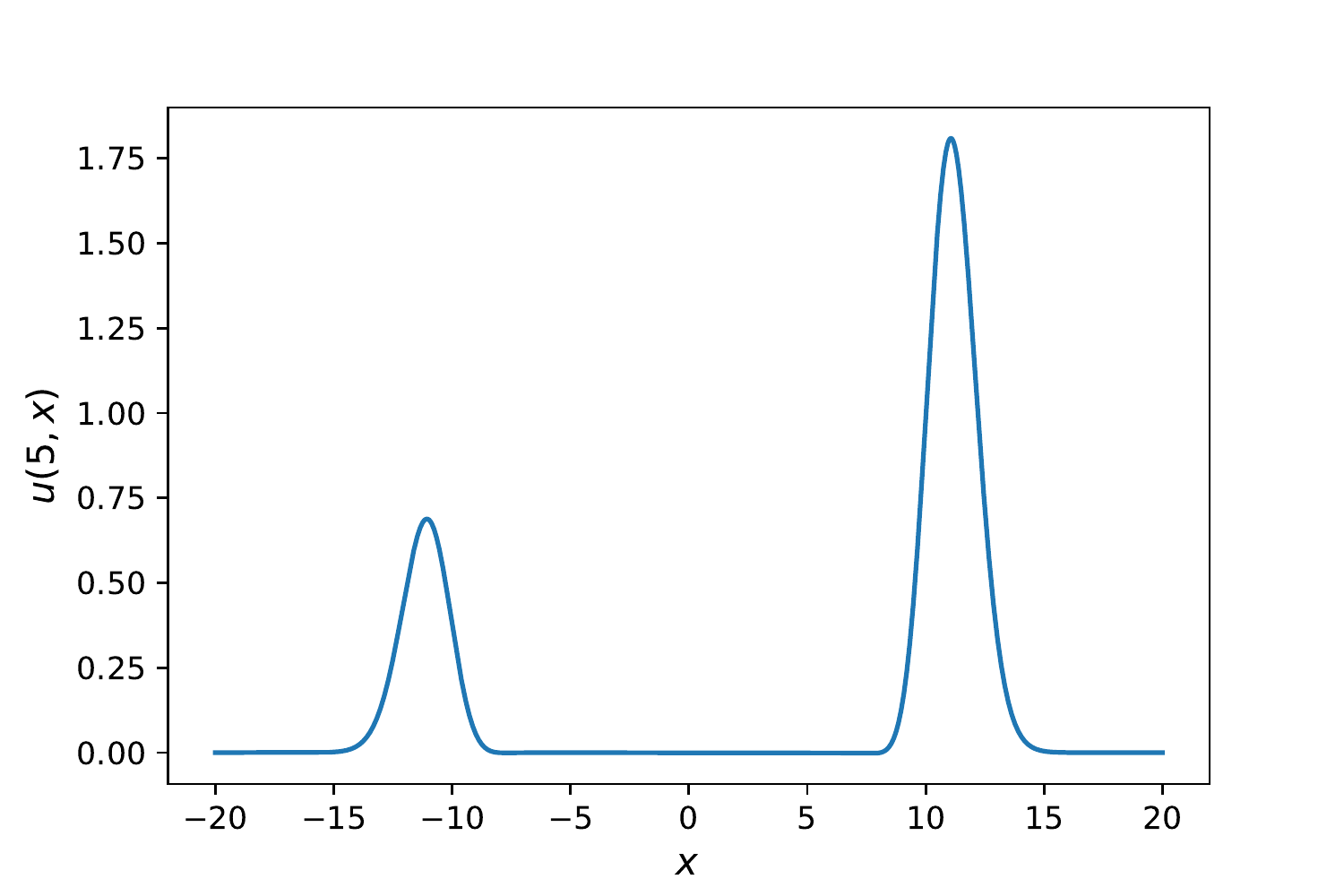}
\includegraphics[scale=0.4]{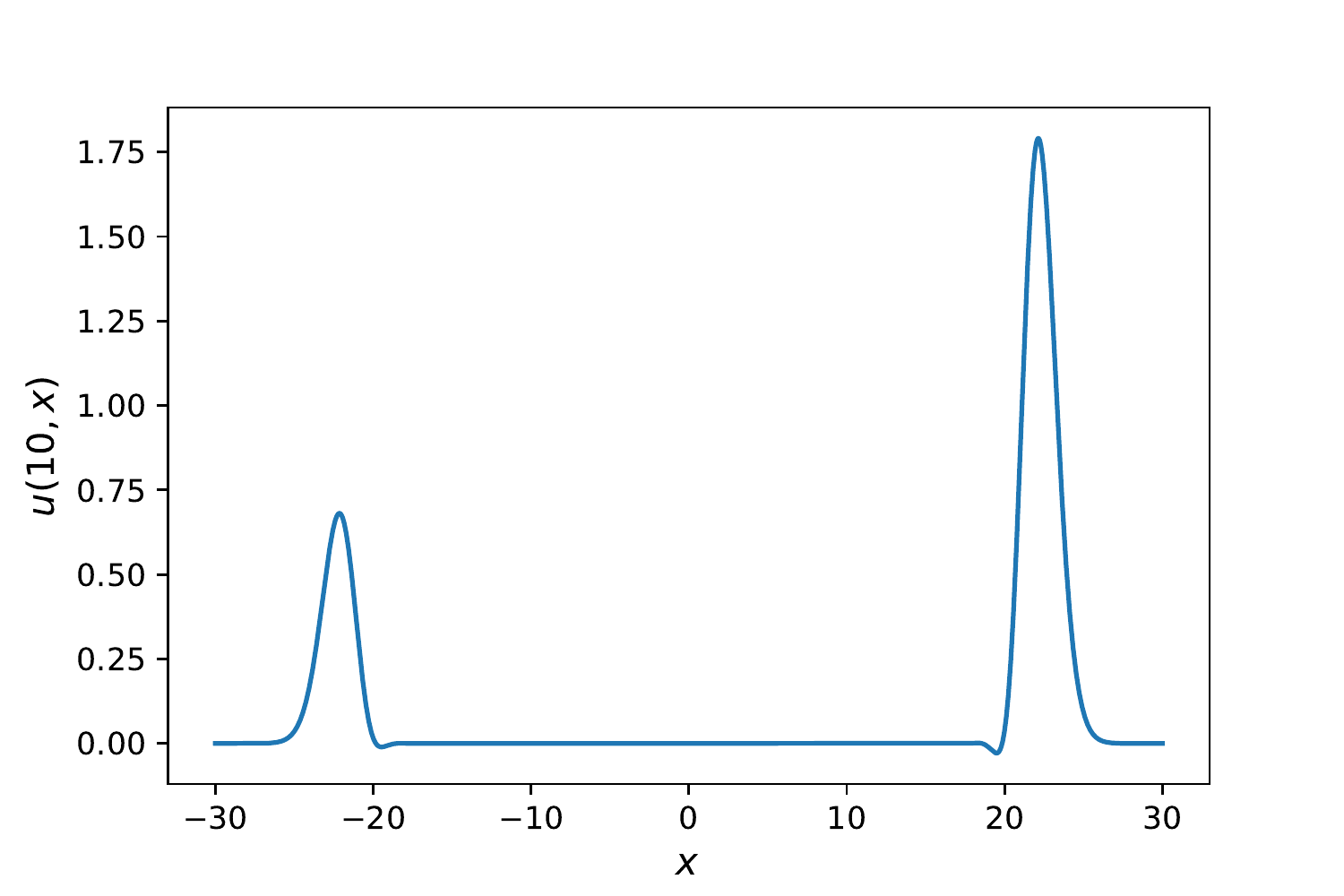}
\end{figure}

\subsection{Case~$\boldsymbol{\sigma\gg\delta}$} 
\begin{figure}[ht!]
\centering
\includegraphics[scale=0.4]{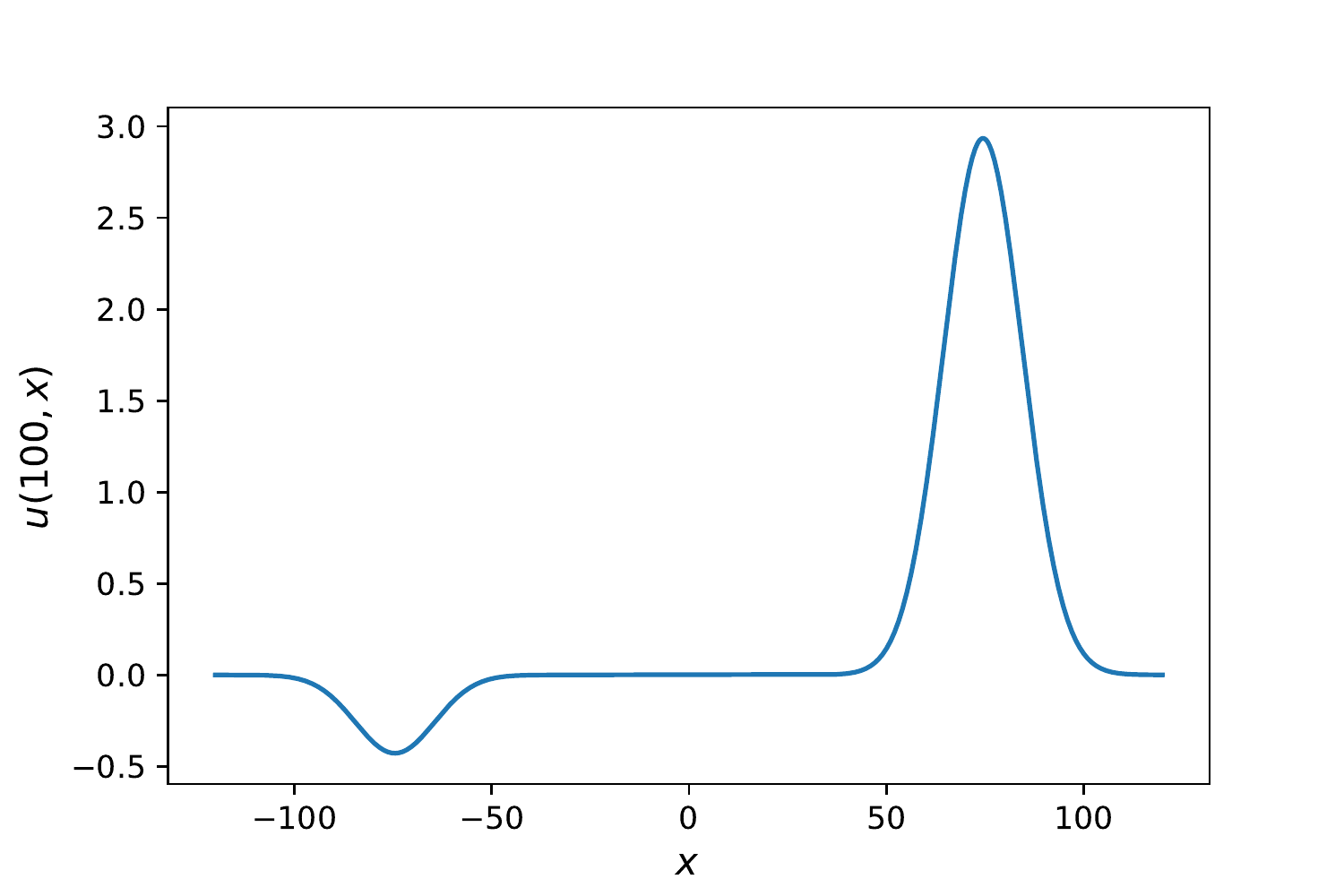}
\includegraphics[scale=0.4]{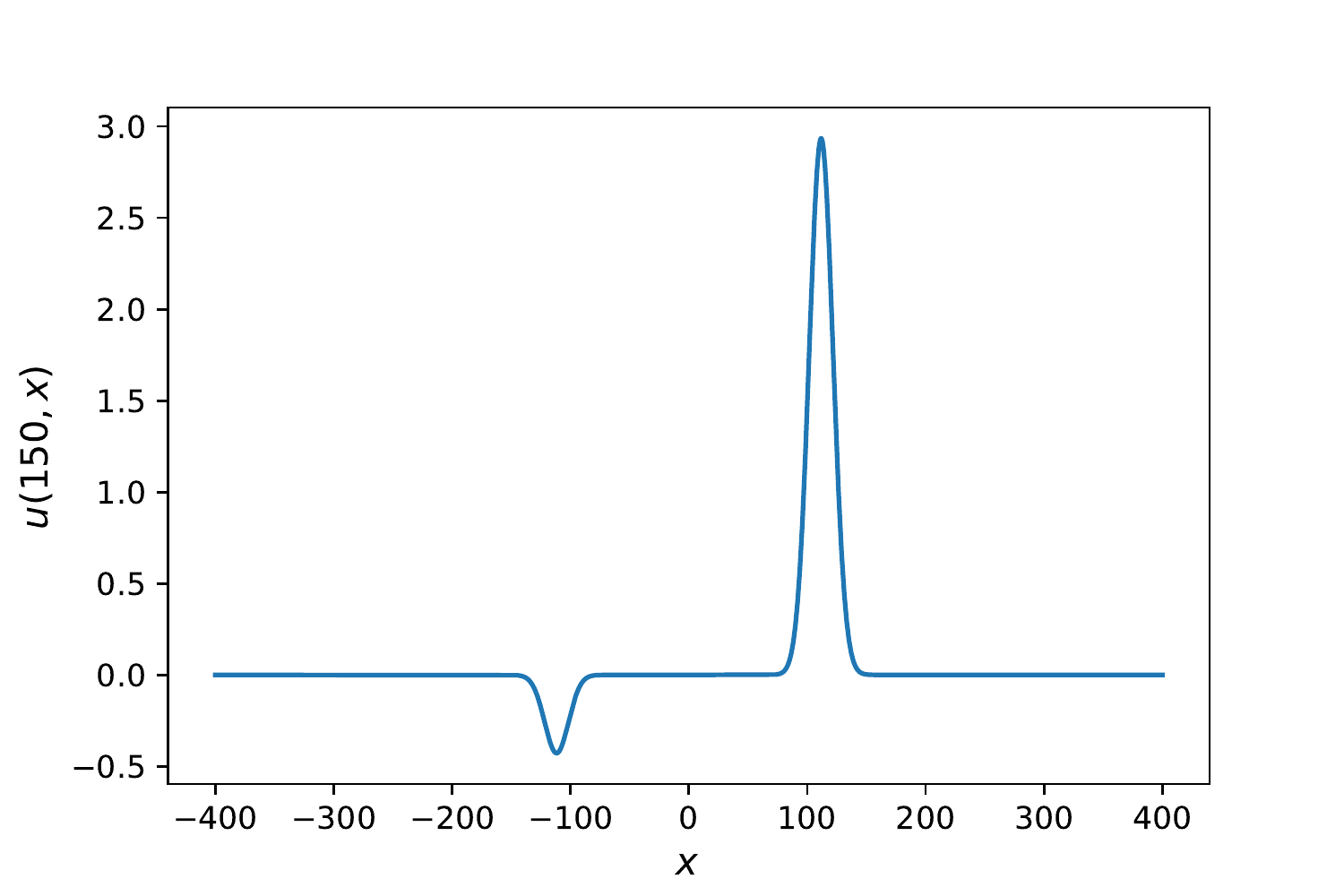}
\caption{Numerical simulations of \eqref{eq:linear_per}-\eqref{eq:ini} at different times in correspondence of $\sigma=10> 1=\delta$, $v=1$ and $\alpha=10^{-1}$.}
\label{fig:7}
\end{figure}

This subsection is devoted to the choice
\begin{equation}
\sigma=10\quad\text{and}\quad\delta=1.
\end{equation}
In order to support our claim about the pure hyperbolic propagation and (almost)
total absence of dispersive behavior, with respect to
the value of $\alpha$, we show different cases regarding three possible choices of $\alpha$.
\newpage
In Fig. \ref{fig:7}
we consider
\begin{equation}
\alpha=10^{-1}.
\end{equation}
\begin{figure}[ht!]
\centering
\includegraphics[scale=0.4]{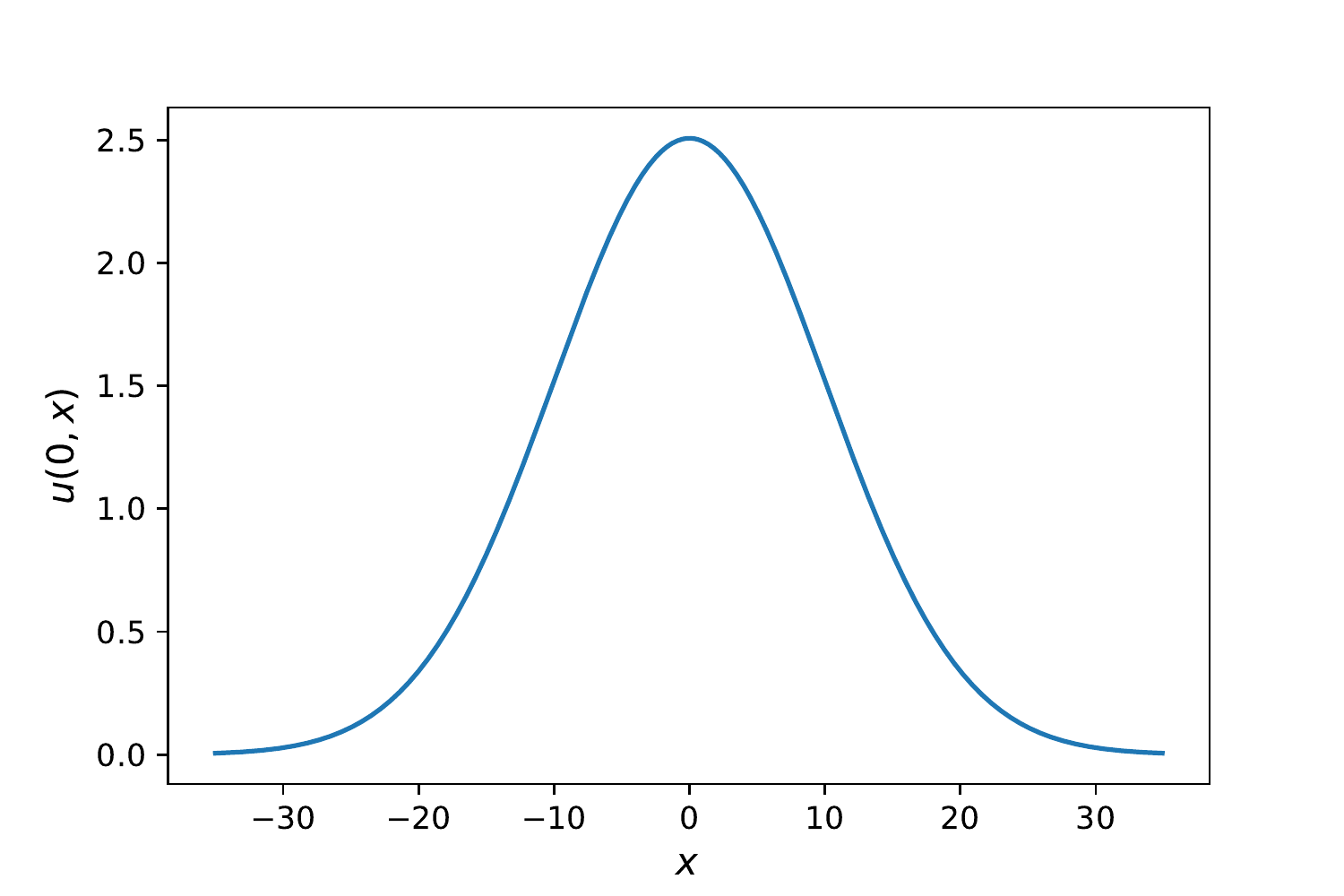}
\includegraphics[scale=0.4]{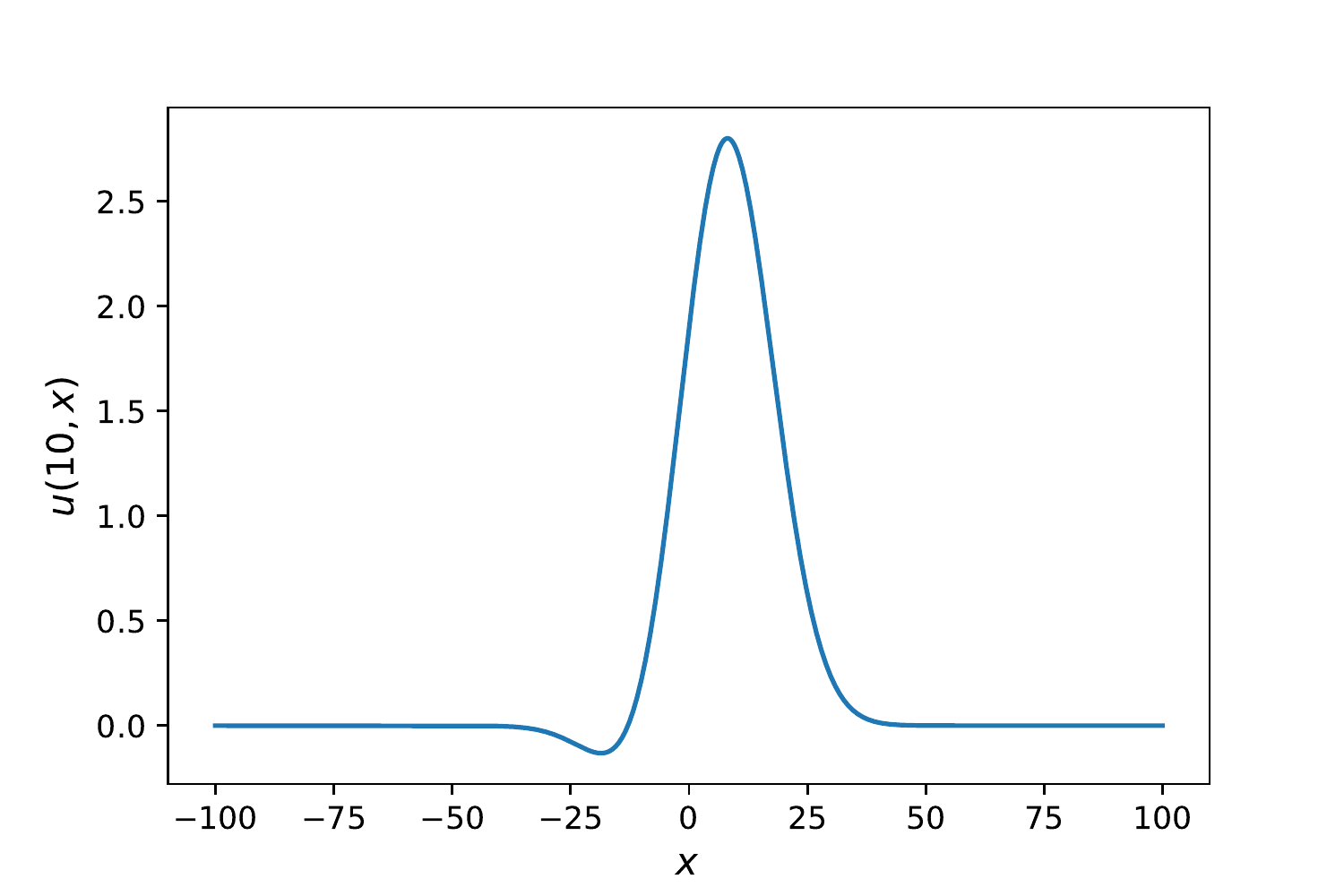}
\end{figure}
\begin{figure}[ht!]
\centering
\includegraphics[scale=0.4]{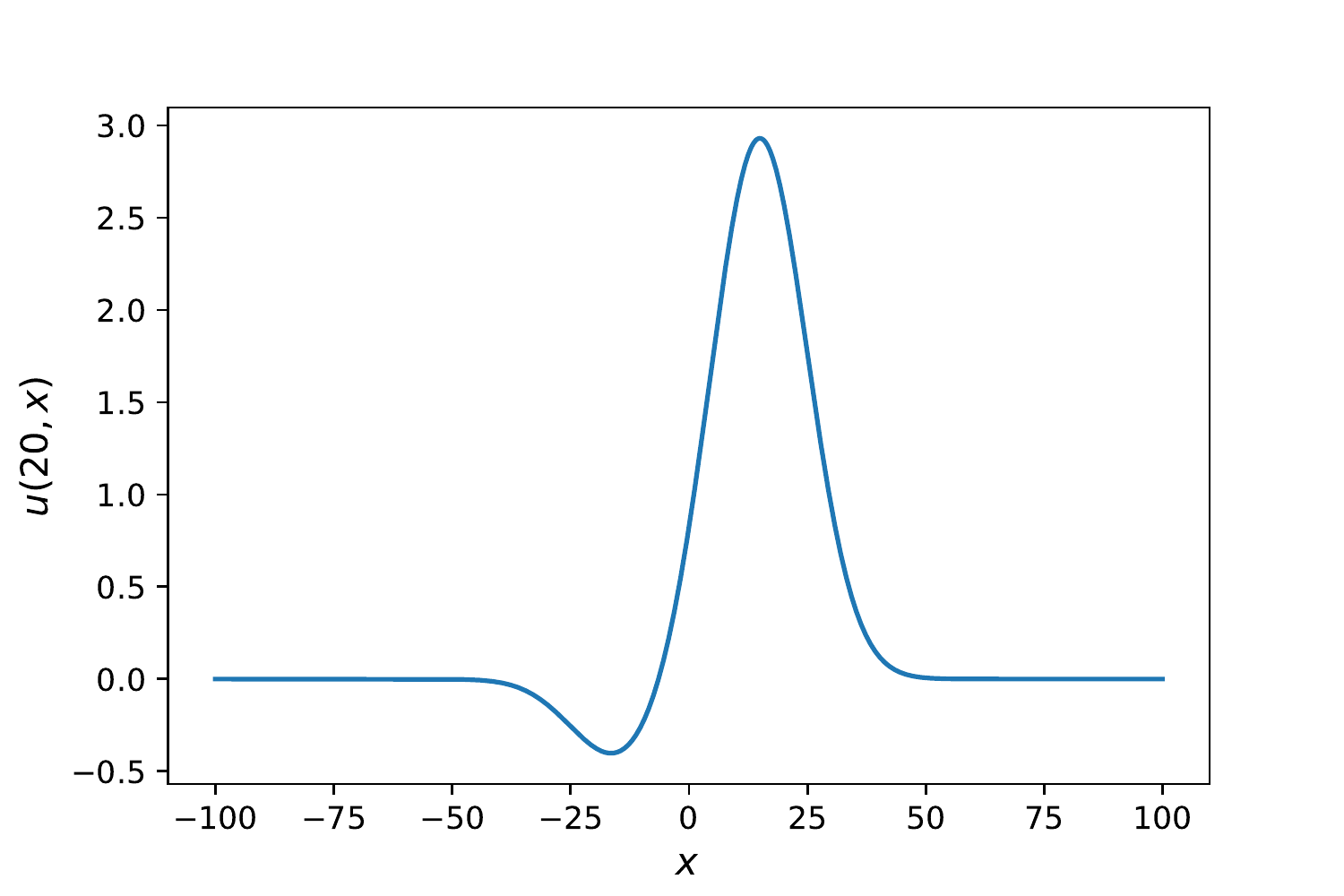}
\includegraphics[scale=0.4]{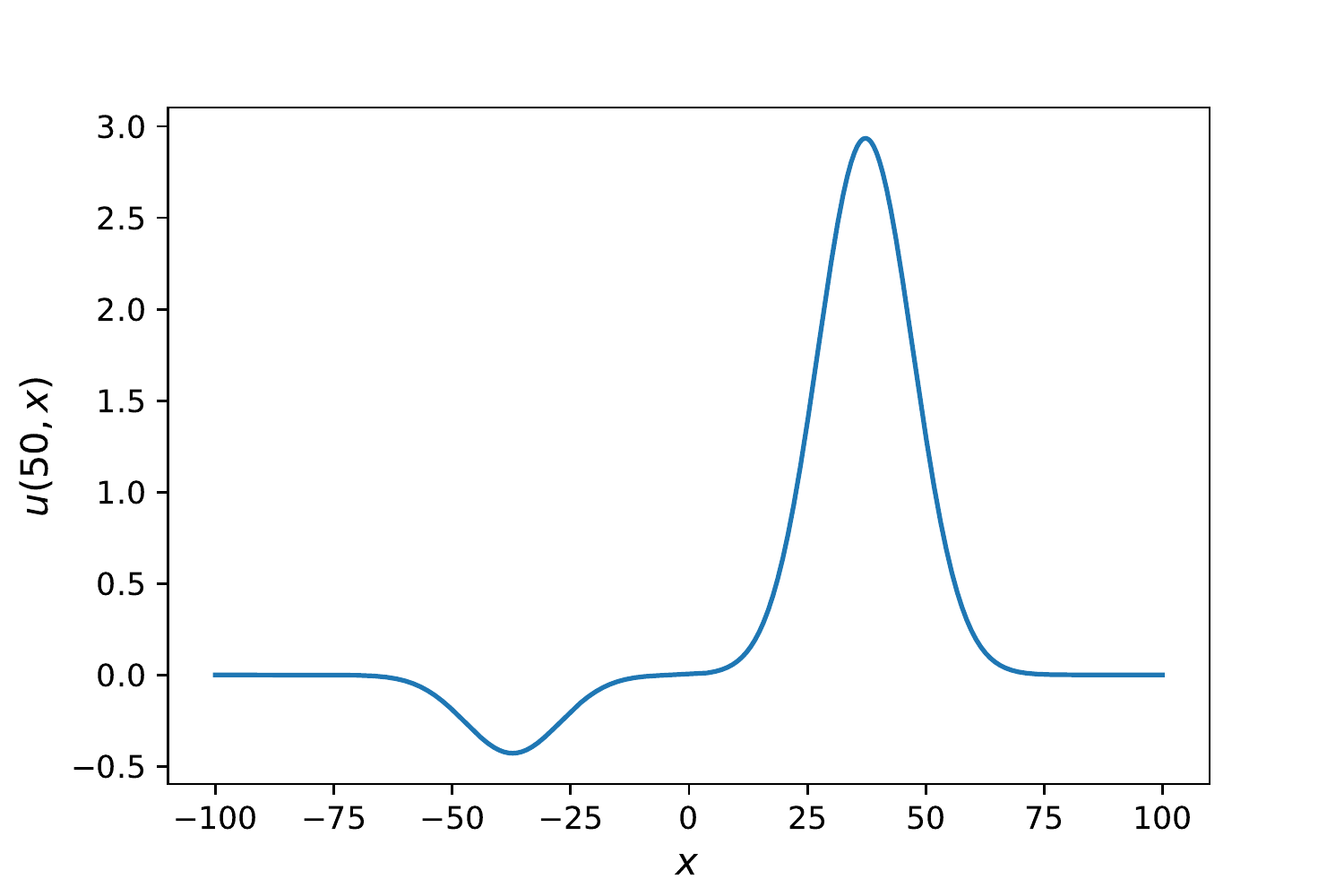}
\end{figure}

\begin{figure}[ht!]
\centering
\includegraphics[scale=0.4]{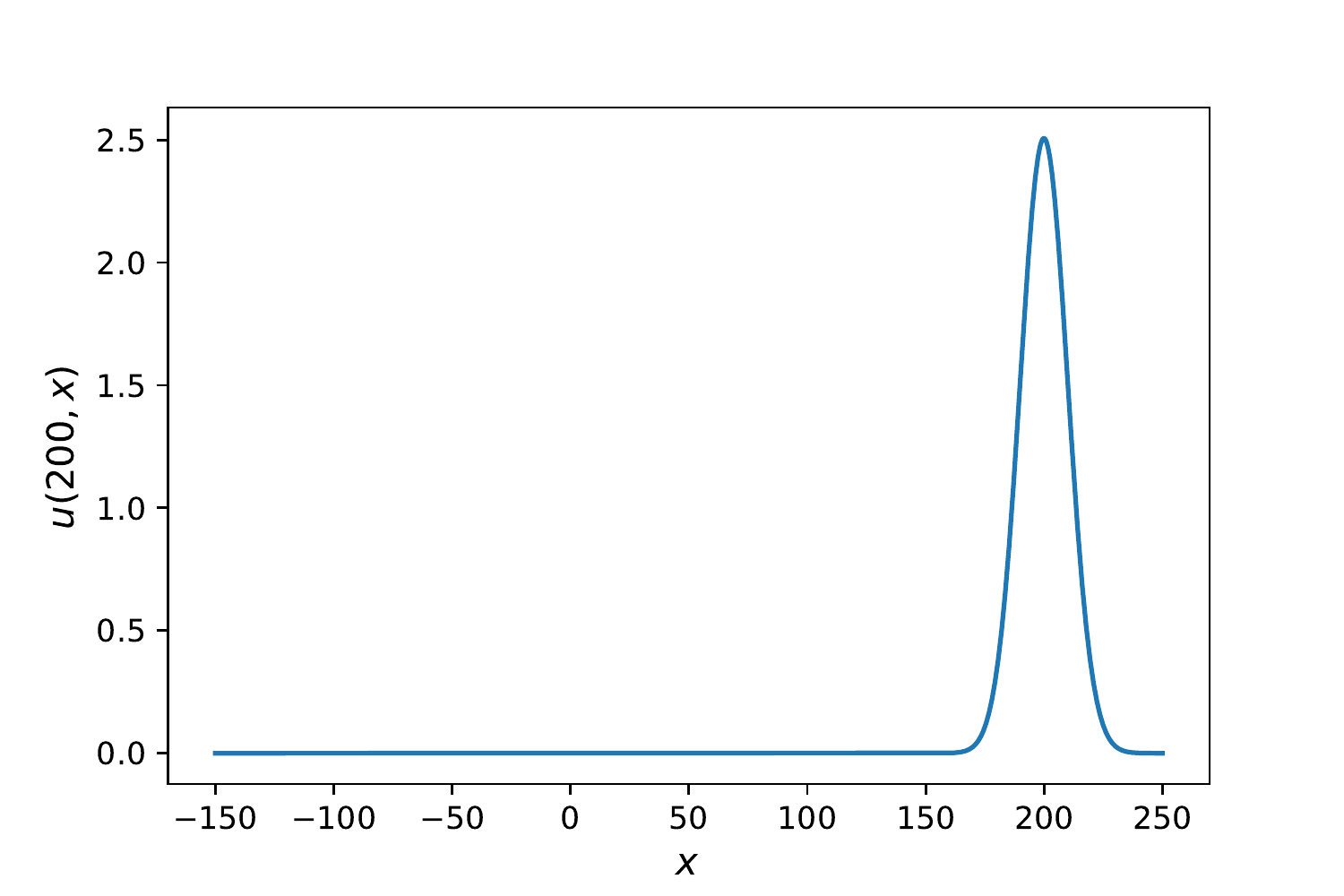}
\includegraphics[scale=0.4]{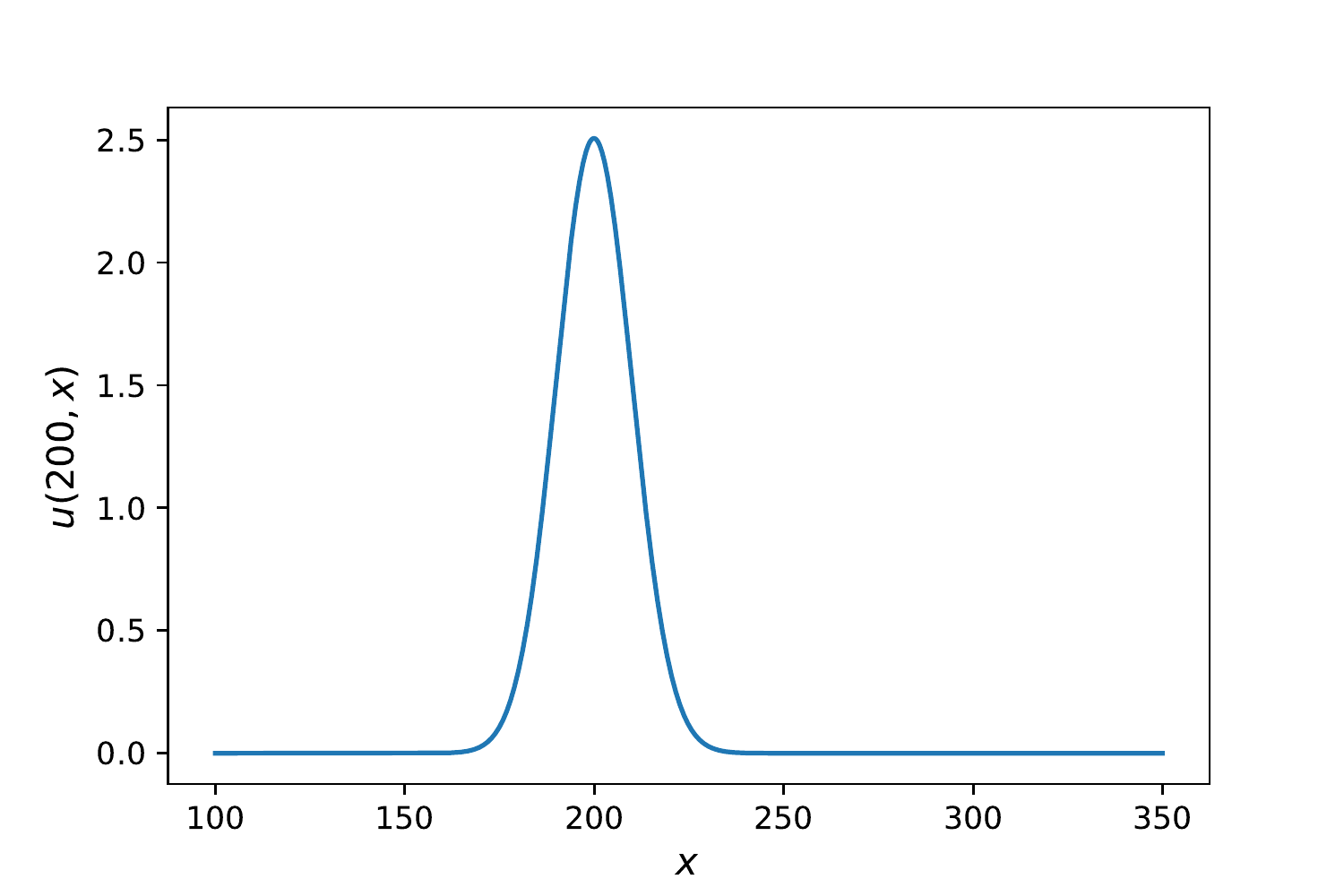}
\caption{Numerical simulations of \eqref{eq:linear_per}-\eqref{eq:ini} at different times in correspondence of $\sigma=10>1=\delta$, $v=1$ and $\alpha=1/2$.}
\label{fig:8}
\end{figure}

\begin{figure}[ht!]
\centering
\includegraphics[scale=0.4]{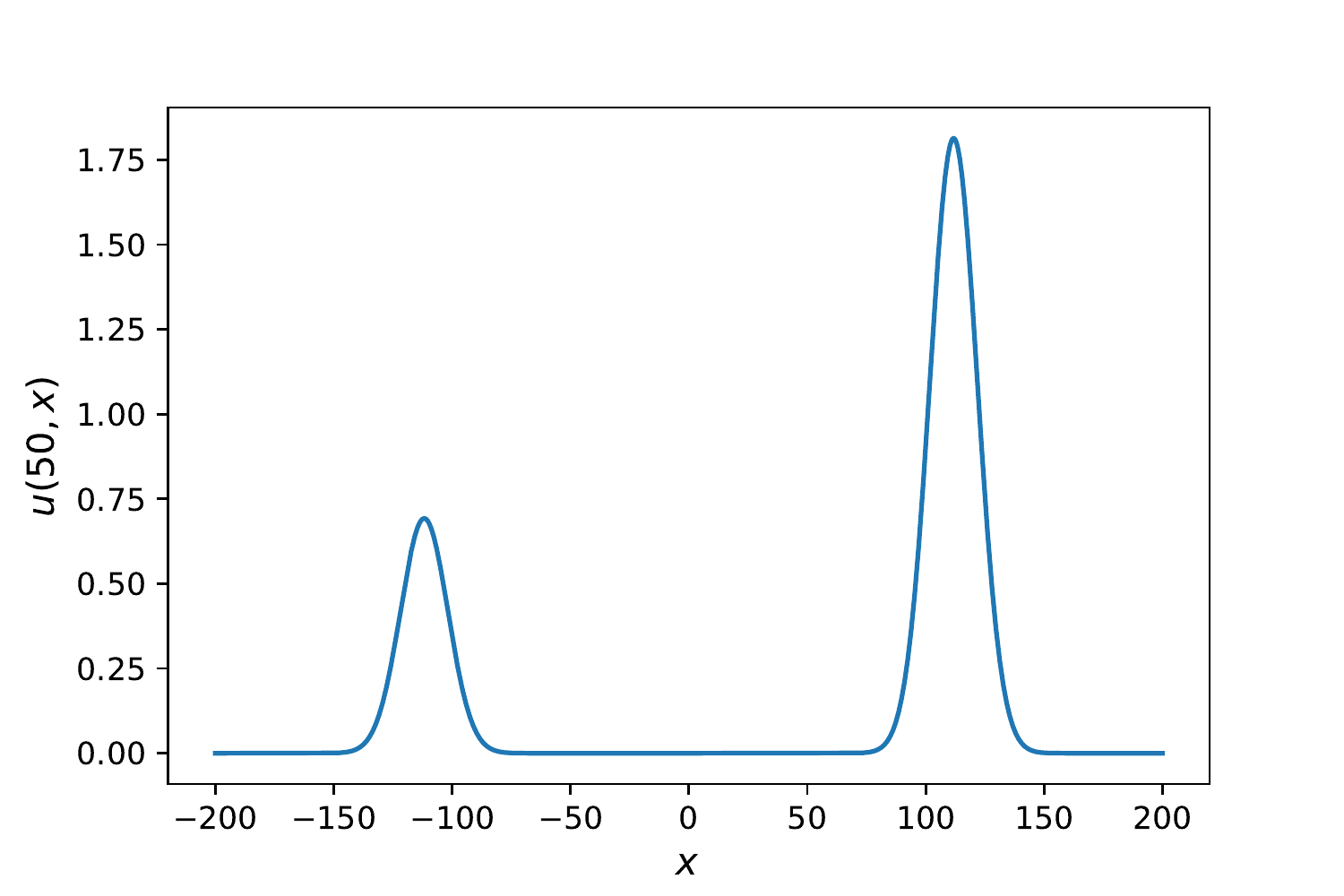}
\includegraphics[scale=0.4]{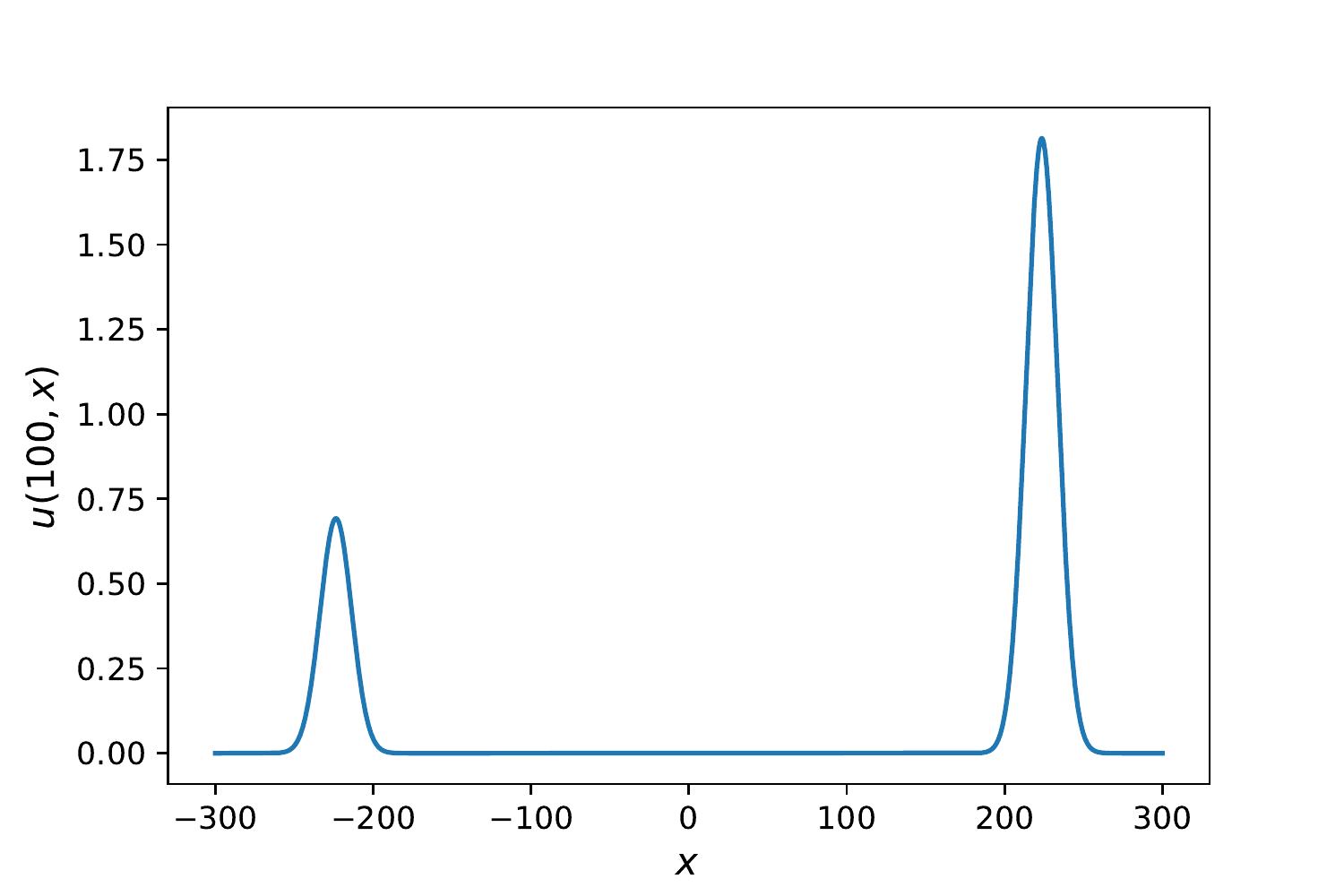}
\caption{Numerical simulations of \eqref{eq:linear_per}-\eqref{eq:ini} at different times in correspondence of $\sigma=10>1=\delta$, $v=1$ and $\alpha=9/10$.}
\label{fig:9}
\end{figure}

\newpage
In Fig. \ref{fig:8}
we consider
\begin{equation}
\alpha=1/2.
\end{equation}
\begin{figure}[ht!]
\centering
\includegraphics[scale=0.4]{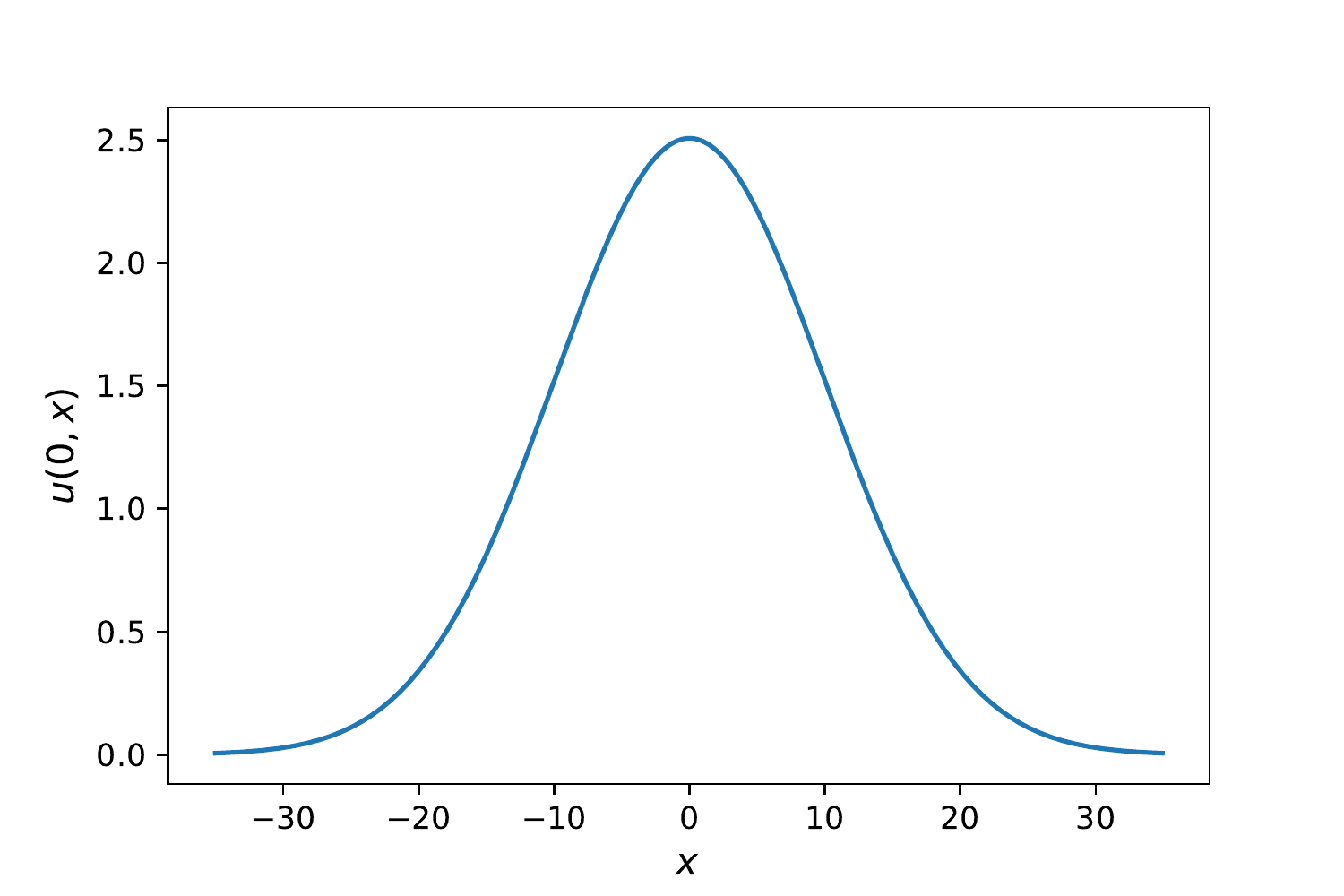}
\includegraphics[scale=0.4]{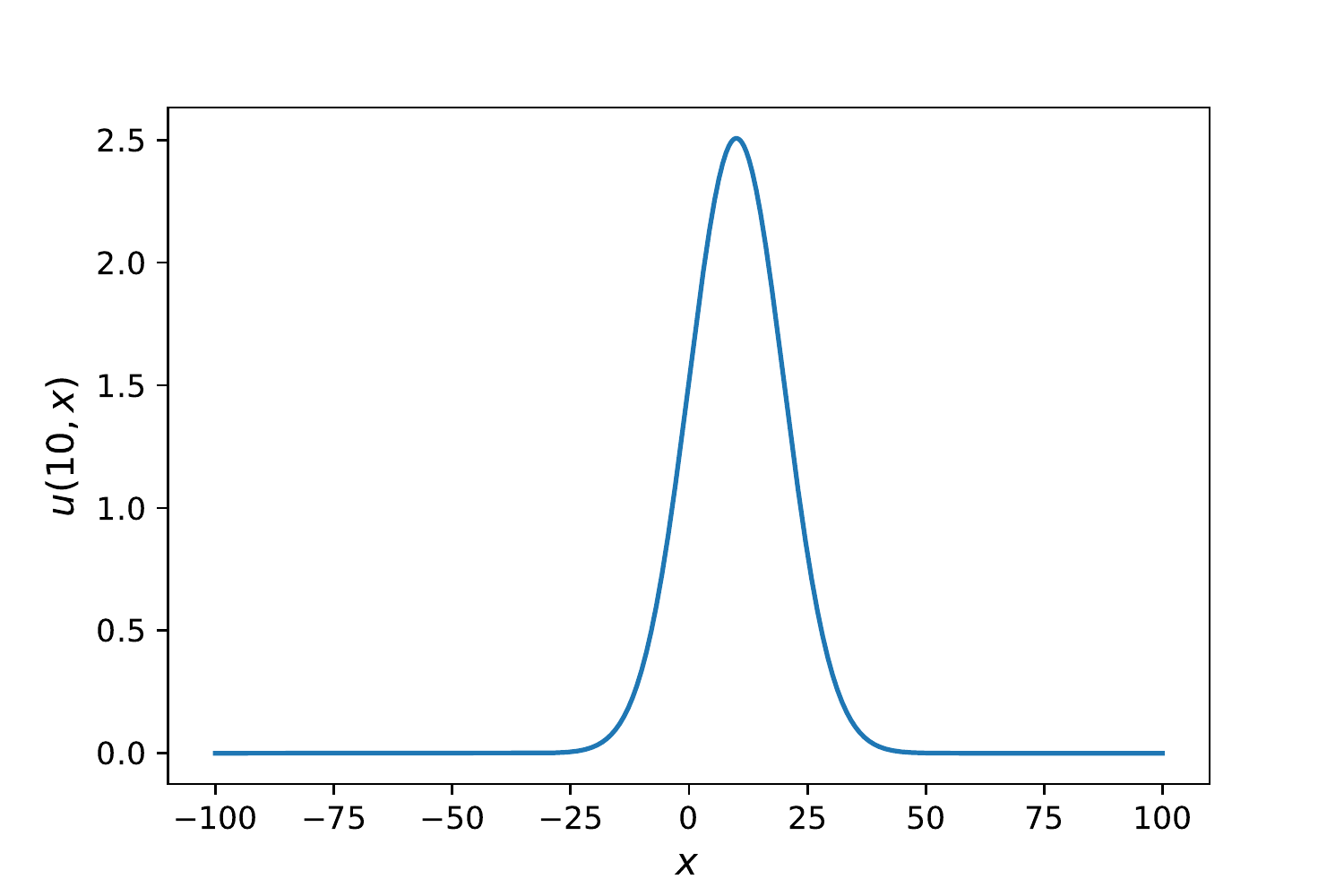}
\end{figure}
\begin{figure}[ht!]
\centering
\includegraphics[scale=0.4]{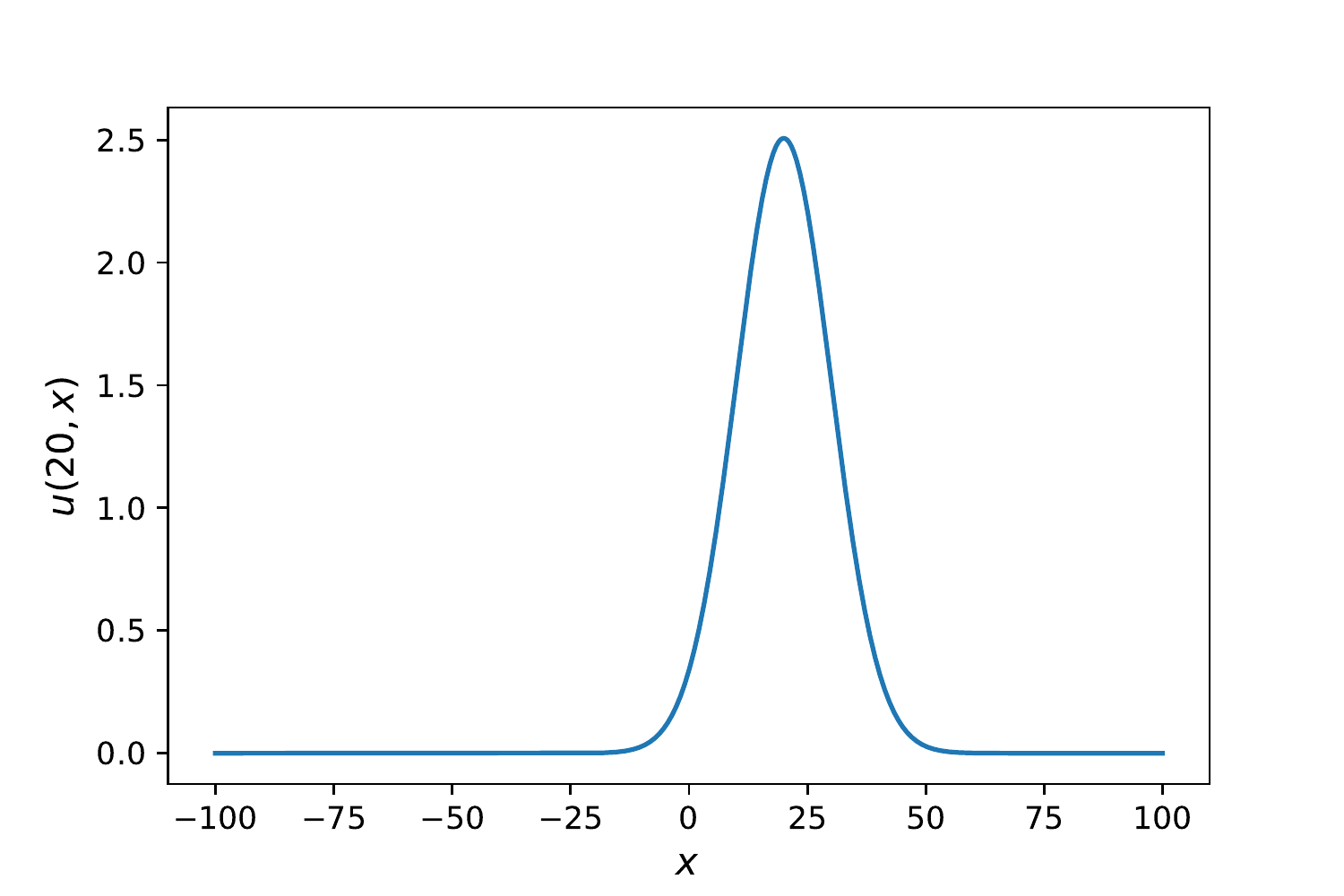}
\includegraphics[scale=0.4]{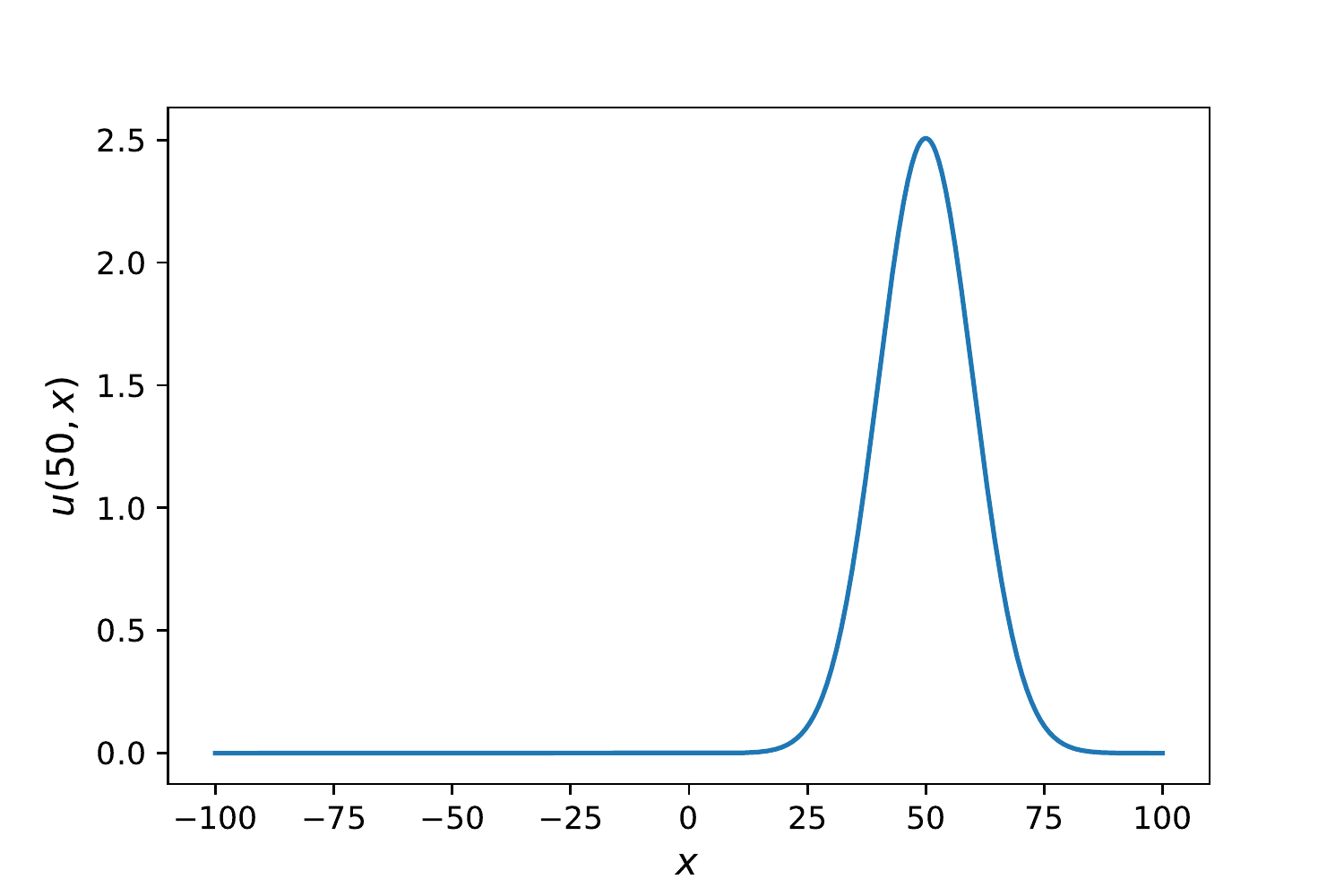}
\end{figure}
\begin{figure}[ht!]
\centering
\includegraphics[scale=0.4]{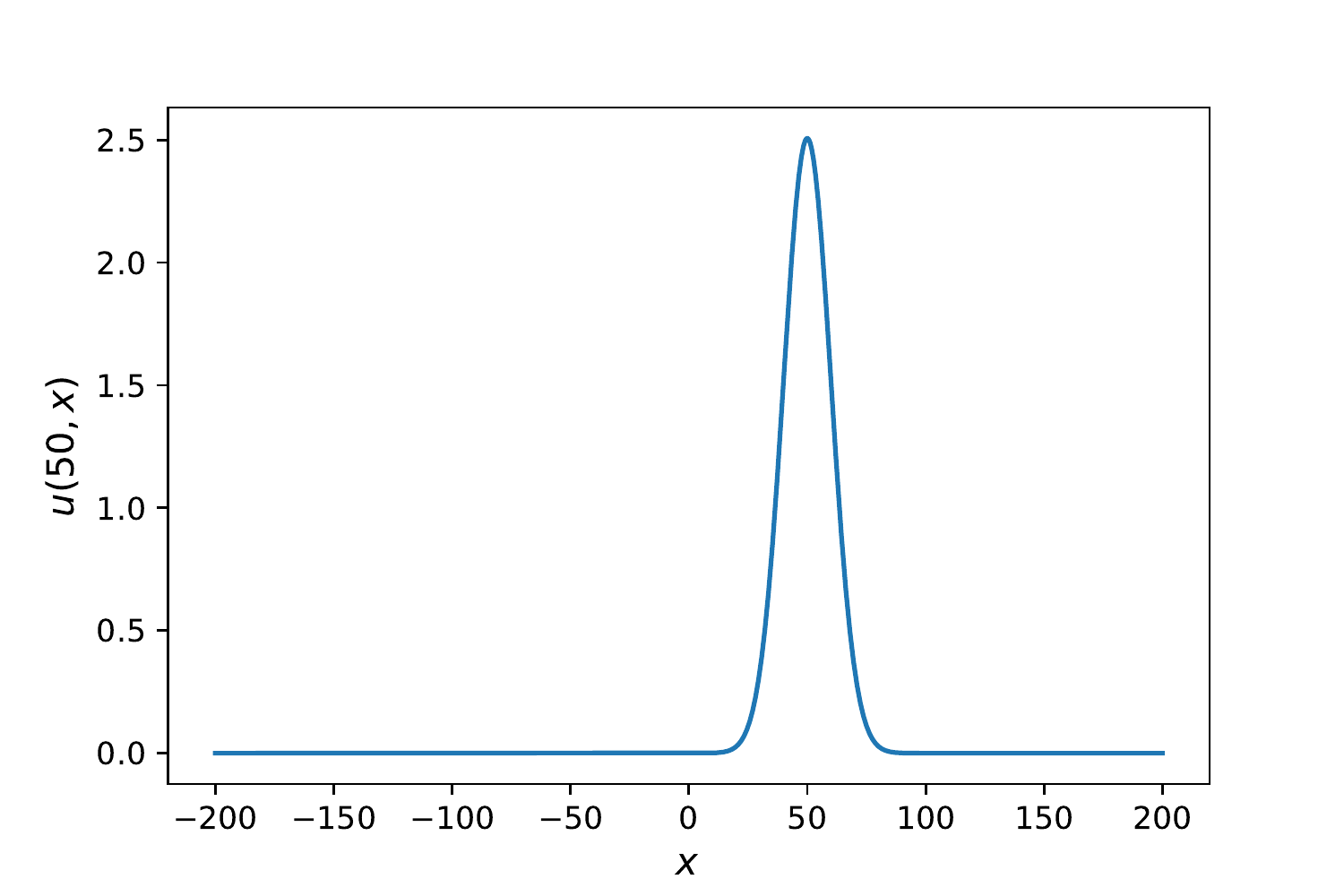}
\includegraphics[scale=0.4]{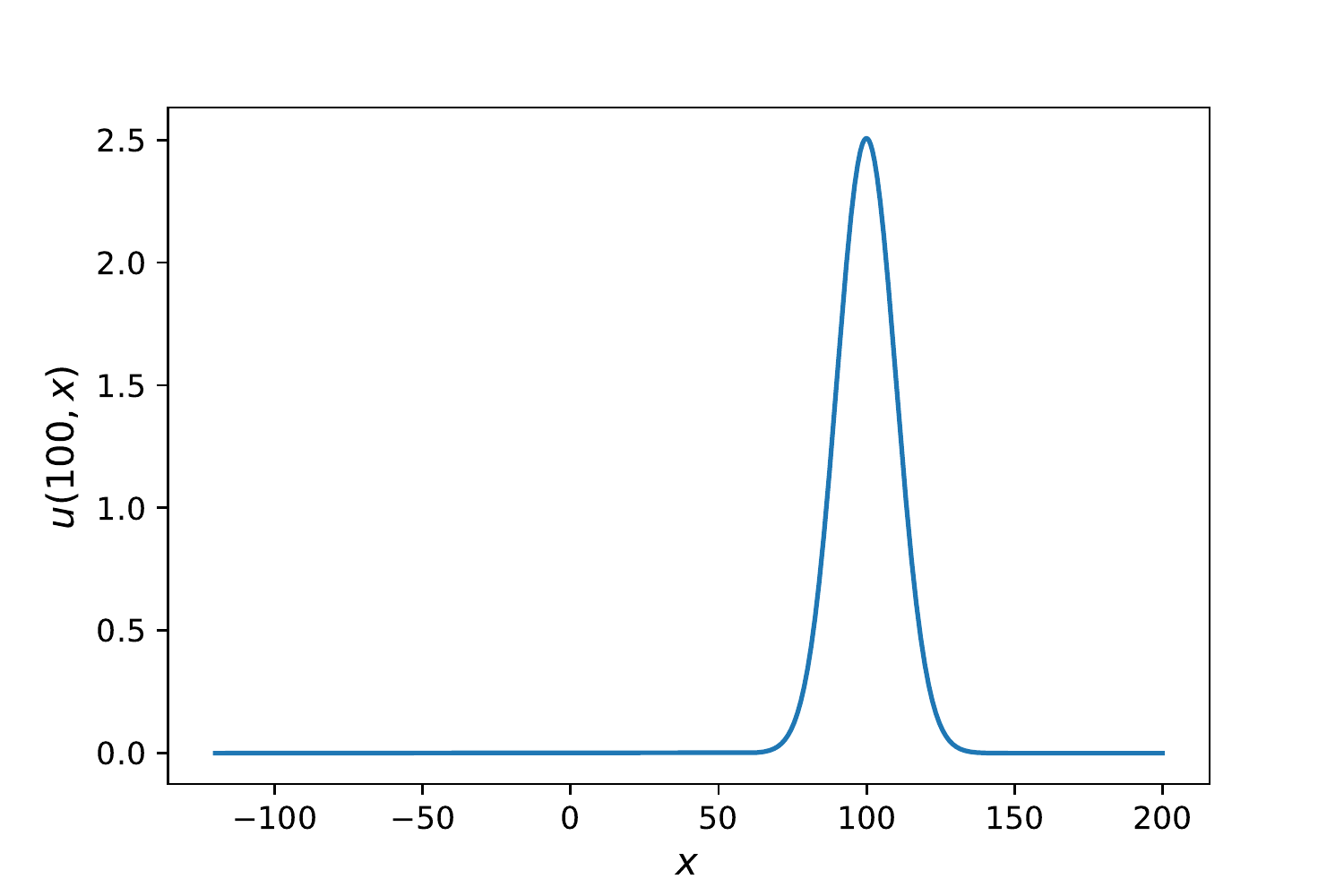}
\end{figure}
\newpage
In Fig. \ref{fig:9}
we consider
\begin{equation}
\alpha=9/10.
\end{equation}
\begin{figure}[ht!]
\centering
\includegraphics[scale=0.4]{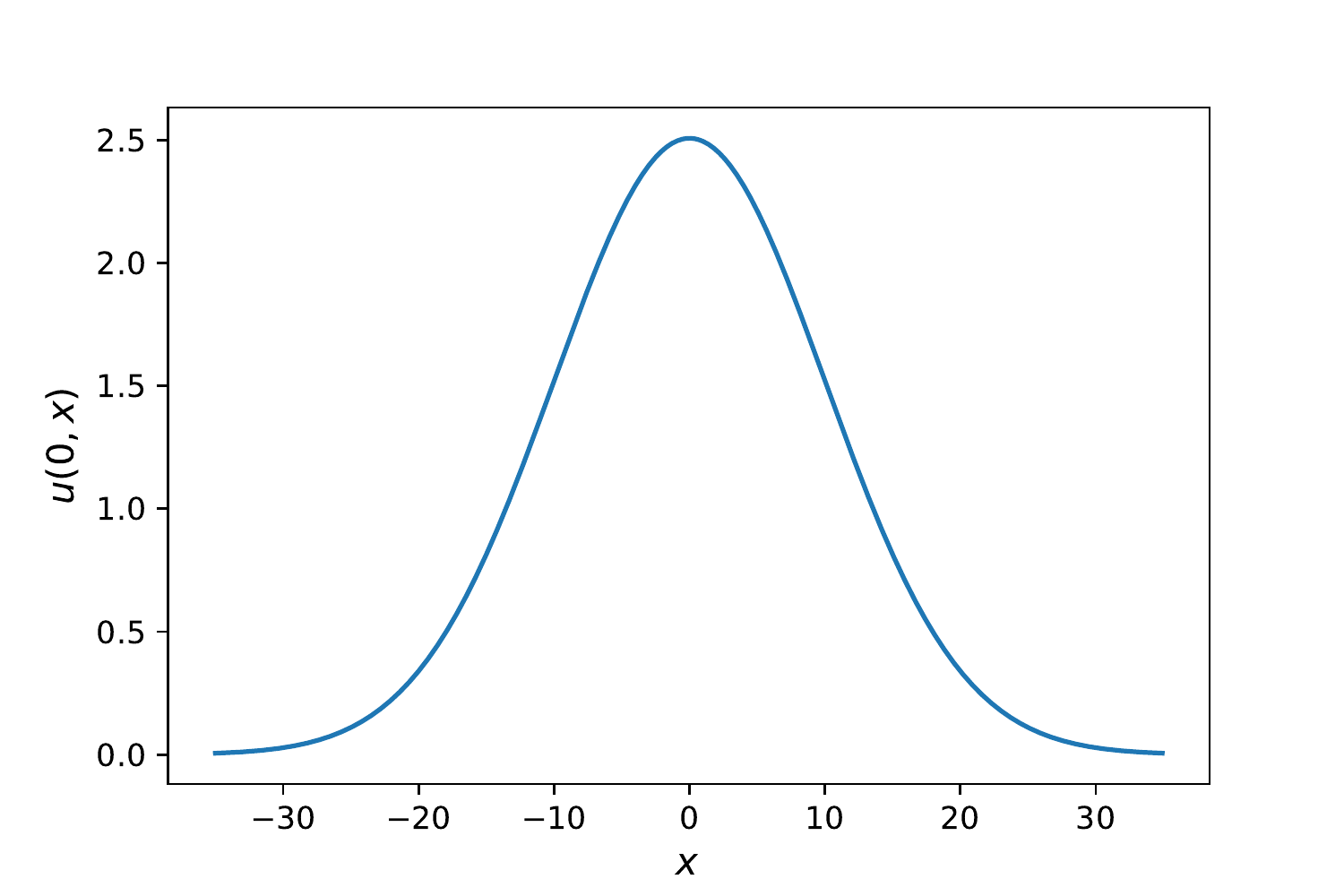}
\includegraphics[scale=0.4]{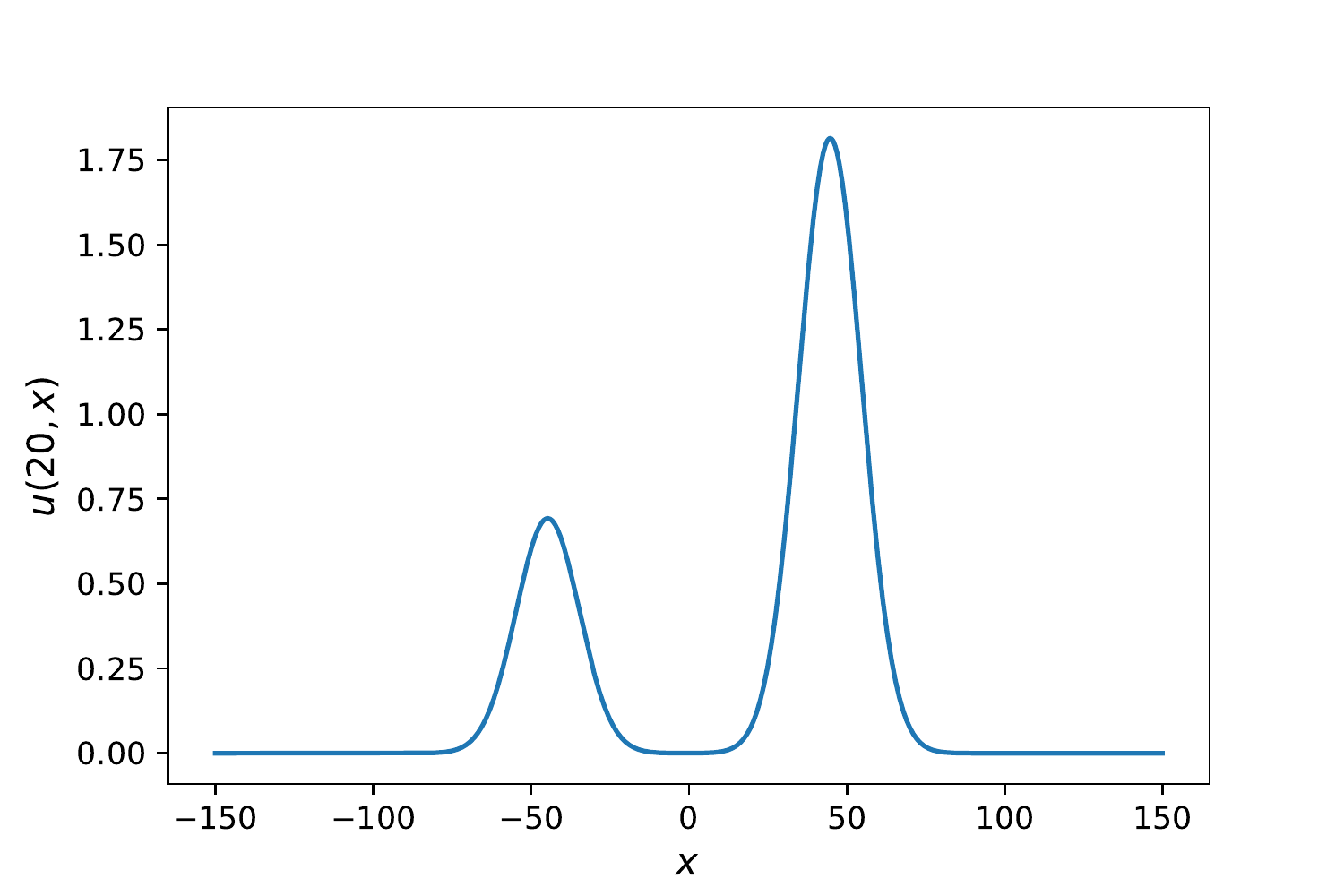}
\end{figure}

From the previous numerical analysis, we emphasize an important feature affecting the propagation problem under study. Indeed, all the previous plots clearly put in evidence the existence of a propagation velocity $v_p$ bounded by
\begin{equation}
v_p\le \frac{\delta^{1-\alpha}}{\sqrt{2\left( 1-\alpha \right)}}=:v_{max}\,.
\end{equation}
Heuristically, one can appreciate that for every $t$, the solutions live {\it almost entirely} within the region $|x|\le v_{max} t$. The physical reason behind the value of $v_{max}$ stands in the fact that this is the upper bound of the group velocity $v_g$, namely no frequency can travel at a propagation velocity higher than $v_{max}$ itself. In the next section, we will investigate in detail the finite propagation velocity ruling the purely hyperbolic case.

\section{Hyperbolic-like propagation} 
\label{hyperb}
This section is devoted to the special case when the solution is a (possibly approximated)
traveling wave. For this case we choose
\begin{equation}
\sigma\gg\delta\quad\text{and}\quad v=\frac{\delta^{1-\alpha}}{\sqrt{2(1-\alpha)}}.
\end{equation}
In particular, we study the case
\begin{equation}
\sigma=10\quad\text{and}\quad\delta=1.
\end{equation}

\begin{figure}[ht!]
\centering
\includegraphics[scale=0.4]{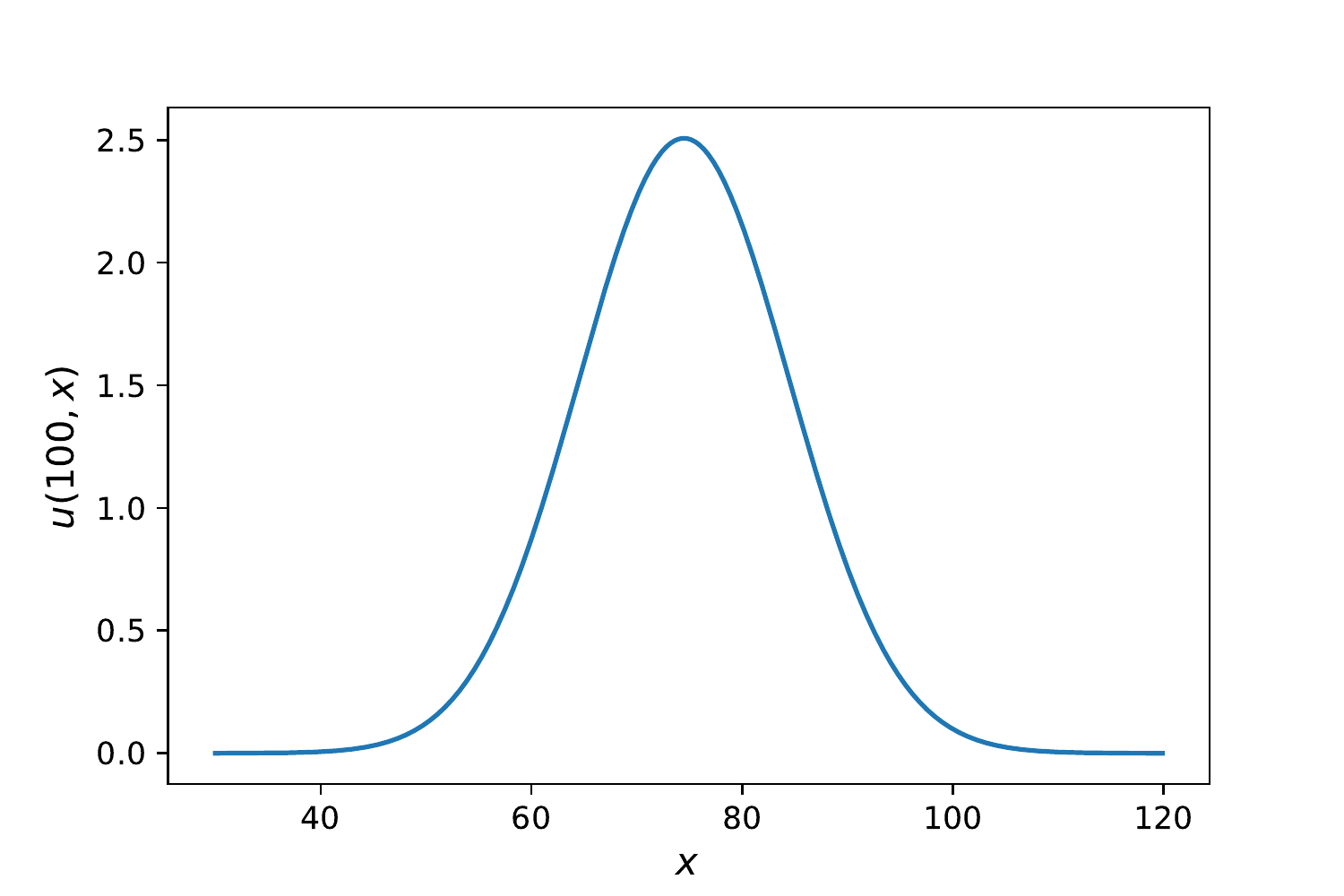}
\includegraphics[scale=0.4]{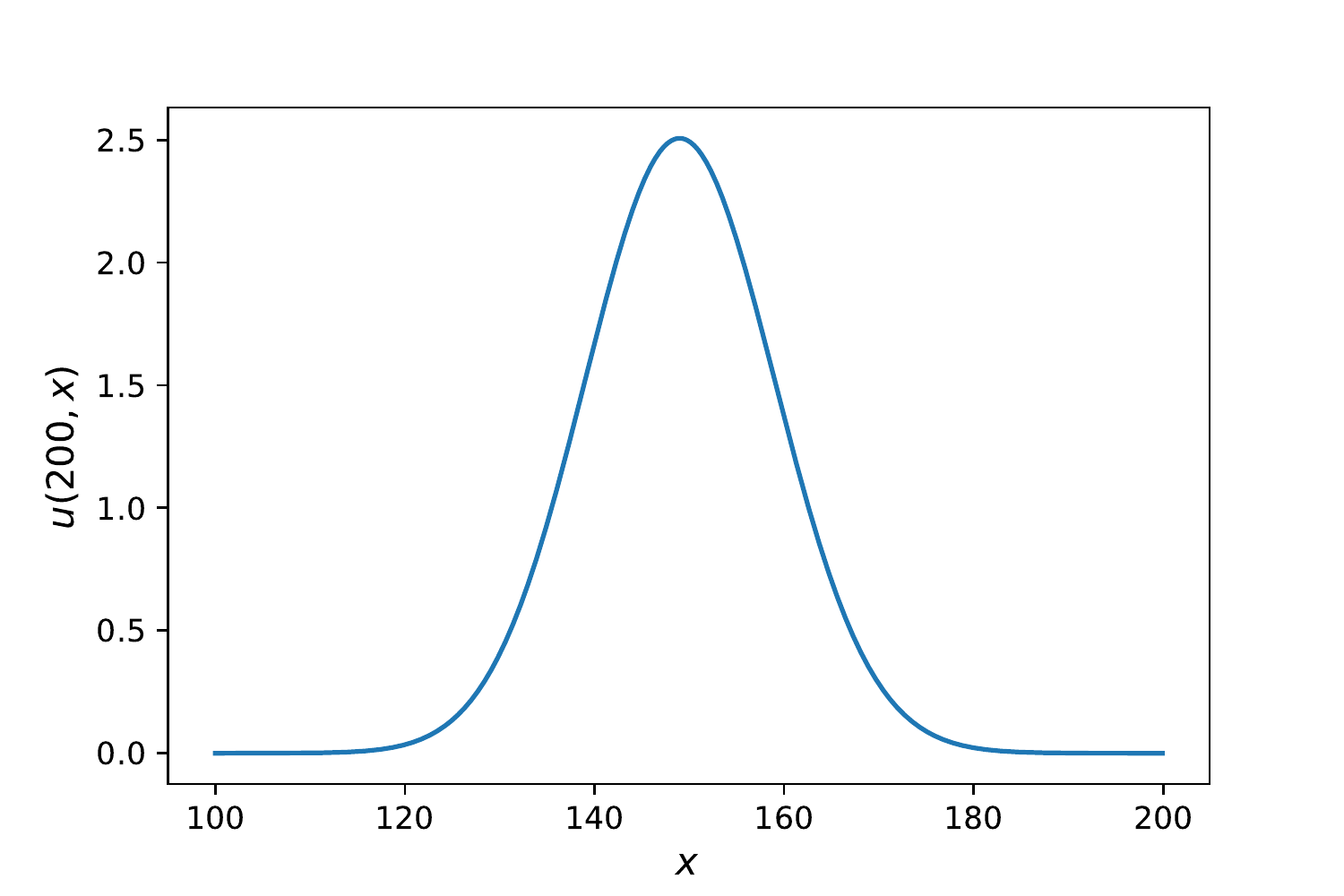}
\caption{Numerical solution at different times. We imposed Gaussian initial conditions with $\sigma=10\,\delta$ and initial velocity $v=\delta^{1-\alpha}/\sqrt{2(1-\alpha)}$. This case refers to the parameters $\alpha=1/10$ and $\delta=1$.}
\label{fig:10}
\end{figure}

In Fig. \ref{fig:10}
we consider
\begin{equation}
\alpha=10^{-1}.
\end{equation}
\begin{figure}[ht!]
\centering
\includegraphics[scale=0.4]{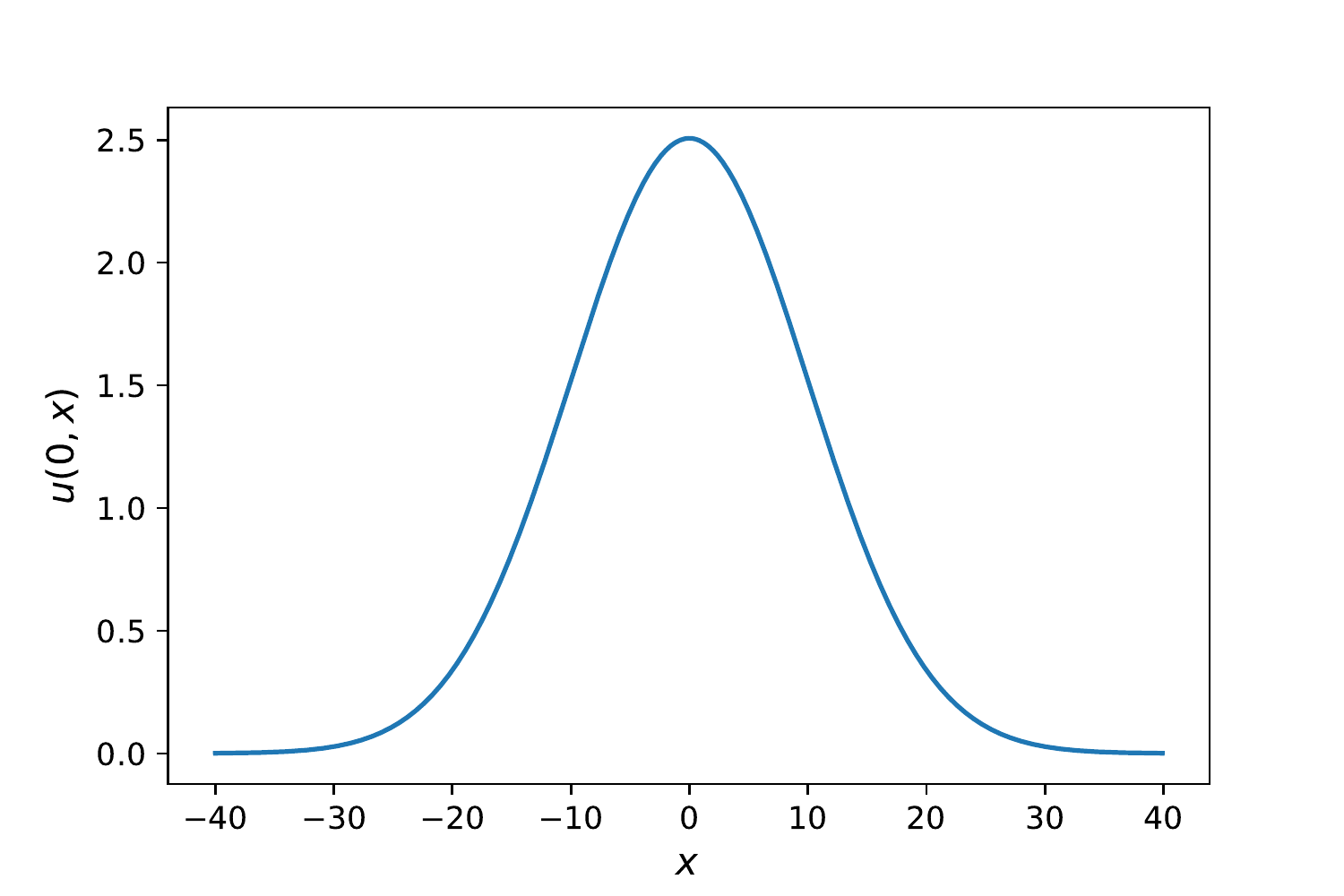}
\includegraphics[scale=0.4]{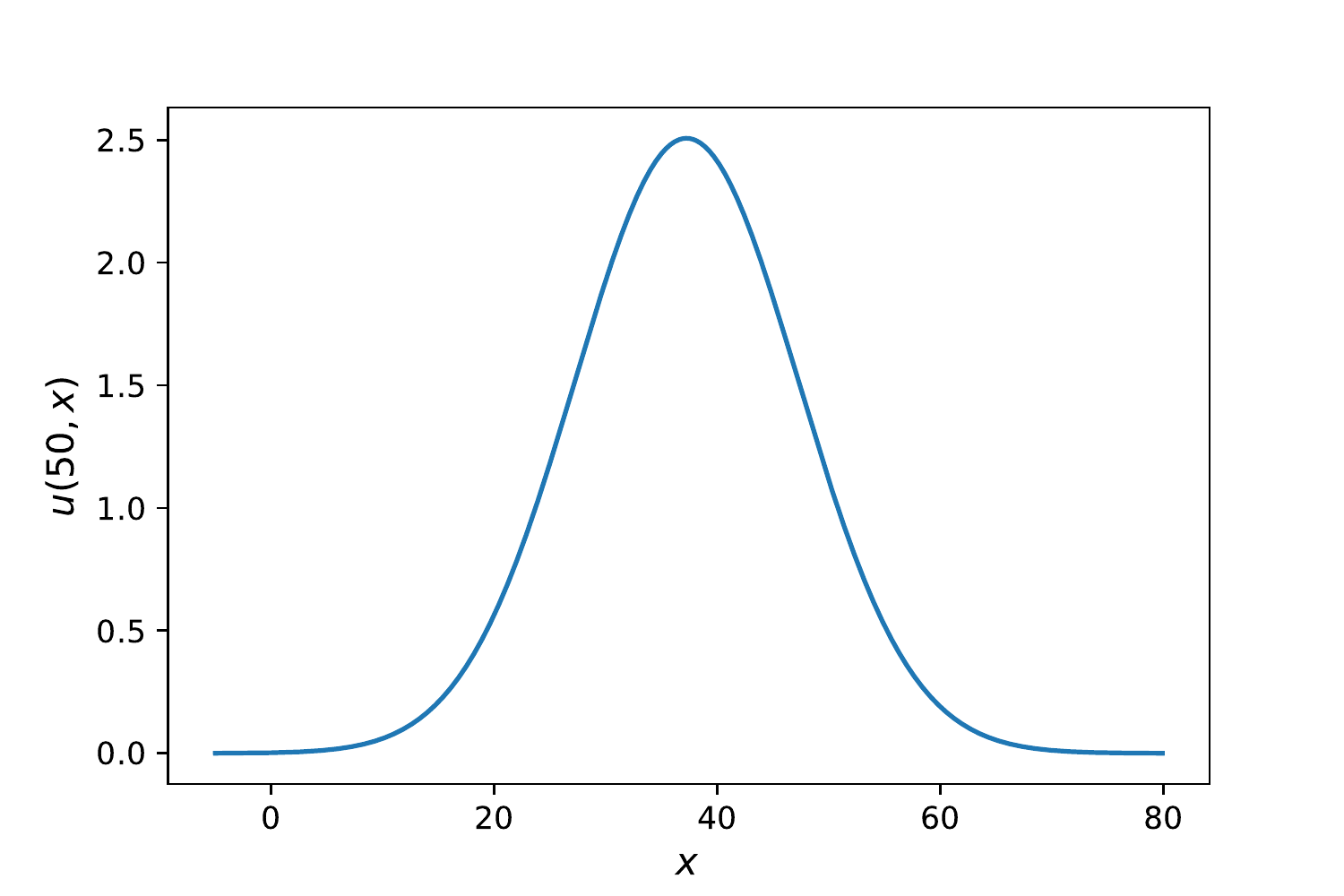}
\end{figure}

\begin{figure}[ht!]
\centering
\includegraphics[scale=0.4]{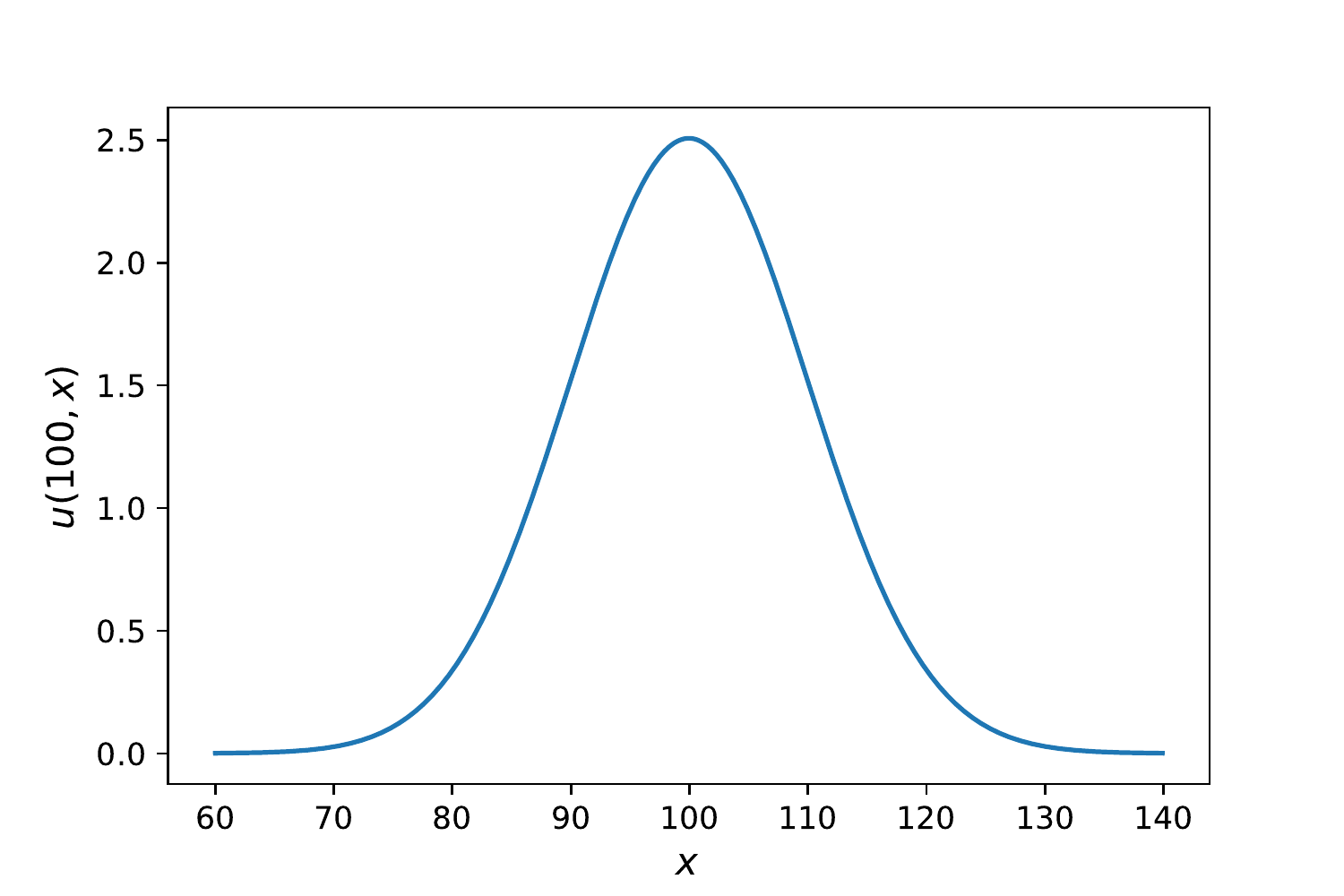}
\includegraphics[scale=0.4]{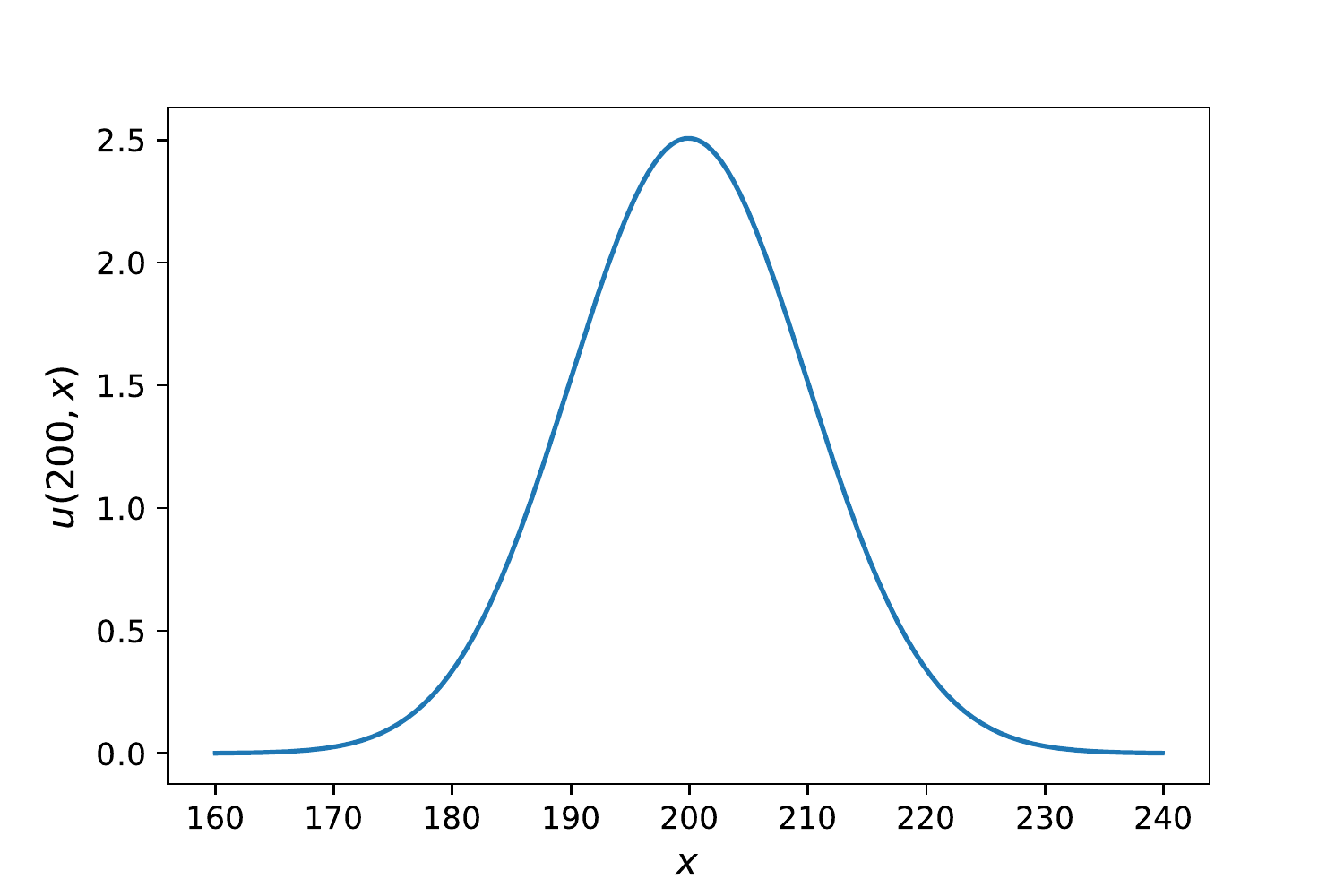}
\caption{Numerical solution at different times. We imposed Gaussian initial conditions with $\sigma=10\,\delta$ and initial velocity $v=\delta^{1-\alpha}/\sqrt{2(1-\alpha)}$. This case refers to the parameters $\alpha=1/2$ and $\delta=1$.}
\label{fig:11}
\end{figure}

In Fig. \ref{fig:11}
we consider
\begin{equation}
\alpha=1/2.
\end{equation}
\begin{figure}[ht!]
\centering
\includegraphics[scale=0.4]{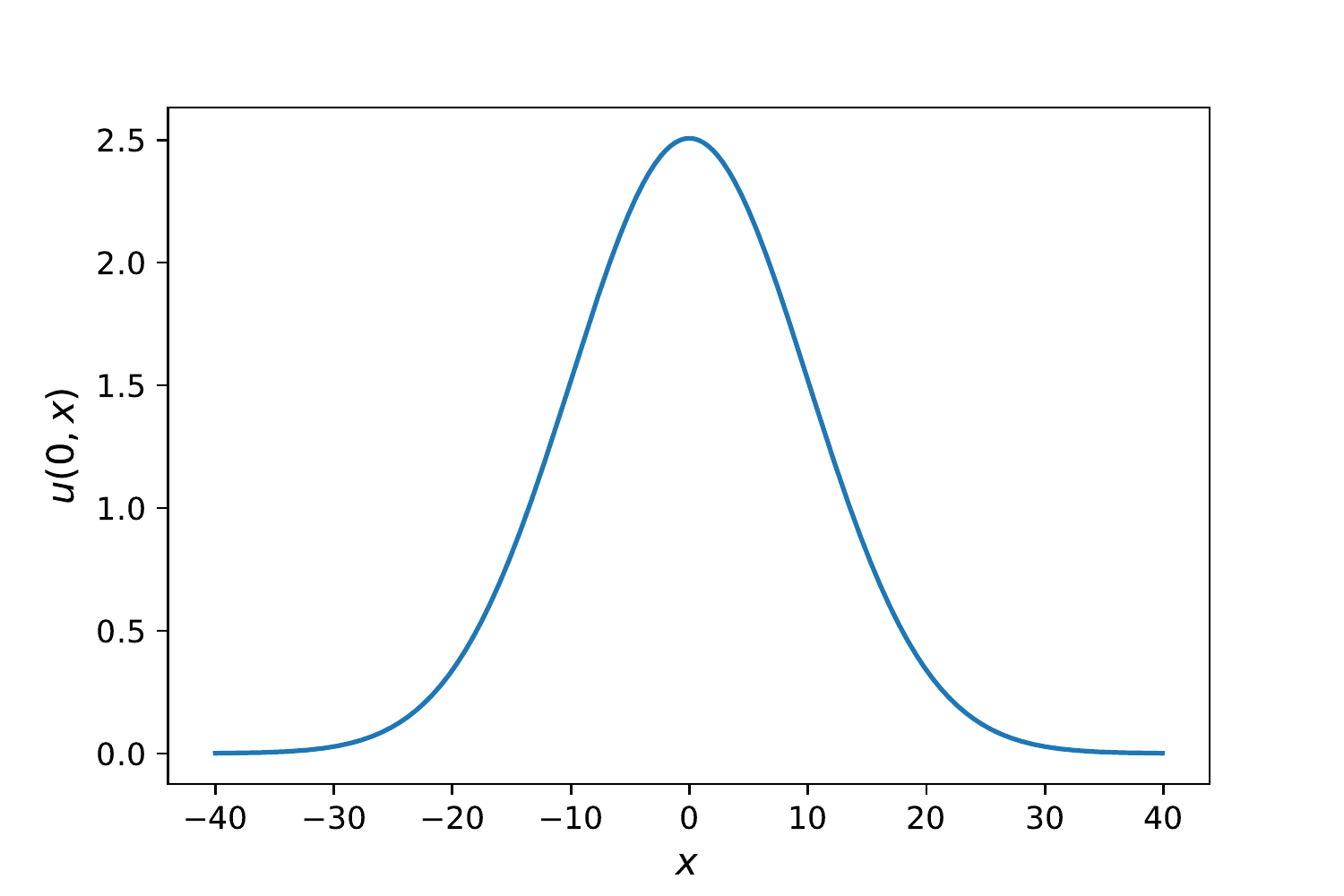}
\includegraphics[scale=0.4]{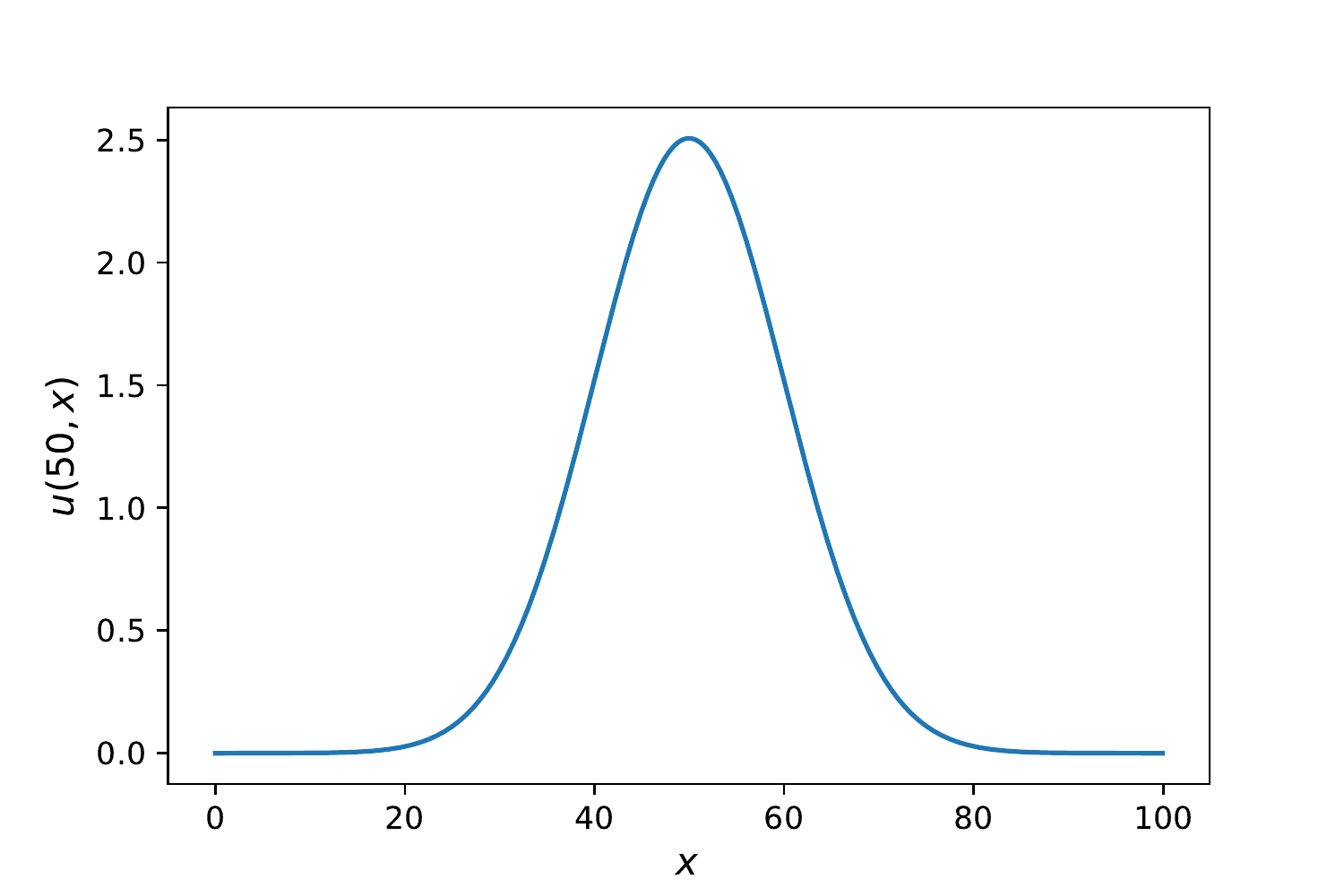}
\end{figure}

\begin{figure}[ht!]
\centering
\includegraphics[scale=0.4]{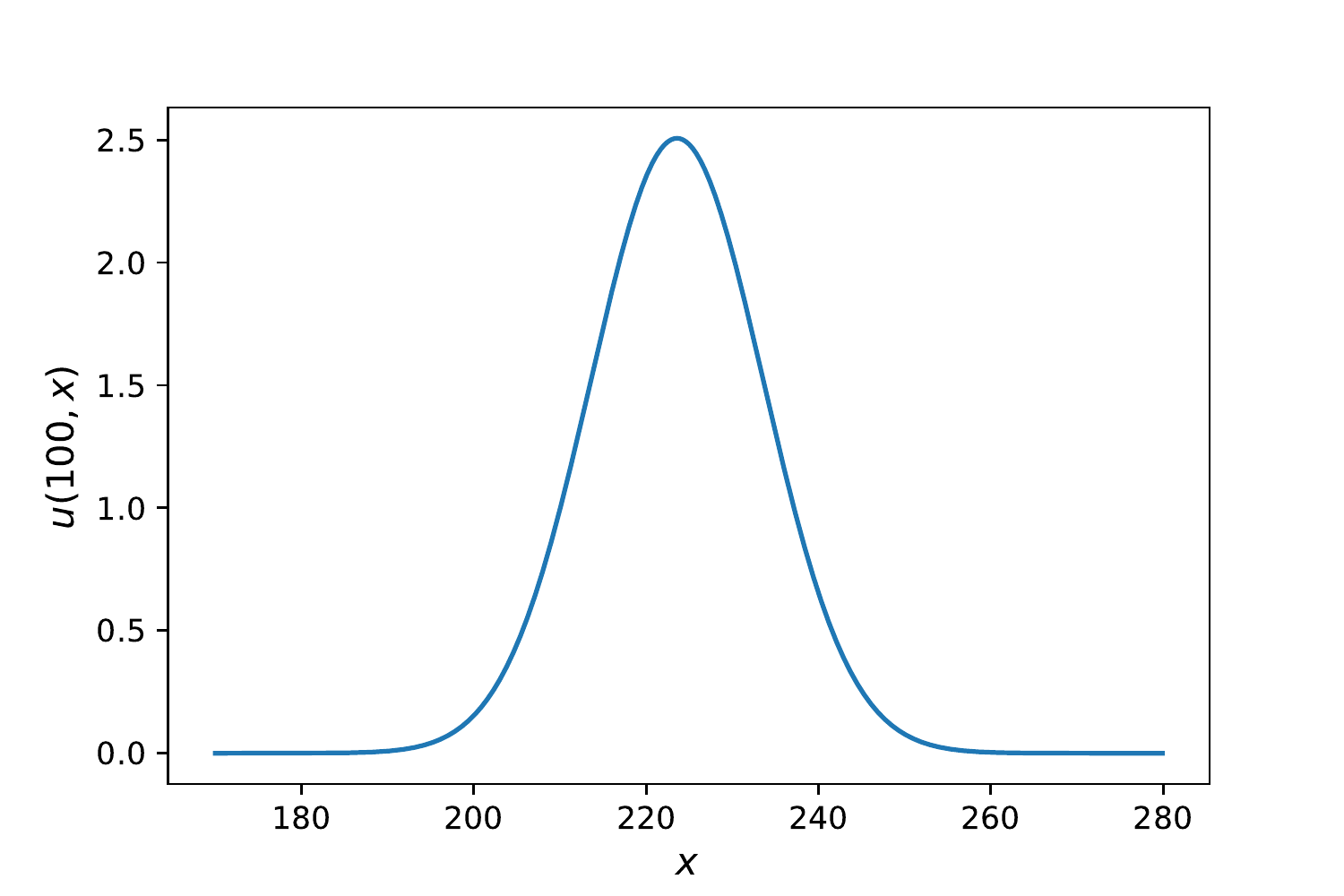}
\includegraphics[scale=0.4]{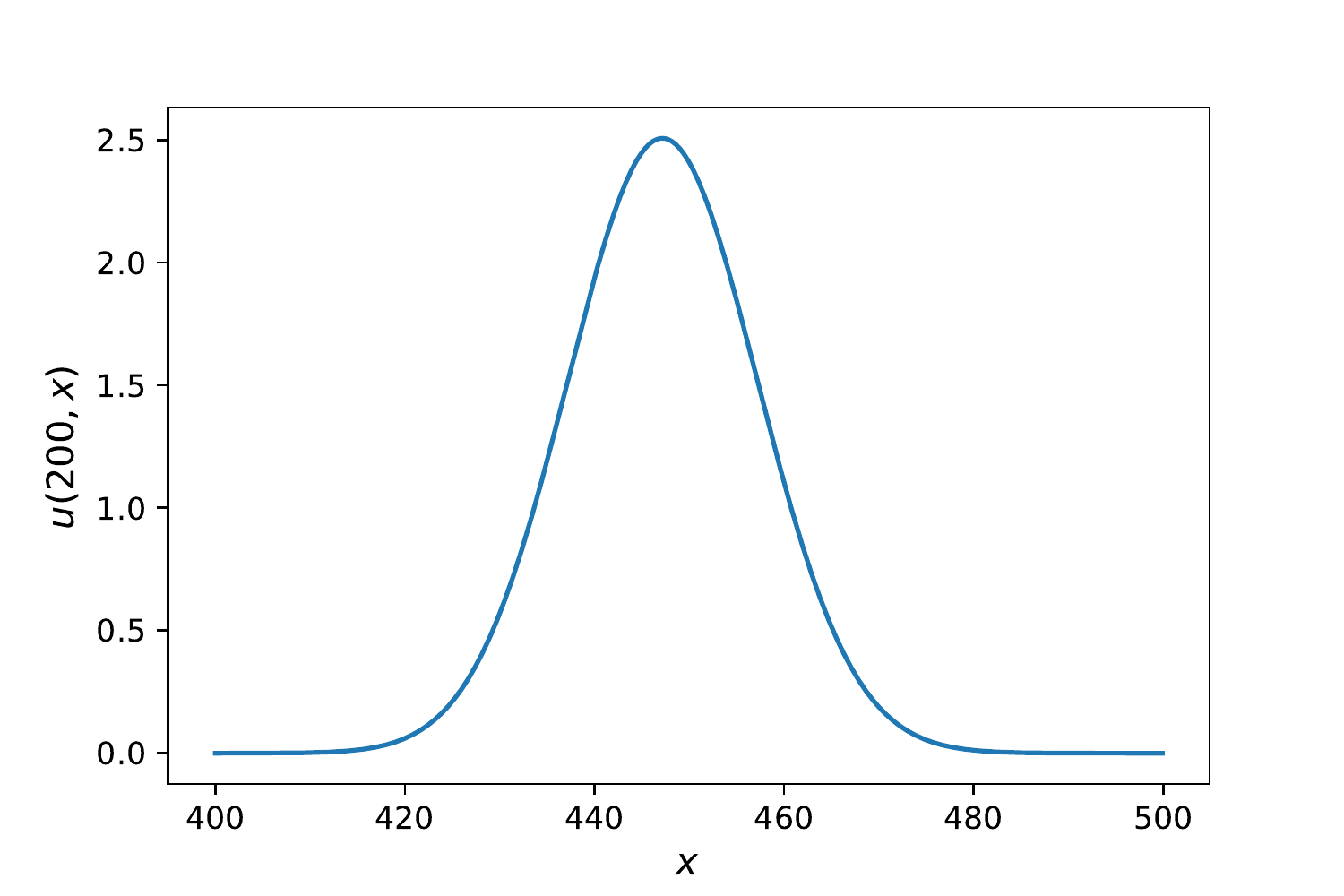}
\caption{Numerical solution at different times. We imposed Gaussian initial conditions with $\sigma=10\,\delta$ and initial velocity $v=\delta^{1-\alpha}/\sqrt{2(1-\alpha)}$. This case refers to the parameters $\alpha=9/10$ and $\delta=1$.}
\label{fig:12}
\end{figure}

In Fig. \ref{fig:12}
we consider
\begin{equation}
\alpha=9/10.
\end{equation}
\begin{figure}[ht!]
\centering
\includegraphics[scale=0.4]{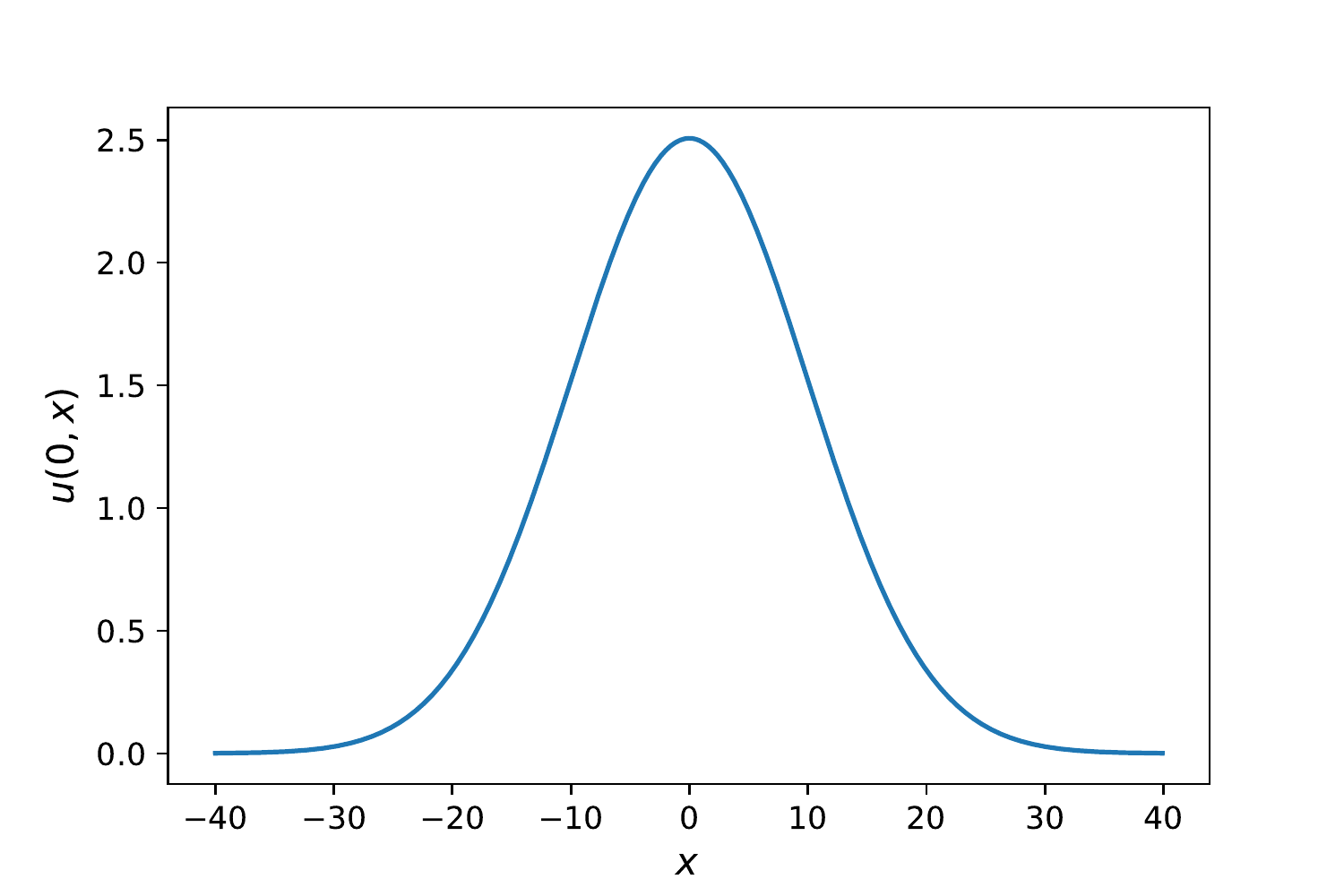}
\includegraphics[scale=0.4]{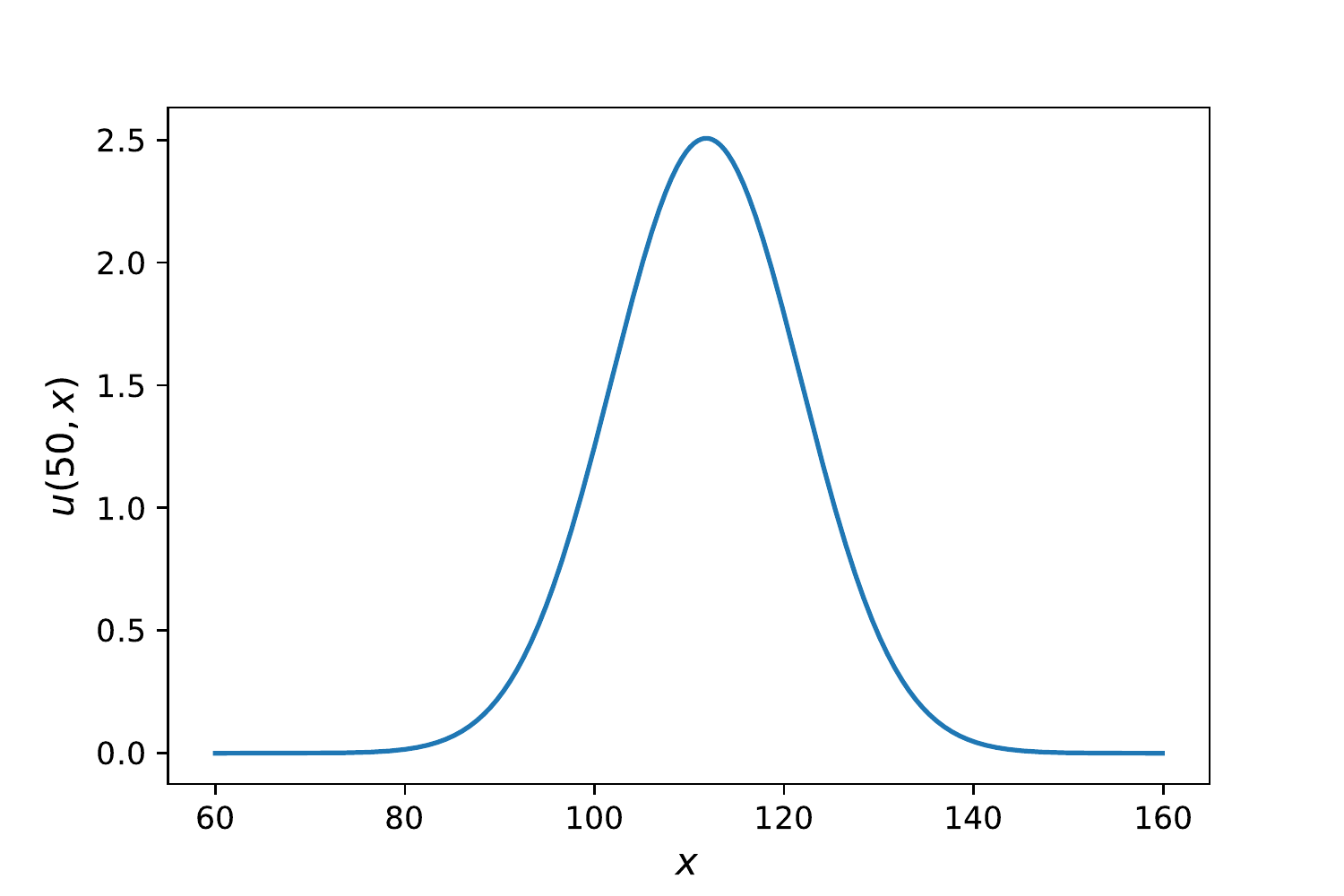}
\end{figure}

\section{Singularity formation}
\label{sing}
In this section we explore the possible spontaneous creation of singularities.
Following the analysis performed in  \cite{AB1, AB2} one can try to study the following Riemann-like problem augmenting \eqref{eq:linear_per} with the following initial conditions
\begin{equation}
v_0(x)=0\qquad\text{and}\qquad
v_1(x)=\begin{cases}
v_+\qquad\text{if}\qquad x\ge 0\\
v_-\qquad\text{if}\qquad x< 0.
\end{cases}\,
\label{eq:41}
\end{equation}
Due to the fact that the Fourier transform of $v_1$ 
\begin{equation*}
\widehat{v_1}(\xi)=v_+\left(\pi\,\delta_0(\xi)+\frac{1}{i\xi}\right)+v_-\left(\pi\,\delta_0(-\xi)-\frac{1}{i\xi}\right)
\end{equation*}
exhibits a strong singularity at $\xi=0$,
rather than using the representation formula \eqref{eq:sol} in this case,
we confine our analysis to some numerical experiments.
For this, we truncate \eqref{eq:41} and choose $v_-=0$ and $v_+=1$, namely
\begin{equation}
v_0(x)=0\qquad\text{and}\qquad
v_1(x)=\chi_{[0,1]}(x)\,.
\label{eq:49}
\end{equation}
Now we have
\begin{equation*}
\widehat{v_1}(\xi)=\frac{i}{\xi}\left(1-e^{i\xi}\right),
\end{equation*}
which does not exhibit  the Dirac delta at $\xi=0$.

\begin{figure}[ht!]
\centering
\includegraphics[scale=0.8]{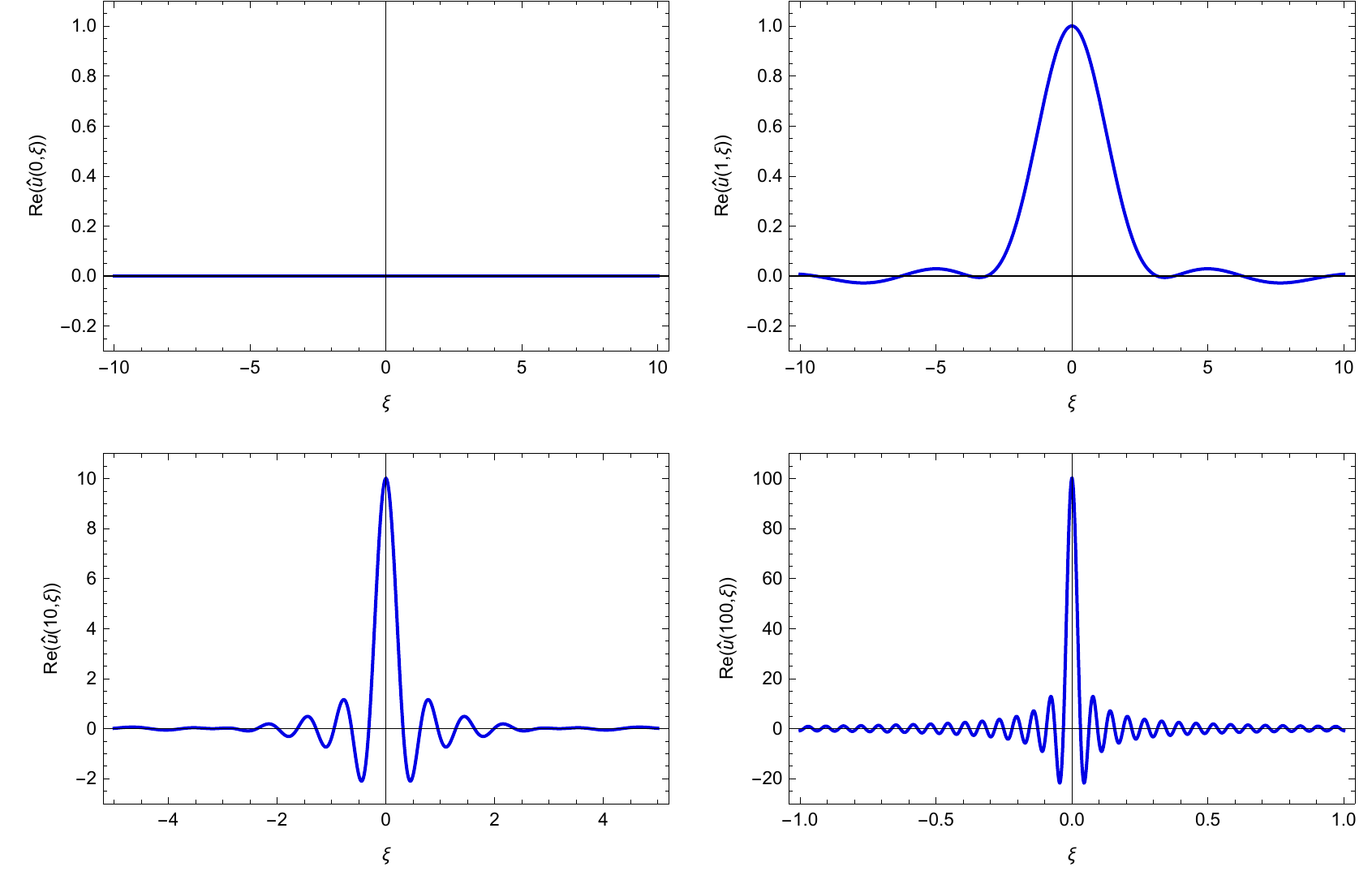}
\caption{Real part of the Fourier transform of the solution with initial conditions \eqref{eq:49} at different times. This case refers to the parameters $\alpha=1/2$ and $\delta=1$.}
\label{fig:re}
\end{figure}

\begin{figure}[ht!]
\centering
\includegraphics[scale=0.8]{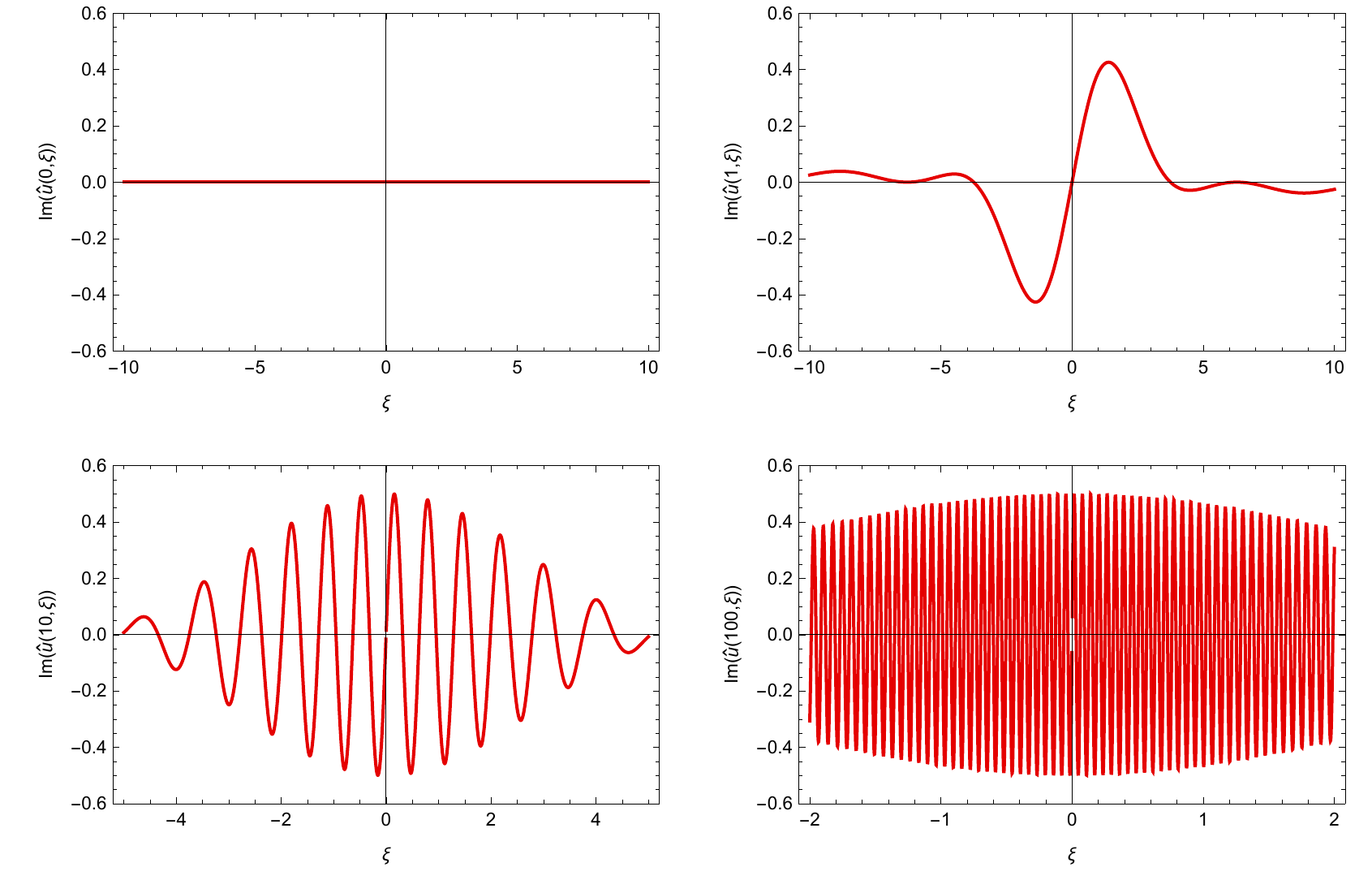}
\caption{Imaginary part of the Fourier transform of the solution with initial conditions \eqref{eq:49} at different times. This case refers to the parameters $\alpha=1/2$ and $\delta=1$.}
\label{fig:im}
\end{figure}

In Figures \ref{fig:re} and \ref{fig:im} we plot the real and imaginary parts of the Fourier transform of the solution expressed in \eqref{eq:sol}
with initial data given in  \eqref{eq:49}.
We can appreciate that as time evolves the real part develops a peak at $\xi=0$ whereas the imaginary one becomes highly oscillating.
Inspired by the analysis carried out
in \cite{AB2} we observe numerically
a possible development of a stationary singularity at $x=0$.

\section{Conclusion}
In this paper, we have explored the features of the peridynamic model proposed in \cite{Coclite_2018,CDFMV} as a viable generalization of the wave equation for the study of nonlocal phenomena in the mechanics of continuous media. We have focused our phenomenological analysis to the 1-D linear scenario. Despite its simplicity, this case study already exhibits deep differences with respect to the standard wave equation. In particular, dispersive propagation can occur when small scales dynamics is non-negligible.

This competitive behavior between classical hyperbolic propagation and dispersive phenomena is intimately related to the multiscale essence of the peridynamic framework and the crossover from one regime to the other one is dictated by the interaction length $\delta$ and the interaction scaling $\alpha$. Moreover, we have also shown that the energy propagates with the group velocity, regardless of the scale lengths involved in the dynamics.

From these general features, we have then exploited the phenomenology of the case of Gaussian initial condition with non-trivial velocity. In this scenario, our numerical investigation has shown that dispersive phenomena are relevant when the width $\sigma$ of the Gaussian initial profile is smaller than the peridynamic radius $\delta$. On the contrary, when  $\sigma$ is much greater than $\delta$, our numerical investigation exhibits a hyperbolic-like propagation. This regime can be interpreted as a direct consequence of the multiscale nature of the dispersive relation $\omega(\xi)$. This is nearly linear when $\xi\ll\delta^{-1}$ and grows with a sub-linear rate when $\xi\gg\delta^{-1}$.

As a final comparison between the hyperbolic and dispersive behaviors, for the case $\sigma\gg\delta$, we have shown the existence of a typical initial velocity of the Gaussian initial condition leading to traveling wave-like solutions. Once again, the cornerstone to understand this behavior is the dispersive relation $\omega(\xi)$: indeed this fine-tuned choice for the initial velocity is given by the limit for $\xi\to 0$ of the group velocity $\omega'(\xi)$, which is a constant, as a consequence of the linearity of $\omega(\xi)$ on large scale. This limit is also responsible for the almost
 finite propagation of the solutions over time.

Finally, in the last section, we have discussed the possible spontaneous formation of singularities developed in an initial-value problem with discontinuous velocity datum.

\bibliographystyle{abbrv}
\bibliography{CFP-ref}

\begin{thebibliography}{10}

\bibitem{Bi}
M.~A. Biot.
\newblock General theorems on the equivalence of group velocity and energy
  transport.
\newblock {\em Phys. Rev.}, 105:1129--1137, Feb 1957.

\bibitem{CDFMV}
G.~M. Coclite, S.~Dipierro, G.~Fanizza, F.~Maddalena, and E.~Valdinoci.
\newblock Dispersive effects in a peridynamic model, 2021.

\bibitem{Coclite_2018}
G.~M. Coclite, S.~Dipierro, F.~Maddalena, and E.~Valdinoci.
\newblock Wellposedness of a nonlinear peridynamic model.
\newblock {\em Nonlinearity}, 32(1):1--21, Nov 2018.

\bibitem{CDMV}
G.~M. Coclite, S.~Dipierro, F.~Maddalena, and E.~Valdinoci.
\newblock Singularity formation in fractional {B}urgers' equations.
\newblock {\em J. Nonlinear Sci.}, 30(4):1285--1305, 2020.

\bibitem{CFLMP}
G.~M. Coclite, A.~Fanizzi, L.~Lopez, F.~Maddalena, and S.~F. Pellegrino.
\newblock Numerical methods for the nonlocal wave equation of the peridynamics.
\newblock {\em Appl. Numer. Math.}, 155:119--139, 2020.

\bibitem{EP}
E.~Emmrich and D.~Puhst.
\newblock Survey of existence results in nonlinear peridynamics in comparison
  with local elastodynamics.
\newblock {\em Comput. Methods Appl. Math.}, 15(4):483--496, 2015.

\bibitem{E1}
A.~C. Eringen.
\newblock {\em Nonlocal continuum field theories}.
\newblock Springer-Verlag, New York, 2002.

\bibitem{E2}
A.~C. Eringen and D.~G.~B. Edelen.
\newblock On nonlocal elasticity.
\newblock {\em Internat. J. Engrg. Sci.}, 10:233--248, 1972.

\bibitem{G}
M.~E. Gurtin.
\newblock The linear theory of elasticity.
\newblock In {\em Linear theories of elasticity and thermoelasticity}, pages
  1--295. Springer, 1973.

\bibitem{Kr}
E.~Kr\"oner.
\newblock Elasticity theory of materials with long range cohesive forces.
\newblock {\em Internat. J. Solids Structures}, 3(5):731--742, 1967.

\bibitem{Ku}
I.~A. Kunin.
\newblock {\em Elastic media with microstructure. {I}}, volume~26 of {\em
  Springer Series in Solid-State Sciences}.
\newblock Springer-Verlag, Berlin-New York, 1982.
\newblock One-dimensional models, Translated from the Russian.

\bibitem{StrG1}
C.~Lim, G.~Zhang, and J.~Reddy.
\newblock A higher-order nonlocal elasticity and strain gradient theory and its
  applications in wave propagation.
\newblock {\em J. Mech. Phys. Solids}, 78:298--313, 2015.

\bibitem{LP}
L.~Lopez and S.~F. Pellegrino.
\newblock A spectral method with volume penalization for a nonlinear
  peridynamic model.
\newblock {\em Internat. J. Numer. Methods Engrg.}, 122(3):707--725, 2021.

\bibitem{Sill4}
S.~Silling and R.~Lehoucq.
\newblock Peridynamic theory of solid mechanics.
\newblock In H.~Aref and E.~van~der Giessen, editors, {\em Advances in Applied
  Mechanics}, volume~44 of {\em Advances in Applied Mechanics}, pages 73--168.
  Elsevier, 2010.

\bibitem{Sill}
S.~A. Silling.
\newblock Reformulation of elasticity theory for discontinuities and long-range
  forces.
\newblock {\em J. Mech. Phys. Solids}, 48(1):175--209, 2000.

\bibitem{Sill1}
S.~A. Silling.
\newblock Linearized theory of peridynamic states.
\newblock {\em J. Elasticity}, 99(1):85--111, 2010.

\bibitem{Sill2}
S.~A. Silling, M.~Epton, O.~Weckner, J.~Xu, and E.~Askari.
\newblock Peridynamic states and constitutive modeling.
\newblock {\em J. Elasticity}, 88(2):151--184, 2007.

\bibitem{Sill3}
S.~A. Silling and R.~B. Lehoucq.
\newblock Convergence of peridynamics to classical elasticity theory.
\newblock {\em J. Elasticity}, 93(1):13--37, 2008.

\bibitem{AB1}
L.~Wang and R.~Abeyaratne.
\newblock A one-dimensional peridynamic model of defect propagation and its
  relation to certain other continuum models.
\newblock {\em J. Mech. Phys. Solids}, 116:334--349, 2018.

\bibitem{AB2}
O.~Weckner and R.~Abeyaratne.
\newblock The effect of long-range forces on the dynamics of a bar.
\newblock {\em J. Mech. Phys. Solids}, 53(3):705--728, 2005.

\bibitem{Wh}
G.~B. Whitham.
\newblock {\em Linear and nonlinear waves}.
\newblock Pure and Applied Mathematics (New York). John Wiley \& Sons, Inc.,
  New York, 1999.
\newblock Reprint of the 1974 original, A Wiley-Interscience Publication.

\end{thebibliography}

\end{document}